\documentclass[11pt, notitlepage, twoside]{report}
\usepackage{latexsym}
\usepackage[dvips]{graphics,color}
\usepackage{latexsym}
\usepackage[dvips]{graphics,color}
\def\be{\begin{displaymath}}
\def\ee{\end{displaymath}}
\def\bne{\begin{equation}}
\def\ene{\end{equation}}
\def\bee{\begin{eqnarray*}}
\def\eee{\end{eqnarray*}}
\def\bnee{\begin{eqnarray}}
\def\enee{\end{eqnarray}}

\font\openface=msbm10 at10pt
\def\Z{\openface{Z}}
\def\R{\openface{R}}

\def\C{\mathcal{C}} % used for set of causets
\def\M{\openface{M}} % Minkowski space
\def\qed{\; \; \Box}
\def\past{\mathrm{past}}
\def\fut{\mathrm{future}}
\def\lab{\mathrm{label}}
\def\nc2{{N \choose 2}}
\def\Nc2{{N \choose 2}} % too wasteful?

\def\half{\frac{1}{2}}
\def\quart{\frac{1}{4}}
\def\thquart{\frac{3}{4}}
\def\E#1{<\!#1\!>}
\def\interval{\mathrm{int}}

\def\iff{\Leftrightarrow}
\def\poscau{{\cal P}}
\def\pp{\varpi}
\def\nexists{\not\hskip-2pt\exists\,} 
\def\Pr{{\rm Pr}}

\def\implies{\Rightarrow}
\def\since{\Leftarrow}

\def\lto{\mathop
        {\hbox{${\lower3.8pt\hbox{$<$}}\atop{\raise0.2pt\hbox{$\sim$}}$}}}
\def\gto{\mathop
        {\hbox{${\lower3.8pt\hbox{$>$}}\atop{\raise0.2pt\hbox{$\sim$}}$}}}

\newtheorem{lemma}{Lemma}

\def\eprint #1 {$\langle$e-print arXive: #1$\rangle$}

\font\openface=msbm10 at11pt

\def\NaturalNumbers{\hbox{\openface N}} 
\def\R{\hbox{\openface R}}

\def\Minkowski{\hbox{\openface M}}

\def\=>{\Rightarrow}
\def\==>{\Longrightarrow}

 \def\dal{\displaystyle{{\hbox to 0pt{$\sqcup$\hss}}\sqcap}}

\def\lto{\mathop
        {\hbox{${\lower3.8pt\hbox{$<$}}\atop{\raise0.2pt\hbox{$\sim$}}$}}}
\def\gto{\mathop
        {\hbox{${\lower3.8pt\hbox{$>$}}\atop{\raise0.2pt\hbox{$\sim$}}$}}}
\def\half{{1\over 2}}

\def\less{\backslash}		% symbol for set-theoretic difference

\def\to{\rightarrow}		% symbol for `approaches' or `maps to'

\def\tilde{\widetilde}		% define tilde to always be the ``widetilde'' 
\def\bar{\overline}		% define bar to always be wide bar
\def\hat{\widehat}		% define hat to always be the ``widehat'' 

\def\ideq{\equiv}		% triple equal sign

\def\interior #1 {  \buildrel\circ\over  #1}     % seems to work
\def\union{\cup}

\def\twoch{\,\,\begin{picture}(0,1) % ``2-chain''
\thicklines
\multiput(0,0)(0,1){2}{\circle*{.6}}
\put(0,0){\line(0,1){1}}
\end{picture}\,\,}

\def\twoach{\,\begin{picture}(1,1) % ``2-antichain''
\thicklines
\multiput(0,.5)(1,0){2}{\circle*{.6}}
\end{picture}\,\,\,}

\def\threech{\,\,\begin{picture}(0,2) % threech ``3-chain''
\thicklines
\multiput(0,0)(0,1){3}{\circle*{.6}}
\put(0,0){\line(0,1){2}}
\end{picture}\,\,}

\def\wedge{\,\,\begin{picture}(2,2) % wedge ``Lambda''
\thicklines
\put(1,2){\circle*{.6}}
\multiput(0,0)(2,0){2}{\circle*{.6}}
\put(0,0){\line(1,2){1}}
\put(2,0){\line(-1,2){1}}
\end{picture}\,\,}

\def\Lcauset{\,\,\begin{picture}(2,2) % Lcauset ``L''
\thicklines
\multiput(0,0)(1.5,0){2}{\circle*{.6}}
\put(0,2){\circle*{.6}}
\put(0,0){\line(0,1){2}}
\end{picture}\,\,}

\def\V{\,\,\begin{picture}(2,2) % V ``V''
\thicklines
\put(1,0){\circle*{.6}}
\multiput(0,2)(2,0){2}{\circle*{.6}}
\put(1,0){\line(-1,2){1}}
\put(1,0){\line(1,2){1}}
\end{picture}\,\,}

\def\threeach{\,\begin{picture}(2,1) % threeach ``3-antichain''
\thicklines
\multiput(0,.5)(1,0){3}{\circle*{.6}}
\end{picture}\,\,}

\def\fourch{\,\,\begin{picture}(0,3) % fourch ``4-chain''
\thicklines
\multiput(0,0)(0,1){4}{\circle*{.6}}
\put(0,0){\line(0,1){3}}
\end{picture}\,\,}

\def\fourach{\,\,\begin{picture}(3,1) % fourach ``4-antichain''
\thicklines
\multiput(0,.5)(1,0){4}{\circle*{.6}}
\end{picture}\,\,}

\def\Y{\,\,\begin{picture}(2,2) % Y
\thicklines
\multiput(0,0)(0,1){3}{\circle*{.6}}
\put(2,2){\circle*{.6}}
\put(0,0){\line(0,1){2}}
\put(0,1){\line(2,1){2}}
\end{picture}\,\,}

\def\iY{\,\,\begin{picture}(2,2) % iY
\thicklines
\multiput(0,0)(0,1){3}{\circle*{.6}}
\put(2,0){\circle*{.6}}
\put(0,0){\line(0,1){2}}
\put(0,1){\line(2,-1){2}}
\end{picture}\,\,}

\def\threecho{\,\,\begin{picture}(1,2) % threecho ``31''
\thicklines
\multiput(0,0)(0,1){3}{\circle*{.6}}
\put(1,0){\circle*{.6}}
\put(0,0){\line(0,1){2}}
\end{picture}\,\,}

\def\Lo{\,\,\begin{picture}(2,1) % Lo ``L1''
\thicklines
\multiput(0,0)(1,0){3}{\circle*{.6}}
\put(0,1){\circle*{.6}}
\put(0,0){\line(0,1){1}}
\end{picture}\,\,}

\def\new{\,\,\begin{picture}(2,2) % new ``nu''
\thicklines
\multiput(0,0)(0,1){3}{\circle*{.6}}
\put(2,1){\circle*{.6}}
\put(0,0){\line(0,1){2}}
\put(0,0){\line(2,1){2}}
\end{picture}\,\,}

\def\inu{\,\,\begin{picture}(2,2) % inu
\thicklines
\multiput(0,0)(0,1){3}{\circle*{.6}}
\put(2,1){\circle*{.6}}
\put(0,0){\line(0,1){2}}
\put(2,1){\line(-2,1){2}}
\end{picture}\,\,}

\def\diamond{\,\,\begin{picture}(2,2) % diamond
\thicklines
\multiput(0,0)(0,1){3}{\circle*{.6}}
\put(2,1){\circle*{.6}}
\put(0,0){\line(0,1){2}}
\put(0,0){\line(2,1){2}}
\put(2,1){\line(-2,1){2}}
\end{picture}\,\,}

\def\flower{\,\,\begin{picture}(2,2) % flower
\thicklines
\multiput(0,2)(1,0){3}{\circle*{.6}}
\put(1,0){\circle*{.6}}
\put(1,0){\line(-1,2){1}}
\put(1,0){\line(0,1){2}}
\put(1,0){\line(1,2){1}}
\end{picture}\,\,}

\def\iflower{\,\,\begin{picture}(2,2) % iflower
\thicklines
\multiput(0,0)(1,0){3}{\circle*{.6}}
\put(1,2){\circle*{.6}}
\put(0,0){\line(1,2){1}}
\put(1,0){\line(0,1){2}}
\put(2,0){\line(-1,2){1}}
\end{picture}\,\,}

\def\wedgeo{\,\begin{picture}(3,2) % wedgeo ``Lambda + 1''
\thicklines
\put(1,2){\circle*{.6}}
\put(3,0){\circle*{.6}}
\multiput(0,0)(2,0){2}{\circle*{.6}}
\put(0,0){\line(1,2){1}}
\put(2,0){\line(-1,2){1}}
\end{picture}\,\,\,}

\def\Vo{\,\begin{picture}(3,2) % Vo ``V + 1''
\thicklines
\multiput(0,2)(2,0){2}{\circle*{.6}}
\multiput(1,0)(2,0){2}{\circle*{.6}}
\put(1,0){\line(-1,2){1}}
\put(1,0){\line(1,2){1}}
\end{picture}\,\,}

\def\pie{\,\,\begin{picture}(1,1) % pie ``Pi''
\thicklines
\multiput(0,0)(0,1){2}{\circle*{.6}}
\multiput(1,0)(0,1){2}{\circle*{.6}}
\put(0,0){\line(0,1){1}}
\put(1,0){\line(0,1){1}}
\end{picture}\,\,}

\def\N{\,\,\begin{picture}(1,2) % N
\thicklines
\multiput(0,0)(0,2){2}{\circle*{.6}}
\multiput(1,0)(0,2){2}{\circle*{.6}}
\put(0,0){\line(0,1){2}}
\put(1,0){\line(0,1){2}}
\put(0,0){\line(1,2){1}}
\end{picture}\,\,}

\def\bowtie{\,\,\begin{picture}(2,2) % bowtie
\thicklines
\multiput(0,0)(0,2){2}{\circle*{.6}}
\multiput(2,0)(0,2){2}{\circle*{.6}}
\put(0,0){\line(0,1){2}}
\put(2,0){\line(0,1){2}}
\put(0,0){\line(1,1){2}}
\put(2,0){\line(-1,1){2}}
\end{picture}\,\,}

\def\No{\,\begin{picture}(3,2) % No ``N + 1''
\thicklines
\put(3,0){\circle*{.6}}
\multiput(0,0)(2,0){2}{\circle*{.6}}
\multiput(0,2)(2,0){2}{\circle*{.6}}
\put(0,0){\line(0,1){2}}
\put(0,0){\line(1,1){2}}
\put(2,0){\line(0,1){2}}
\end{picture}\,\,\,}

\sffamily % tried to change the font, but it doesn't seem to do anything
\renewcommand{\baselinestretch}{2}
\def\singlesp{\renewcommand{\baselinestretch}{1}\footnotesize}
\begin{document}
\pagenumbering{roman}
\pagestyle{headings} % have it give section rather than chapter title ?
% I think that I would have to do \markright{sec name} after each
% section declaration!  I won't bother for now.
% Though with twoside option it works well...

\vspace*{25mm}
\begin{center}
\renewcommand{\baselinestretch}{1}
\thispagestyle{empty}
%\LARGE Classical Dynamics of Causal Sets\\[14mm]
\LARGE Dynamics of Causal Sets\\[14mm]
\normalsize
by\\[6mm]
\large
David P. Rideout\\[3mm]
\normalsize
B.A.E., Georgia Institute of Technology,\\
Atlanta, GA, 1992\\[2mm]
M.S., Syracuse University, 1995\\[15mm]
DISSERTATION\\[2mm]
Submitted in partial fulfillment of the requirements\\
for the degree of Doctor of Philosophy in Physics\\
in the Graduate School of Syracuse University\\[1mm]
May 2001\\[17mm]
\end{center}

\begin{flushright}
Approved\underline{~~~~~~~~~~~~~~~~~~~~~~~~~~~~~~~~~~~~~~~~}\\[-3mm]
Professor Rafael D. Sorkin~~~~~\\[9mm]
Date\underline{~~~~~~~~~~~~~~~~~~~~~~~~~~~~~~~~~~~~~~~~~~~~}
\end{flushright}

\newpage
\thispagestyle{empty}

%\documentclass[11pt]{article}
%\begin{document}
%\pagestyle{empty}
%\renewcommand{\baselinestretch}{1.55}
%\Large
%\centerline{\bf Abstract}
%\normalsize

The Causal Set approach to quantum gravity 
asserts that spacetime, at its smallest length
scale, has a discrete structure.
This discrete
structure takes the form of a locally finite %(?)
order relation, where the
order, corresponding with the macroscopic notion of spacetime
causality, is taken to be a fundamental aspect of nature.  

%This thesis constructs and investigates a classical dynamics for the causal set
%approach to quantum gravity.  The dynamics is expressed in terms of a
%stochastic process of sequential growth.  The behavior of the dynamics
%is studied both analytically and numerically.

After an introduction to the Causal Set approach, this thesis %begins by
considers a simple toy dynamics for causal
sets.  %, expressed as a classical probability measure.  
Numerical simulations of the model provide evidence 
for the existence of a continuum limit.  % for the model.
While studying this toy
dynamics, a picture arises of how 
the dynamics can be generalized
%one would like to generalize %it
%the toy dynamics 
in such a way that the theory could hope to produce %predict?
more physically realistic causal sets.
By thinking in terms of a stochastic growth process, and positing some
%very basic 
fundamental principles,
we are led almost uniquely to a family of dynamical
laws (stochastic processes) parameterized by a countable sequence of
coupling constants.  % $q_n$ (or equivalently the $t_n$).
This result is quite promising in that we 
%Promising that 
now know how to speak of dynamics for a theory with discrete time.
% Are there other theories with discrete time?  e.g. Ambjorn?
%now in this previously
%UN-understood regime, for a theory with discrete time.
In addition, these dynamics can be expressed in terms of state models
of Ising spins living on the relations of the causal set, which
indicates how non-gravitational matter may arise from the theory
without having to be built in at the fundamental level.

%In addition, 
These results are encouraging in that there exists a
natural way to transform this classical theory, which is expressed in
terms of a probability measure, to a quantum theory, expressed in
terms of a quantum measure.  A sketch as to how one might proceed in
doing this is provided. % in \S \ref{quantum}.
%formulation of QM provides framework for expressing quantum theory
Thus there is good reason to expect that Causal Sets are close to
%constructing
providing
a background independent theory of quantum gravity.

\newpage
\Large
\centerline{\bf Acknowledgements}
\normalsize
%''Preface: (Optional) The preface page briefly indicates the purpose
%of the thesis.  The author may choose to make acknowledgments to
%publishers and persons who have provided special assistance with the
%preparation of the thesis.  [ Since no one ever does preface, I will
%skip it.]
%usually: more professional to less

I would like to express deep gratitude for my 
advisor, Rafael Sorkin,
for his patient teaching and support throughout my graduate career.
%Through him I learned not just physics itself, but also how to 
%approach the process of constructing a physical theory.
%think about teaching me how to think about physics.  
% Mention his making me aware of global politics?  ``things outside
% of physics'' or something?
His depth of insight into fundamental issues in physics is
%into the issues that lie at the core of physical problems, such as quantum
%gravity, 
%has been 
extremely helpful and illuminating.
I also would like to thank Peter Saulson, for acting as my advisor at
a critical stage in my graduate career, providing much needed support
and encouragement.
%I would also like to thank 
Let me also express my appreciation to Fatma Husein
for some very illuminating conversations, 
%Peter Saulson for
%encouragement and support at 
%a critical stage,
%some transitional times, 
Scott Klasky for teaching me
% Also Peter Saulson for sticking up for me at my research oral ``defense''!
how to write efficient code, Asif Qamar for introducing me to
GNU/Linux at a critical time, and Saul Teukolsky for kind hospitality
at Cornell.  
%I would also like to thank 
Thank you to
my 
fellow students who helped me
in many ways throughout my time at Syracuse
%at early stages of my career in physics 
--- Bill Kahl, Eric Gregory,
Asif Qamar, Rob Salgado, Arshad Momen, Jim Javor, and many others.
%that I am forgetting.

%Thank you to so and so...

% More professional acknowledgements may be good, and less mention of
% perspective.  Let me just leave it for now, though, and maybe come
% back at the end.

%Rafael Sorkin -- below, in a different sense, for teaching me how to
%	think about physics, and making me aware of many things
%	outside of physics...  opening my eyes to truths beyond physics
%a teacher, a mentor
%Fatma Husein... -- for some very illuminating discussions
%Saul Teukolsky -- for kind hospitality at Cornell (and John Miner)
%Peter Saulson -- for helpful discussions during some transitional
%times...
%Scott Klasky -- for all he taught me about coding
%Asif Qamar -- for introducing me to GNU/Linux, which made the
%	computational aspect of this work possible (in effect)

%Eric Gregory?  Bill Kahl(?)
%Eric and Asif in the beginning...
%Arshad Momen
%Roberto Salgado

In addition, %I would like to 
let me thank Lawrence Lyon and Nelson Mead, 
%and Rob DeRoos, 
for standing with me in prayer, and especially Lawrence for
many discussions both about the physics and the larger perspective of life.
Let me also thank Wayne Lytle, for teaching me about C++, and the
rest of
%Also I
%appreciate %my
%thank you
%to 
%Thank you to
the people of Covenant Love Community Church/School, 
%who are too numerous to mention, 
especially Kathy Mead,
%and our pastor Ken Negvesky,
for their support in prayer, meals, babysitting, raw labor, and more,
%kindness, and support 
while I was finishing this work.  
%Let me mention especially Kathy Mead
%for providing meals and support through two extended illnesses and a
%pregnancy, allowing me to continue this work during difficult stages.

%I should add Ruth Tempel here in high priority, maybe, and maybe also
%reduce priority of some of these others.  I dunno.
Finally, I would like to thank my family, for their endless patience,
and especially my wife, Yvonne Tempel, for her love, exemplary patience,
continuous pressure, and reminders of a broader perspective.  
My daughter Kendra deserves special thanks for her encouraging hugs. %and kisses 
%when necessary.
%And my
%daughter Kendra, for hugs when necessary.

%Donald Rideout -- for ``painting my desk''
%Yvonne Tempel -- support, pressure, perspective -- helping me to
%remember the bigger context.
%Kendra Rideout -- for hugs at critical times...
%To my lovely wife, all she put up with, and my beautiful daughter
%Kendra...  And my daddy too (in memory of). and mommy and bro \& sis?

%others that I am forgetting.

%For Kendra Joy, and her generation....

%relatively speaking, probably negligible, but glory of the living God...(?)  My hope is that we may understand the Creator through understanding his Creation.
%%
%If I may an editorial comment --- My feeling, my hope, my expectation,
%is that the quantum theory of spacetime will be so awesome as to blow
%us away.  Knowing how awesome is our creator, Elohim, how awesome will
%be His creation.  What a privledge to be one of the few who gets to
%uncover nature's mysteries, and perhaps in the process get some
%further glimpse of the glory of our God...

This is dedicated to the memory of my father, Donald C. Rideout, whose support
in the beginning made all of this possible.

{\renewcommand{\baselinestretch}{1} \normalsize
\tableofcontents}

% If I put \pagenumbering in thesis.tex then the page numbering of the contents gets messed up.

%There may be any number of good things that are sitting in old
%versions of cosacc or contlim that Rafael deleted in the final form.
%Go back through all these to flesh this stuff out.  (But how?  It
%seems so hard to find it that I don't know what to do besides put it
%off and hope for the best...)

%I should probably insert all my questions for R back in here before
%giving him the draft, if they have not been addressed yet.  (10/24/000)

% [[ These don't work.  Please print header pages first, as separate
% ps file, then print remainder of document
%\newpage
%\newpage

\chapter{Introduction to Causal Sets}
\label{introchap}
\pagenumbering{arabic}

\section{Quantum gravity in general}

%Why are we interested in it?
%Why discrete?   did briefly
%Some discrete approaches...   did very briefly
%What is the promise of it?
%interesting points
%
% Embed my problem into physics as a whole -- motivations for discrete
% quantum gravity, etc.
%
% Scott's major comment:  I need to have a point, a thesis sentence.
% Say it verbally a few times.  The reason why I am writing this paper is ...

The quest for a theory of quantum gravity arises from the fundamental
inconsistencies of quantum mechanics and general relativity.  Due to
the philosophical differences between the two theories, it appears that
a successful marriage of the two will occur only with a radical
reformulation of both theories.  In thinking about how to approach
such a reformulation, one must decide which aspects of nature are
fundamental, and which arise as ``emergent'' structures
\cite{forks}.

There are strong indications that nature, at its smallest length
scale, has a discrete structure.  The ultraviolet divergences of quantum field theory and the singularities of
general relativity are two examples.
%The discreteness is manifested by
%infinities which occur in nature, for example the ultraviolet
%divergence in quantum field theory or 
The infinite ``entanglement entropy'' of a black hole in
semi-classical gravity is another indication.
(In fact, a continuum is not experimentally verifiable by any finite
experiment, even in principle.)

In the case of causal sets, two aspects are regarded as fundamental
--- discreteness and causality.  
The choice of
causality as a fundamental notion is partly aesthetic, and partly due to
the great success one has in recovering other aspects of spacetime
geometry from a causal order.  The Causal Set program %asserts 
postulates that
spacetime is a macroscopic approximation to an underlying discrete
causal order.
The other familiar properties
of a spacetime manifold, such as its metrical geometry and Lorentzian
signature, arise as ``emergent'' properties of the underlying discrete
order.

% Look at Taketani's writing to make this better
Mituo Taketani, a Japanese physicist and philosopher of science, has described the
process of physical theory construction in terms of three distinct
stages \cite{taketani}. The three stages repeat cyclically, except
after each cycle the theory is understood at a deeper level.  The
three stages, using physics terminology, may be called the phenomenological,
kinematical, and dynamical.

% Probably just use GR for all examples.  But phenomenology for GR had
% more to do with causality violation in Newtonian gravity than actual
% physical observations.
% hadrons --> QCD?
% classical mechanics?
% classical electrodynamics?
The phenomenological stage concerns itself with what physical
phenomena the theory seeks to address.  For what observational results
should the theory provide explanation?  For the Copernican model of
the solar system, an example of the phenomena would be the retrograde motion of the
planets with respect to the fixed background stars. % correct?
%% phenomena, experimental results
%simply an accumulation of knowledge
%Tycho Brahe

In the kinematical stage one decides in what language the theory will
be expressed.  What are the basic elements of the theory, the
``substance'', and how do they interrelate?  For the case of general
relativity, one chooses a manifold with a Lorentzian metric.  This
stage is very important, as in it one decides ``what really exists''
in nature.  It determines the mathematical and philosophical construct
on which the theory will be based.
%describing motion in Newtonian
%mechanics, one studies the interrelationships among position,
%velocity, and acceleration, and how they can be used to describe the
%motion of an object.
%properties of substance, substantialistic stage
%concept of the substance
%he speaks in terms of models of the solar system
%Kepler

Once a language is set in place, and the basic constructs of the
theory are chosen, then the remaining task is to determine how these
objects behave.  What are the ``equations of motion'' of the theory?
In the case of General Relativity, this would be the Einstein
equations.  It selects which of the kinematical possibilities (in this
example spacetime manifolds) will be realized by nature.
%essentialistic stage
%Newton
After finishing the third stage, then one will observe new phenomena
which are still not explained by the theory, which then begins the
phenomenological stage of the next cycle.  Each successive theory will
contain those of the earlier cycles, generally as some limiting form
of the latest theory.
% of theory development.  

This thesis presents a step towards understanding the full quantum
dynamics of spacetime. % (or at least of causal sets).

Prior to this work, most knowledge about causal sets was %has been
kinematical in nature.  This involved questions such as when a causal
set is well approximated by a continuum spacetime, and what
characteristics of a causal set one can ``measure'' to extract
information about the spacetime into which it might faithfully embed.
However, little was understood about the dynamics of causal
sets.  One of the primary difficulties was that most of our
experience with dynamics was for continuum theories, where one
can easily write down Lagrangians with differential operators.  Causal
sets unfortunately seemed to repel attempts to construct an action
using simple analogy with existing continuum theories.  An entirely
new approach was needed to express the dynamics of the theory.

There exist other discrete approaches to quantum gravity which involve
a causal ordering.  Some examples are discussed in
\cite{finkelstein,amblol,fotini}.

\section{Kinematics of Causal Sets}

%Causal set theory postulates that spacetime, at its most fundamental
%level, is discrete, and that its macroscopic geometrical properties
%reflect a deep structure which is purely order theoretic in nature.
%This deep structure is taken to be a partial order
%and called a causal set (or ``causet'' for short). 

In the case of Causal Sets, the causal set is the kinematical
``substance'' of the theory.  The kinematical stage of causal sets
then concerns itself with understanding how the spacetime manifold
\emph{emerges} from the underlying discrete causal order.\footnote{
For a more extensive introduction to Causal Sets, see
\cite{bombelli,causets2,reid,sor90,causets1}.}
% These are just real general references.  There are of course lots of
% more specific ones.

Note that a causal structure is a natural choice for the ``substance''
of the theory, because it encodes all the information of a continuum
spacetime (metrical geometry, topology, differential structure) save a
conformal factor.  Thus all that is missing is the volume information.
However, since the theory is discrete, this arises naturally as well,
from counting.  So a discrete causal structure has sufficient
information to encode all the kinematical framework of general
relativity.

\subsection{Mathematical Definitions}

A \emph{partially ordered set}, or \emph{poset},
\footnote
{%I follow the usual convention of 
%The distinction between the
%pair $(S,\prec)$ and the set $S$ alone (sometimes called the ``ground
%set'' in the mathematical literature) is blurred, referring to each as the
%``partial order'', or simply ``order''.  
\singlesp There are a number of
synonyms for this mathematical object: partially ordered set, poset,
(partial) order, transitive acyclic digraph, ...}
is a set $S$ with
order relation $\prec$\footnote
{\singlesp At times I will employ the usual abuse of notation, referring to both
 the set $S$ and the poset $(S,\prec)$ by the same symbol $S$.}
which is \emph{irreflexive} ($x\not\prec x \;\;\forall x\in S$) and
\emph{transitive} ($x\prec y$ and $y\prec z \implies x \prec z
\;\;\forall x,y,z \in S$).
An \emph{(induced) subposet} of a poset $P$ is a subset $P' \subset P$
whose order relation is determined by the condition that $x \prec y$
in $P'$ iff $x \prec y$ in $P$.
An \emph{interval} (sometimes called an Alexandrov set or Alexandrov
region) $\interval[x,y]$ in a poset $P$ is the induced subposet $P'$
defined by the set of all elements $\{z|x \prec z \prec y\}$.
%order of presentation??  It may be much better to give some intuitive
%notion of causal set before embarking on all this discussion.  How
%does Alan Daughton present it?
A \emph{causal set} is defined to be a
partial order for which every interval $\interval[x,y] \;\forall x,y$
has finite cardinality.\footnote
{\singlesp
 \emph{Causet} is an abbreviated form of the term causal set.  I will
 also use the term sub-causal set, or subcauset, which have the obvious
 meaning.}
An order which obeys this condition on the
cardinality of all intervals is called \emph{locally
finite}.
% prob reword
% probably discard next two sentences!
%I will generally use the term causal set (or causet for short) in
%cases where the physical interpretation is important.  In cases where
%the order is not meant to be interpreted as a causal order in the
%sense of spacetime causality I will use the term poset.
%  Modulo local
%finiteness they can be used interchangably.% sortof.

%Some more terms that need to be defined...
In a causal set $(S,\prec)$, a pair of elements $x,y\in S$ such that
$x\prec y$ form a \emph{relation}; one writes that $x$ and $y$ are
\emph{related}, that $x$ \emph{precedes} $y$, $y$ \emph{succeeds} $x$
($y \succ x$), $x$ is an \emph{ancestor} of $y$, and $y$ is a
\emph{descendent} or \emph{successor} of $x$.
The \emph{past} of an element $x\in{S}$ is the subset
$\past(x)=\{y\in{S}\,|\,y\prec{x}\}$, i.e. the set of all ancestors
of $x$.  
The past of a subset of
$S$ is the union of the pasts of its elements.  
The \emph{future} of an element, $\fut(x)$, is the set of all
descendents of $x$.
A \emph{link} %$x \link y$ 
is a relation for which there is no $z$ such
that $x \prec z \prec y$.  A link 
is an irreducible relation, that is, one not
implied by other relations via transitivity.\footnote%
{Links are often called ``covering relations'' in the mathematical
 literature.}

A \emph{Hasse diagram} is a pictorial representation of a poset.  In
the diagram, each element of the poset is represented as a dot.  A
line connects any two elements $x\prec y$ related by a link, such that
the preceding element $x$ is drawn below the succeeding element $y$
(as is the case in spacetime diagrams).

A \emph{chain}, or total order, is a set of elements for which each
pair is related by $\prec$.  An \emph{$n$-chain} is a chain with $n$
elements.
A {\it path} in a poset is an increasing sequence of elements, each
related to the next by a link, i.e. it is a ``chain made of links''.
%The length of a path is the number of links, i.e. one less than the
%number elements in the path.
More precisely, a path $C$ between two elements $x \prec y$ is a chain
which has $x$ as its past endpoint and $y$ as its future endpoint, for
which there exists no element $z$ comparable with all elements in $C$
such that $x \prec z \prec y$.  The length of a path is its number of
elements.
An \emph{antichain} is a set of elements in which no pair are
related by $\prec$.  Note that this corresponds to an ``achronal'' or
``acausal'' set in spacetime.
An \emph{$n$-antichain} is an antichain with $n$ elements.
A \emph{maximal antichain} $A$ is an inextendible antichain, i.e. there
does not exist any element of the poset which is unrelated to every
element of $A$.
A maximal antichain would correspond roughly to an ``edgeless achronal
set'', or Cauchy surface
assuming global hyperbolicity.
The \emph{height} of an order is the length of the longest chain in
that order (or, equivalently, the length of the longest path), while
the \emph{width} is the size of the largest antichain.
An \emph{$N$-order} is a partial order defined on $N$ elements.
A \emph{covering graph} of an order is the order's Hasse diagram,
regarded as an undirected graph.
A \emph{connected component} of an order is subset of elements which
form a connected (sub)graph in the covering graph of the order. % clear?
i.e. it is a subposet for which any two points are connected by a set
of paths.
A \emph{maximal element} is one which has no successors, i.e. an
element $x$ for which $\nexists y \in S \,|\, x \prec y$.
A \emph{minimal element} is one which has no ancestors, i.e. an
element $x$ for which $\nexists y \in S \,|\, y \prec x$.
%width?
%ordering fraction (only if needed before its definition in tranperc)
An \emph{automorphism} is a one-to-one map of $S$ onto itself that
preserves $\prec$. % Would this go better in dynamics chapter?

\subsection{Faithful Embedding --- Correspondence with the Continuum}
\label{embedding}
%Perhaps at least *some* mention of Malament (and Hawking?) should be
%made here, to provide some critical motivation!  Otherwise this will
%seem extremely strange to the reader who is not already familiar with
%causal sets -- my reader!  (See Meyer's intro.)

Consider a (Hausdorff, paracompact) manifold $M$ with a globally
hyperbolic, time orientable, Lorentzian, smooth
%($C^\infty$, say) 
metric $g_{ab}$.  Consider a map $\phi : S \to M$ from a causal set
$C=(S,\prec)$ into a spacetime $(M,g_{ab})$.  We call this map a
\emph{conformal embedding} if $x \prec y \iff f(x) \in J^-(f(y))
\;\;\forall x, y \in S$.  Consider an Alexandrov region of
$(M,g_{ab})$, $J^+(p)\cap J^-(q)$ for every $p,q \in M$.  The
map $\phi$ is called a \emph{faithful embedding} if it has two
further properties.  
Firstly, the number of points $n$ mapped into an
Alexandrov region is equal to its spacetime volume $V$, up to
% [[Rafael suggested replacing ``proportional'' with ``equal'' (in
% fundamental units), thereby setting all \rho's to 1.  In the case of
% coarse graining have \rho < 1, which may cause a difficulty, so let
% me try it this way for now.]]
Poisson fluctuations, i.e. the probability of getting $n$ points in
the region is
\be
P(n) = \frac{(\rho V)^n e^{-\rho V}}{n!} \,,
\ee
where $\rho$ is the density set by the fundamental length
scale. 
%mean volume density of points in the image of
%$\phi$.  
Second, the mean spacing between points in the embedding,
$\rho^{-1/d}$ in $d$ dimensions, must be everywhere much less than
the characteristic length scale $\lambda$ over which the continuum
geometry varies.  
(Admittedly these properties are not entirely well defined yet, but they
provide a good heuristic notion.)
%The notion of a faithful embedding provides a
%correspondence with the continuum in the sense that
We say that a spacetime $M$ \emph{approximates} a causal set $C$ if
there exists a faithful embedding of $C$ into $M$.  See
\cite{bombelli} for more discussion on this issue.\footnote
{\singlesp Actually the requirement of a strict conformal embedding may be too
stringent, as we require only a probabilistic reproduction of the
volume information.  Why should the light cones be exactly fixed?  One
way to loosen this requirement is to remove relations in the embedding with
probability $p$.  (If $p\ll (R-L)/R$, for a causet with $L$ links and $R$
relations, this will have negligible effect, so $p$ must be at least
$\sim (R-L)/R$.)} % probably promote this from a footnote status, but it
	      % works o.k. for now like this
% [[ Should I mention somewhere that the notion of faithful embedding
% is not rigorous?  Even the precise formulation of the Hauptvermutung
% is not really nailed down.  The conditions that I mention for a
% faithful embedding are not really precisely defined, they are still
% on a hand waving level.  Maybe I'll skip this commentary for now,
% though.]] 

The notion of faithful embedding gives a correspondence between a
causal set and a continuum spacetime.  The belief is that a given causal
set embeds faithfully into a unique spacetime.  
% [[ There is a lot of evidence supporting this.]]
Expressed more
precisely, if $\phi:C\to M$ and $\phi':C\to M'$ are two faithful
embeddings into two spacetimes $M$ and $M'$, then there exists an
approximate isometry $g:M\to M'$ such that $\phi'=g\circ \phi$.  The
isometry will only be approximate because the condition of faithful
embedding is not sufficient to fix the entire continuum geometry ---
there will remain small variations in the geometry, with length scale
$>\lambda$, which still admit a faithful embedding of $C$.
%That is, for any given causal set which
%admits a faithful embedding into two spacetimes $M$ and $M'$, $M$ and
%$M'$ will be related by an (approximate) isometry.
In general the precise formulation and proof of this conjecture, which
is called the \emph{``Hauptvermutung}'' of Causal Sets, is a difficult
mathematical problem.

%Does this paragraph fit in well here?
Given an arbitrary (past and future distinguishing) spacetime, it is
easy to construct a causal set which will admit a faithful embedding
into that spacetime.
The method, suggestively called \emph{sprinkling}, is to simply select
at random, via a Poisson distribution, a finite set of events of the
spacetime.  These will correspond to the elements of the causal set.
The causal relations of the causal set are simply those inherited
directly from spacetime causality.

% [This comment needs to be put in somehow:]
Note that in a faithfully embedded causal set most links will then be
``almost null'', due to the Lorentz invariance of (local) spacetime.
There can exist two events which are spatially very distant, perhaps
lying in different galactic clusters, which are nevertheless ``nearest
neighbors'' in the sense that they are connected by an (almost) null
geodesic.
%non-compactness of the Lorentz group
%An issue with Lorentz invariance is is there enough spacetime to allow
%the boost to take place?  In this way causal sets provides a coupling
%of the infrared with the ultraviolet ``catastrophes''.
%yes, infrared-ultraviolet blurring should be mentioned somewhere...
(Also note in this connection that causal sets naturally suggest a sort
of ``blurring'' between the ultraviolet and the infrared, in that a
field that is highly boosted gets blueshifted, while at the same time it
makes connection with remote parts of the universe.  The extent to
which a field can be boosted (ultraviolet cutoff) is limited by the
size of the universe (infrared cutoff), as increasing large boosts
require the existence of links connecting increasingly ``distant''
elements of the causal set.)  % Is this clear?!

% ``restored'' -- explain!
% differential structure?
So the causal set is a ``discrete manifold'', using the words of
Riemann, from which the macroscopic properties of the continuum
spacetime: topology, differential structure, metrical geometry, and
causal structure, are emergent.  Note that by keeping the causal
structure as the substance of the theory, and then ``counting'' to get
volume information, we have been able to recover the full metrical
geometry.  
%The fact that this succeeds is demonstrated by the fact that
%the causal structure of a spacetime completely determines its metric
%up to a conformal factor \cite{malament}.  
This succeeds because
the causal structure of a spacetime completely determines its metric
up to a conformal factor \cite{malament}.  
The discreteness,
``number'', by providing the missing volume information, restores the
full metrical geometry.  Note also that topology is restored, by
inheritance from the topology of the manifold into which the causal
set will faithfully embed.  Note that the Lorentzian signature also
arises naturally from this scheme, as the unique signature which
maintains the distinctness of past and future directions in the causal
order.

%This is the ``Hauptvermutung'' of causal sets, that a causal set can
%(approximately) uniquely specify a spacetime manifold.
%% [ This needs to be elaborated more!  See e.g. p. 8 ``First Steps''.]
% No, already mentioned above.

% It may be useful to discuss physical interpretations to the various
% terms defined above.  Perhaps most of that ``string of definitions''
% section should be placed here instead?
% It may be good to do this, but let me skip it for now.  It is not
% completely necessary.

% Myrheim -- continuum picture is only valid if the spacetime is
% homogeneous on a sufficiently large scale.

A considerable amount of work has been done in understanding the
kinematics of causal sets.  For example it is understood how to
extract spacetime dimension and proper time from only the discrete
causal order.  The dynamics, however, was much less developed
%more poorly understood 
before this work.  Below is a brief sketch of some of these earlier
results.

\subsection{Causal Set Dimension}

There are many different indicators of the dimension of a causal set,
that is, ways to estimate the dimension $d$ of the spacetime into
which a given causal set might faithfully embed.  The actual value of
this dimension is the \emph{physical dimension}.  To determine
directly the
physical dimension of a causal set one would need a faithful embedding,
which is difficult to achieve.  The hope, then, is to deduce what this
dimension will be by looking at certain simpler 
indicators which depend only
on more easily accessible features of the causal order.
% by looking at certain simpler indicators.
% and not on any representation of it as an embedding.
%We wish to compose a notion of dimension of a poset in a manner which
%is useful for their interpretation as a causal set.

\subsubsection{Integral dimension indicators}

The following dimension indicators use only ``conformal information'' of
the causal set, i.e. they only consider conformal embeddings in their
construction, rather than faithful embeddings.

\paragraph{Linear dimension}
Linear dimension (also called combinatorial dimension) is a definition
of dimension for a poset generally used by mathematicians.  It is not
quite appropriate for causal sets, however, as the ``light cones'' it
uses to define the order are ``square''
%(more precisely they are a the shape of the first
%``octant'' of $\R^n$), 
and thus will not be a good dimension indicator
for embeddings into Minkowski space.  It is defined as the minimum
dimension $d$ such that the poset $P$ can be realized as points in
$\R^n$ with the order given by $x \prec y \iff x_i < y_i \;\;\forall
i$.  In spite of its unphysical character, there are many results
known for this notion of dimension.

\paragraph{Flat conformal embedding dimension}
% called Minkowski dimension by Meyer
The flat conformal embedding dimension (also called Minkowski
dimension, or causal dimension) is the minimum
dimension Min\-kow\-ski space into which the causal set can be
causally embedded.
% (at all, without regard to the
%distribution of the embedded points).  
Since in general one would expect the causet to embed into Minkowski
space only locally (spacetime is locally flat),
%A more practical application of the notion of causal dimension would
%be to 
the more appropriate question to ask is that
%only ask that 
an (appropriately chosen) sub-causal set embed into Minkowski space,
where the subcauset is chosen such that it is ``small enough'' to
``not see the curvature'' of the larger spacetime.
% into which it might
%faithfully embed as a whole.  
These subsets, then, should contain the dimensionality information of
the causet.  Some conjectural bounds have been placed on causal
dimension, in part by using results known for linear dimension.
%\cite{bombelli} 
Some \emph{pixies} can be identified, 
%at least for small dimension, 
which, if present as subposets, establish a lower
bound on flat conformal embedding dimension.  See \cite{bombelli} and \cite{pixies}.

There are a number of difficulties with the notion of causal
dimension.
%, however.  
One is the problem of deciding how to choose a ``local'' subset which
may represent a ``small'' region of the spacetime.  For causal sets
locality is very difficult to define, because the ``unit balls'' of
the Lorentzian metric contain spatially very distant points.  There
are indications that a notion of locality is maintained in causal
sets \cite{daughton,salgado}, but it remains difficult to see how to
use this to construct local subsets. % of a causal set.
Another shortcoming is that it is not unreasonable to imagine that the
dimensionality of spacetime will vary with length scale.  Thus the
dimensional information encoded in one of these subsets should have
length scale dependence.  If the appropriate subset can be chosen only
after some probabilistic process of coarse graining (see Section
\ref{coarse_graining}), the arguments used to establish bounds on
flat conformal embedding dimension will no longer be valid.
Lastly, as alluded to in an earlier footnote, it may be too strict to
demand an exact conformal embedding.  Then, for example, one cannot use
pixies to establish firm bounds on dimension.

These difficulties illustrate how in general it is difficult to
construct a global, exact indicator of dimension.  The more useful
notions are quasilocal, probabilistic, and fractal in nature, taking
advantage of the volume information of a faithful embedding.

%[pixies are useful in a proof that Luca sketches along the lines of the hauptvermung of causets]  Probably don't bother with this importance boost, not worth the trouble...

\subsubsection{Fractal dimension indicators}
There are other indicators of causal set dimension which depend on the
volume information as well, i.e. they consider faithful embeddings in
their definition.  Consider an Alexandrov region $\interval(x,y)$ of a
causal set $C$.  One can derive causal set invariants by counting the
occurrence of various substructures, such as the number of relations,
number of elements, number of chains, number of links, etc.,
and compare the result with what is known for sprinklings into
$n$-dimensional Minkowski space, obtaining a dimension estimate for
(that region of) the causal set from which the region was taken.
Since the number (of relations, say) counted will never come out
exactly as that which would arise from sprinkling into an integral
dimension Minkowski space, the dimension which obtains from this
procedure will always be fractional, as in computing the effective
dimension of fractals.

This fractal dimension will be most meaningful if there exists a large
region of $C$ covered by many Alexandrov sets of
(approximately) the same
cardinality $V$, which all yield (approximately) the same effective
dimension $n_\mathrm{eff}$.
In general different
pairs of points $(x,y)$ will not yield the same $n_\mathrm{eff}$.
This will occur for a number of reasons:
\begin{itemize}
\item random fluctuations.
\item sampling a different region of the causal set.
\item curvature of causet/spacetime:  Because the region of spacetime
enclosed within the interval $\interval(x,y)$ in general will not be 
flat, the relationship between number of relations, say, and dimension
will not quite be the same as that for flat spacetime.
For this reason it is important to choose intervals which are much
smaller than the curvature scale.
\item scale dependence of topology, though this would vanish if we
constrain $V$ to be same in each region $\interval(x,y)$.
\end{itemize}
%If also independent of V, have stronger evidence for dimension being
%$n_eff$ -- silly in light of scale dependent topology?
%
%[ Somewhere in here some problems arise with the idea of using
%Alexandrov intervals as ``patches'', e.g. to measure curvature,
%consistence of MM dimension over a larger region, etc.  This seems
%like it will fail because most all intervals are extremely null, so
%their interconnectivity, covering, etc., may not be what intuition
%implies.  Maybe I should just put this comment itself!]
In addition, it should be noted that the notion of ``covering a region
of $C$ with many Alexandrov sets'' may be relying on an
intuition which is not valid for causal sets.  The important point is
that in any given reference frame, almost all Alexandrov
sets ``look extremely null''.  Thus it is difficult to
``tile'' a region with such sets, a task which looks
something like trying to cover a two dimensional region with a
collection of thickened diagonal lines.  In general each Alexandrov set
will overlap a very large number of ``neighboring tiles'', and will
not cover the region in the same manner as one's intuition from
Riemannian signature geometries suggests.

\paragraph{Volume -- length scaling}
In $d$-dimensional Minkowski space, an Alexandrov set of
``height'' $T$ (proper time between its two end points) has spacetime volume
%\cite{meyer}:
\bne
V = \frac{2V_{d-1}}{d} \left(\frac{T}{2}\right)^d
= \frac{2\pi^{(d-1)/2}}{((d-1)/2)! \:d} \left(\frac{T}{2}\right)^d
\label{vl-scaling}
\ene
where $V_{d}$ is the volume of a unit $d$-ball.
%(This is simply the volume of a double cone of height T.)
For a given Alexandrov set $A=\interval[x,y]$, measure $T$ by
finding the longest chain connecting $x$ and $y$ (see
sec. \ref{proper_time} below), count the number of elements $z$ in $A$,
%such that $x \prec z \prec y\}$, 
and invert (\ref{vl-scaling}) to get
a dimension.  This is perhaps the most obvious measure of dimension
--- to simply determine the exponent with which the volume of a region
scales with length.

A caveat about this scheme is the unknown coefficient $m_d$ relating
length of the longest chain to proper time (\ref{m_d}).

\paragraph{Counting chains}
\label{longest_chain_dim}
% Meyer calls this Hausdorff dimension (I think).
Consider an interval $\interval[a,b]$ which
contains $N$ points and $C$ chains.  Then in the limit of large $N$, 
\bne
d = \frac{\ln N}{\ln \ln C} \label{doC}
\ene
is a measure of the causal set's dimension.  
This measure is useful because it can be written explicitly, but is
perhaps impractical because of the $\ln \ln$ in the denominator.
((\ref{doC}) follows from the fact that
%arises because 
the number of chains $C$ in an interval grows exponentially
with its height $T$, while its volume ($N$) grows as $T^d$ in $d$
dimensions.)
% I could cite Meyer, but I don't think that he ever states this
% explicitly.  It is implicit between he and Rafael...  (arrgh!)

\paragraph{Myrheim-Meyer dimension}
\label{MMdim}
For a causal set $(S,\prec)$ 
% Abuse notion here?
define $R$ to be the number of related pairs of elements, i.e. the
number of pairs $(x,y)$ such that $x \prec y$ or $y \prec x\;\:\forall
x,y \in S$, and (following Myrheim \cite{myrheim}) define the
\emph{ordering fraction} $r$ to be the
fraction of pairs of elements which are related, i.e. $r=R/{N \choose
2}$.  A causal set which is formed by sprinkling $N$ points into an
interval of $d$ dimensional Minkowski space will have an ordering
fraction given by
\be
r = \frac{3}{2} \frac{d!(d/2)!}{(3d/2)!}\,,
\ee
which decreases monotonically with dimension \cite{meyer}.
%where $R$ is the number of pairs of elements which are causally
%related.  
Inverting this relationship (numerically) yields a fractal dimension
for any given $r$, called the Myrheim-Meyer dimension.
% the formula:
%This relationship can be inverted, providing a measurement of the
%dimension of a causal set simply by counting relations in an
%%(sufficiently small, so as to not see curvature) 
%interval.  
%For this dimension indicator to work well, the interval
%should be chosen small enough so that it represents an approximately
%flat region of spacetime.  
Since this measure is based on measuring a large number ($R$), the
random fluctuations will be smaller than those arising from similar
dimension estimators which count other quantities, making this a
computationally efficient method to estimate causal set dimension.

Because this measure of dimension associates a dimension to any
ordering fraction $r$, it is sometimes used heuristically to specify the
``dimension'' of a causal set as a whole, without regard
to whether it represents an Alexandrov set or whether the
region is small enough not to see the spacetime curvature.

\paragraph{Midpoint scaling dimension}
Consider an Alexandrov set $\interval[x,y]$ of volume $V$.
A midpoint, $z$, between $x$ and $y$, will subdivide $\interval[x,y]$
into 3 regions $\interval[x,z]$, $\interval[z,y]$, and the remainder.
% (plus some extra outside either of these two intervals).
If this causet is faithfully embeddable into $\M^d$, then the volume $V_R$
of the first two of these regions will be $(1/2)^dV$.
Inverting this gives a dimension estimate of $d=\log_2(V/V_R)$.
%Since $V_R/V$ depends exponentially on the dimension, this method will be quite
%efficient for large $d$.
A convenient definition for the midpoint 
%(due to R. Sorkin) 
is to maximize the minimum of $V(x,z)$ and $V(z,y)$.

%\paragraph{Nugget dimension}
%Can I put this here?  We mention it in our papers but never expound on
%it.  Why not do so somewhere?!
%See ~ Oct. 998 of my notes, top of right hand page
%Let's skip it for now -- it is Rafael's and it is not coarse graining
%invariant...
% Also Rafael does not think too highly of it.

\subsection{Geometry}
The previous section discussed briefly how to extract some topological
information from the discrete order.  Here I mention some ideas on how
to extract geometrical information.

\subsubsection{Proper time} %% Timelike intervals ?
\label{proper_time}

% [[ Causal structure (Weyl tensor...) IS geometry, so these next two
% sentences don't really make sense. ]]
%So far we have seen how a causal set reproduces many aspects of the
%continuum spacetime, for example its causal structure and dimension.
%Obviously one would also like to construct geometrical information from
%the discrete order as well.  
The causal set should tell us not only whether
two events are related, but ``how much to the future'' one occurred after
the other.

Consider two elements in a causal set $x \prec y$.  The longest chain
connecting them will be a path, which may be called a (timelike)
\emph{geodesic}.  Note that this corresponds directly to the notion of
timelike geodesic in continuum spacetime --- it is an extremal
chronological curve connecting $x$ and $y$.
%, where in this case extremal means longest.  
For a causal set which embeds faithfully into a spacetime, there will
usually be an extremely short path 
%(almost) always be a path of length two (one that contains two links)
between any two related elements, e.g. one composed of two links, because one can always go as far out
along the light cone as one wishes (``following a link'') and likely find an element which is
linked to $x$.  
Recall, however, that for Lorentzian geometry the
appropriate (timelike) extremal path is the \emph{longest} path.
%have triangle inequality -- why is this here?!
Note that in general there may be multiple longest chains passing through
two elements.  In this sense the discrete notion of geodesic departs
from the continuum (in a small region),
%, so that a unique
%geodesic curvature to be topologically trivial), 
but still a unique path length is assigned to the pair $x$
and $y$.  In fact, this path length is proportional to the proper
time interval of Minkowski spacetime, in the following sense.

Consider a causal set which arises from sprinkling into
$d$-dimensional Minkowski space.  Brightwell and Bollob\'as
\cite{boxspace,ruth} have shown that for an interval $\interval[x,y]$
of volume $\rho V$, the length $L$ of the longest chain satisfies
$L(\rho V)^{-1/d}\to m_d$ in probability, where $m_d$ is an unknown
constant which depends on the dimension of the Minkowski space.
Fairly tight bounds can be placed on $m_d$.  For $d\geq 3$
\bne
1.77\leq \frac{2^{1-1/d}}{\Gamma(1+1/d)} \leq m_d 
\leq \frac{2^{1-1/d}e\Gamma(d+1)^{1/d}}{d} \leq 2.62
\label{m_d}
\ene
and it is known that $m_2 = 2$.
Assuming that this correspondence remains in the presence of
curvature, this provides a simple explicit method to extract timelike
distances from the discrete order.
For simplicity, we define proper time between two related elements to
be the length of the longest chain connecting them.

Quite a bit is known about the fluctuations in this length as well,
see e.g. \cite{boxspace2}.  Owing to the fact that a sprinkling into
an interval of $\M^2$ is isomorphic to a random permutation, and that the
length of the longest increasing subsequence (which is equivalent to
the length of the longest chain) has been studied extensively, much is
known about the 2 dimensional case \cite{rains1,rains2}.

\subsubsection{Spacelike distance}
It is difficult to construct a notion of spacelike distance on a
causal set, in part because of the non-compactness of the Lorentz
group.  For example, an early proposal by 't Hooft \cite{thooft} went
essentially as follows, for $\M^d$.  For two unrelated elements $(x,y)$, find the
pair of elements $(a,b)$, such that $a\prec x,\:a\prec y,\:b\succ x$,
and $b\succ y$, which minimizes the proper time (as computed in the
previous section) between $a$ and $b$.  Unfortunately this will always
turn out to be zero (for sprinklings of $\M^d$ for $d>2$), because there will always be a pair $(a,b)$ which, by
a statistical fluctuation, are linked.  To see this, consider in
$\M^d$ every
pair $(a,b)$ where $a$ is chosen from the intersection of the past
light cones of $x$ and $y$ (this is where the maximal elements of
$\past(x)\cap\past(y)$ will lie) and $b$ is chosen from intersection
of the future light cones of $x$ and $y$.  Since these regions are
noncompact, there will be an infinite number of (approximately)
statistically independent pairs to consider,
%thus 
leading to certainty of finding a linked pair.

However, there is a proposal for finding the
spacelike distance between a %(timelike) geodesic 
maximal chain and a point \cite{ruth}. 
%\footnote{
%c.f. figure 38 on page 81 of \cite{geroch}.}
%This is not unreasonable, because o
% below two sentences already appear above:
%A natural definition for a (timelike) \emph{geodesic} is a maximal
%path between two related elements.  Note that this correctly
%corresponds to the extremal path in continuum spacetime.
The construction is simply this: for a geodesic $\gamma$ and a point $x$
(the construction assumes that $\gamma$ is chosen such that
$\fut(x)\cap\gamma\neq\emptyset$ and
$\past(x)\cap\gamma\neq\emptyset$) find the minimal point $b$ in
$\fut(x)\cap\gamma$ and the maximal point $a$ in $\past(x)\cap\gamma$,
and take the spacelike separation between $\gamma$ and $x$ to be half
the timelike distance between $a$ and $b$ (i.e. half the number of
links in $\gamma$ between $a$ and $b$.
% (for simplicity we neglect the
%factor $m_d$ in the definition of proper time from section
%\ref{proper_time})).
% Is all this correct?  Probably put this embedded parenthetical
% remark into the above section --- use it to define proper_time on a
% causet.  done.

%In fact, it appears that this definition can be generalized to give
%the squared interval between any two elements in a ``local'' region of
%a causal set.  The method follows from simply applying the above
%definition of proper time to the construction in Chapter 15 of
%\cite{geroch985} (or, for more discussion of this construction, see
%chapter 5 of \cite{geroch978}).
%Here the region of the causal set under consideration is assumed to be
%small enough to be flat, so any (timelike) geodesic passing through
%one of the two points to be compared will give (approximately, up to
%statistical fluctuations of course) the timelike or spacelike distance
%between two elements of the causet.
%
% It fails due to fluctuations again.

\subsubsection{Curvature}
\label{curvature}
One way to extract curvature information from the causal set is generalize
Equation (\ref{vl-scaling}), as done in \cite{myrheim}, 
%from section \ref{longest_chain_dim} 
%can be generalized 
to the case of non-zero curvature.  
From this one can extract information about the Ricci tensor \cite{bombelli}.
%expanding the right hand side in powers of $T$. ??
%, to get a measure of curvature.  Then 
The smaller intervals could be used to measure dimension, and then
larger ones could be used to estimate curvature.

\subsection{Closed Timelike Curves}
\label{ctcs}
The irreflexivity of the definition of a partial order used here is
simply a
convenient convention.  One could just as well have chosen to define
the poset to be reflexive ($x\prec x \;\:\forall x \in S$), but then an
added condition of acyclicity would be required: $x \prec y$ and $x\neq y
\implies y \not\prec x$.  Without this extra condition the order would
allow cycles, %closed causal loops, 
corresponding (one might think) to closed causal curves in
continuum spacetime.  Note, however, that such an order would be sick
in the sense that all the elements in such a cycle are
indistinguishable from each other in terms of the order relation, so
they might as well be regarded as a single element.\footnote
{\singlesp 
%If the elements of the order are regarded as distinguishable in some
%innate sense, then 
Perhaps this suggests that we should attach a
positive integer to each element of our causal set, encoding the
cardinality of a closed causal loop which that single element
represents.}
In this sense causal sets ``predict'' that there do not exist closed
causal loops in spacetime.

Evidence indicates that the failure of closed causal curves may
already be encoded into quantum field theory, in the form of Hawking's
chronology protection conjecture \cite{cronprot}, which prohibits
closed causal curves from forming via a divergence of quantum field
energy density at a chronology horizon (a horizon which separates a
region of spacetime which admits closed causal curves from one which
does not).  In order for our definition of a causal set to be
consistent, %with observation, 
something like the chronology protection
conjecture must hold.
%Then this 
%%constraint imposed by the 
%definition of the partial order says nothing new about the physics of
%spacetime --- it already existed at the level of semiclassical
%gravity.

\subsection{Coarse graining and Scale dependent topology}
\label{coarse_graining}

In practice it will be extremely rare that a given causal set
faithfully embed into \emph{any} spacetime.  Somehow the dynamics must
select four dimensional, ``spacetime-like'' causal sets.
%, in the quantum theory by constructive
%interference of histories of the quantum measure, or, for a classical
%dynamics, by simply by being most probable within the classical
%(stochastic) measure.  (See below for more discussion of the histories
%formulation of causal set dynamics.) % Where?
%% This interference discussion conflicts with that in the manifoldness
%% section.
However, it is important to note that one would not expect 
the topology of spacetime to be four dimensional all the way down to
the Planck scale.  It is reasonable to expect some extra
compact, Kaluza-Klein-like dimensions at small length scales.
In addition, it is likely that even the continuum approximation itself
will break down at Planck distances,
%spacetime's smallest scales, 
leaving %perhaps 
something like a ``spacetime foam''.
%to get a
%continuum-like structure at all length scales, or even if quantum
%spacetime is well approximated by a continuum manifold at all length
%scales, it is still likely that its topology will be very different near
%the planck scale.
% as compared to that at larger scales.
%, for example it
%may have 
%many more dimensions and a much more complicated topology
%near the planck scale.
%%, e.g. a
%%``spacetime foam''. 
%%It is reasonable to imagine that at small length
%%scales the continuum approximation will break down, leaving a
%%non-smooth structure such as a spacetime foam, small compactified
%%Kaluza-Klein dimensions, etc. near the Planck scale.
%Thus it is likely that the continuum approximation will be valid
%only after some appropriate coarse graining of the causal set, or even
%if spacetime does remain approximately continuous down to very small
%length scales, its topology will likely be very different from the
%four dimensional spacetime that exists at macroscopic scales.  
Thus
some form of coarse graining will %most likely 
probably be necessary to make
connection with macroscopic spacetime.
%Of course 
%Given that 
However, even after coarse graining, it is very possible that no
causal set will be precisely faithfully embeddable into a spacetime.  A notion
of an approximate embedding will likely be required, as alluded to in
\S \ref{embedding}.
%which is not yet understood in detail.
%dynamics selects embeddable ones - not really, because:
%  (success of Kaluza-Klein)
%    if scale of KK dims ~ l_p  ``fuzzy'' at l_p, manifold not right at
%smallest scales, fields are remnants of this ``non-geometrical''
%structure...
%%    if scale of KK >> l_p, are o.k. with embedding into manifold
%    (Though probably both true...!  exists KK-like regime and more
%  broken structure at smaller scales.  KK is approximation to reality,
%  seeing that spacetime is not 4d at small scales...
% [ Since all this is quite speculative anyway, let's not say too
% much about it... ]

%An advantageous feature of causal sets is that there exists for them a
%simple yet precise notion of coarse graining. 
In general, there are two different approaches to coarse graining.
One is to ``blur'' or ``average'' points.
%, either on a lattice or in a continuum.  
A major difficulty with this method, though, is the
difficulty of maintaining Lorentz invariance, since a blurring which
``looked natural'' in one frame would appear extremely non-local in another.
%Lorentz
%invariant ``blurring'' will extend to arbitrarily distant regions.
%  Naturally one would
%want to blurring of a point to be over a local neighborhood of that
%point.  However, in the case of a Lorentzian manifold
%it is hard to see how this can be made Lorentz invariant.
An alternate method is to use a decimation procedure, wherein some
fraction of the ``lattice sites'' are simply ignored.  This approach
is easier to use than the blurring procedure, and it maintains Lorentz
invariance.  The precise method of coarse graining we use is simply to
select some fraction of the existing elements of the causal set at
random, ignoring the remaining elements, and inheriting the causal
relations directly from those of the ``fine-grained lattice''.  The
random procedure is necessary both to maintain Lorentz invariance, and
because of the ``background-free'' nature of the theory, 
%the absence of a background 
which leads to an absence of any other obvious method with which to select
points for coarse graining.
Stated more precisely, a coarse grained approximation to a causet $C$
can be formed by selecting a sub-causet $C'$ at random, with equal
selection probability for each element, and with the causal order of
$C'$ inherited directly from that of $C$, i.e. $x\prec{}y$ in $C'$ if
and only if $x\prec{}y$ in $C$.
Notice that such coarse graining is a random process, so from a
single causet of $N$ elements, it gives us in general, not another
single causet, but a probability distribution on the causets of $m<N$
elements.

\begin{figure}[htbp]
\center
\scalebox{.5}{\includegraphics{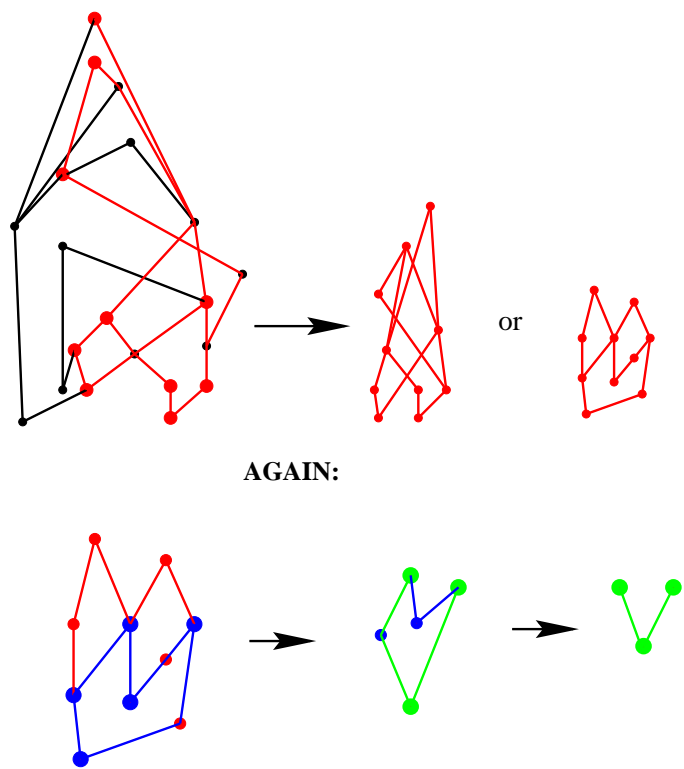}}
\caption{Three successive coarse grainings of a 20-element causet}
\label{coarse_grain}
\end{figure}

For example, let us start with the $20$ element causet $C$ shown in
Figure \ref{coarse_grain},
%(which was percolated using $p=0.25$)
and successively coarse grain it down to causets of 10, 5 and 3
elements.
We see that, at the largest scale shown (i.e. the smallest number of
remaining elements), $C$ has coarse-grained in this instance to the
3-element ``V'' causet.

\section{Dynamics for Causal Sets}
%Here I can put historical ideas about the dynamics of Causal Sets,
%casting my work into its proper perspective.  This might help motivate
%it?

%%Could make some comments regarding old ideas about dynamics, as
%%discussed in the end of earlier theses, but this would be difficult to
%%make sense of...  Low priority --- it is not too important.
%%%
%What we have done is somewhat similar to a suggestion of Alan
%Daughtons \cite{daughton} starting at last paragraph of page 42.
%different guesses for action, etc.
%log number of spanning trees...

% [ Is this discussion here silly, in that it doesn't give a coherent
% picture of causet dynamics to the reader?  Should I just include a
% disclaimer ``early thought, see chap. 3 for more complete description''?]

Because of their discrete character, many issues arise in causal set
dynamics which are not present in the formulation of dynamics for
continuum theories.  Some general principles which need to be
understood in a discrete context are general covariance,
``manifoldness'' (i.e. the emergence of a continuum at macroscopic
length scales), and locality.
This section will merely present some general issues which arise when
attempting to express a dynamics for causal sets.  A precise, detailed
account follows in Chapter \ref{dynamicschap}.
%Here I discuss some issues with constructing a
%dynamics for causal sets.

\subsection{General covariance}
\label{gencov}
%The order in which elements arise is a sort of gauge freedom of the
%theory, roughly analogous to diffeomorphism invariance of GR.  We call
%this gauge invariance condition ``discrete general covariance''.
% Actually invariance under arbitrary labeling is more closely
% analogous to diffeomorphism invariance.

An $N$ element partial order $P$ admits a \emph{natural labeling},
which is an assignment of a non-negative integer $0, 1, 2, \ldots N-1$
to each element of $P$ such that $x \prec~y \implies \lab(x) < \lab(y)
\;\:\forall\, x,y\in P$.\footnote
{\singlesp A natural labeling of an order $P$ is equivalent to what is
called a ``linear extension of $P$'' in the mathematical literature.}
Since a coordinate system in general relativity is simply an
assignment of a ``label'' for
each event of spacetime, 
% perhaps with some differentiability condition, but does this have to
% be mentioned?
a labeling of a causal set corresponds to a
choice of coordinates in general relativity.
The continuum analog of a natural labeling might be a
 coordinate system in which $x^0$ is everywhere timelike (and this in
 turn is almost the same thing as a foliation by spacelike slices).
One can also consider an \emph{arbitrary labeling}, which is an
assignment of integers to the elements as above but in a manner
irrespective of the causal order $\prec$.
This would be more closely 
% One could also consider arbitrary labelings, which would be 
analogous to arbitrary coordinate systems.  
In that case, there would be a
well-defined gauge {\it group} --- the group of permutations of the
causet elements --- and labeling invariance would signify invariance
under this group, in analogy with diffeomorphism invariance
and ordinary gauge invariance.  
However, we have not found a useful way make use of this,
and consider only natural labelings.
%(Invariance under arbitrary labelings
%more closely coincides with the full diffeomorphism invariance of
%general relativity, associating the group of permutations with the
%gauge group of diffeomorphisms of GR.  
%However, we have found no
%useful way to take advantage of this group structure, thus we will
%consider all labelings to be natural labelings in the sequel.)
%Thus, 
In the context of causal sets, then, general covariance will translate into a
statement of label independence of the dynamics.

The dynamics for causal sets will be expressed as a measure defined on
suitably chosen classes of \emph{histories}, which in the context of
causal sets are just the causal sets themselves.  Generally we think
of these histories as having infinite cardinality, i.e. they extend
arbitrarily far into the future.\footnote{\singlesp It is a logical possibility
that the universe ``ends'' after some finite time, i.e. the causal set
has finite cardinality, but we disregard this eventuality 
on purely metaphysical grounds.} % is this correct use of metaphysical?
%somehow this possibility seems unnatural.
%For the classical theory, the measure will take the form of a
%probability measure, 
%defined on a sigma-algebra.  The issue of general
%covariance 
The issue of general covariance then serves to limit the sets of
histories which have a physically meaningful measure, or equivalently,
what are the physically meaningful questions one can ask of the theory.
%for now 
%let us consider %me mention 
%briefly the issue of ``What is a generally
%covariant question?'' which one can ask of the dynamics.  

In discussing this issue, it will be useful to first define the notion
of a stem.  A \emph{full stem} of a causal set $C$ is a finite
subcauset $S$ for which every element not in $S$ %, $C\setminus S$
succeeds a maximal element of $S$.  % explain better?  
A full stem corresponds to
 a completed partial history of the universe.  A \emph{partial stem}
of a causet $C$ is a finite subcauset $S$ which contains its own past,
i.e. if $x\in S$ and $y \in C$ such that $y\prec x$ then $y\in
S$.
%A \emph{partial stem} of $C$ is a finite subset
%which contains its own past.  
%(A full stem is a partial stem such that
%every element of its complement lies to the future of one of its
%maximal elements.)  

An example
of a non-generally covariant question is ``What is the probability
that the universe has a 3-chain as a full stem after $N$ elements
appear.''  This is not generally covariant because a labeling is
implicit in the notion ``after $N$ elements appear.''  A covariant
question could ask ``What is the probability that the universe has a
3-chain as a full stem, after the growth process runs to completion'',
i.e. in the limit as $N\to\infty$.
%``What is probability of getting a diamond (as a subcauset, say.  or
%set N=4) after N elements form is
%not gen cov.  ``What is ... as a full or partial stem \emph{is} g.c.''
%non-covariant question:  What is probability that after make 5000
%elements, R will be within 10-20\% of whatever?\\
%What does ``first 5000 elements mean?''\\
%covariant question: (assuming that there exists a first element)  What
%is the probability that the first element has a unique successor?  One
%must take the limit as $N\to \infty$ to answer this question.
%Another example of a generally covariant question:  What is
%probability of occurrence of any particular poset as a full stem?\\
%
%The dynamics for causal sets can be expressed in terms of a measure
%on classes of histories, where a history in this context is simply a
%(completed) causal set.  
It is conjectured that all physical questions
can be expressed as a logical combination of probabilities of the
occurrence of a given partial stem in the universe.  Can a measure
be formulated that assigns a finite answer to all such questions?
%
%In this manner the answers to questions such as these
%define a $\sigma$-algebra.%, allowing physical questions as any logical 
%
%These questions then serve to partition the set of histories
%(completed, i.e. $N\to\infty$ causal sets).
%Is this a well defined measure?  
%A measure is defined by allowing
%these questions and all logical combinations of questions of this
%form, i.e. any expression involving a finite number of boolean
%operators.  (and, or, not).
%Does this measure converge?  i.e. are the answers to these questions
%finite?  Do all these questions have a well defined answer?
%This measure forms a sigma algebra
%Is it well defined as a classical measure?  The answer is yes!  (For
%appropriate choices of the $t_i$'s.  Is there a choice of the $t_i$'s
%which gives a not-well-defined measure?
%
%Regard classical dynamics as probability theory for causal sets
%i.e. dynamics is rule which assigns a classical measure (probability)
%to (suitably chosen) classes of causal sets.
%(A causal set being a ``history'' in a ``sum-over-histories''.)

\subsection{``Manifoldness''}
% Rename to something like ``approximate continuum''?
Almost every causal set in no way resembles any spacetime manifold.
To get a feel for how extreme this is, it has been shown that, in the
limit of large $N$, the number of partial orders defined on $N$
elements grows as $2^{N^2/4}$ (to leading order).  
In comparison, one may estimate that the number of
``spacetime resembling'' posets on $N$ elements is only $\sim 2^{N \ln N}$.
Somehow the dynamics must select those that (at least at a
sufficiently large length scale) resemble spacetimes.  
%Furthermore,
One might expect this selection to occur at the classical level,
i.e. in the classical limit ``dynamically preferred'' causets should
faithfully embed into spacetimes.  
This limit arises from the
constructive interference of histories, so the ``spacetime resembling
causets'' should lie at ``a stationary point in the causal set
action'', while those that are very unlike continuua should have
``rapidly varying'' amplitudes.  
Of course one needs a precise notion
of how causets are ``close to each other'' to be able to speak of a
stationary point.  Our intuition comes roughly from a notion of
``closeness'' for Lorentzian geometries.\footnote{\singlesp
See \cite{bom00} for an interesting approach to the issue of
defining a distance functional on the space of Lorentzian geometries.}

A related question to the existence of a continuum is why the
cosmological constant is so small.  If it had it's ``natural'' value,
of 1 in Planck units, then spacetime would have curvature on scales of
the Planck length, meaning that there is no continuum.  Thus any theory
of quantum gravity must provide some mechanism for driving $\Lambda$
to zero.  Given that such a (relatively unknown) mechanism exists, 
causal sets provide a heuristic explanation for why $\lambda$ is not
exactly zero, but fluctuates about zero with an amplitude which falls
off as $1/\sqrt{N}$, where $N$ is the volume of the universe in
fundamental units \cite{causets1}.  Given that for the current era
$N\sim 10^{240}$, this predicts an order of
magnitude that is consistent with current observation.

%(We have one, age of universe and cosmological renormalization.)  --
%I'm not sure exactly what I was thinking here.  I'll just skip it for now.

%usually Lambda>0 achieved via fine tuning
%continuum limit also requires fine tuning of parameters
%(in lattice gauge theory, say)
%so there are two fine tunings needed
%Ising model at Tcrit is a continuum limit, need fine tuning to get
%exactly Tcrit
%feeling is that all discrete theories need such fine tuning
%But does the continuum limit of matter require such fine tuning?!
%coarse graining from off Tcrit leads to trivial theory
%
%they have to do Lambda to small fine tuning
%we have that and continuum fine tuning
%but R believes that two coalesce into one, why continuum?
%furthermore, he thinks that this arises from non-locality
%consv. of mass provides long range correlations in gasses?

In addition, this problem is addressed in part by the renormalization type
behavior of the dynamics and the existence of a continuum limit,
discussed in Sections \ref{contlimsec} and \ref{scaling}.

\subsection{Locality}

Whatever the microscopic dynamics for causal sets may look like, we
expect that the continuum approximation will be governed by an
effective Lagrangian
%In the end we will expect the effective action for causal sets to look like the
%integral of a locally defined quantity in spacetime, even though on
%the causal set itself it may appear extremely non-local.  
%This is much
%more difficult with causal sets than in ordinary `Euclidean' type
%lattices, largely due to the non-compactness of the Lorentz group.
%The obvious form for the effective action is
\bne
L_{\mathrm{eff}} \sim -\Lambda + \frac{R}{2\kappa} + (R^2) + \cdots
\label{cont_act}
\ene
where $(R^2)$ represents terms involving the curvature squared, etc.
Dimensional consistency indicates that the coefficient before each term gets
smaller in the expansion, since the curvature $R$ has dimension
$1/\mathrm{length}^2$, so each $R$ must have a coefficient of $l_p^2$,
implying that the higher order terms have negligible contribution.
%Thus high order terms have extremely small coefficients.
%Einstein-Hilbert action
%\be
%S \to \half \int R dV
%\ee
%A problem with this is that it effectively
%requires a manifold for its
%definition.
%  It is hard to define locality for a causet, even though
%this dynamics of GR is local.
However, this expression will be difficult to compute, even using
the method of estimating curvature from counting in an interval
mentioned in \S \ref{curvature},
because almost every Alexandrov set is ``extremely null''.
%Apparently 't Hooft pointed this out(?) (I can't find it.)
%Note that most all links are nearly null in any reference frame, for a
%causet which is resembled by a spacetime manifold.  non-compactness of
%the Lorentz group...
%This problem seems to make the task of formulating a locally defined
%action on a causal set, using only ``obvious'' aspects of the partial
%order, extremely difficult. % in general.

However, the causal set is an inherently non-local object, so it is
not unreasonable to expect that the notion of locality in the
continuum will not carry over in an obvious, direct manner to the
discrete dynamics.  In fact the discrete dynamics, in its current
formulation (see e.g. (\ref{PC0})), appears quite non-local, in that
the ``behavior'' of a ``region of the causal set'' depends
on its entire past. % of a region of the causal set.

Rather than directly trying to reproduce the action of
(\ref{cont_act}) (say with some additional matter terms), an
alternative approach is to get locality later, in the effective,
continuum theory.
Then the objective would be to use the notion of locality as a guide
in choosing the microscopic dynamics, trying to determine what it
means in this context, without worrying necessarily
about getting an action as in (\ref{cont_act}).  If we do manage to
choose an effectively local microscopic dynamics, then the Einstein-Hilbert
action will come out ``for free'', given the dimensional arguments
above (and local Lorentz invariance).  For the case of causal sets,
this approach seems more likely to bear fruit.
% Action will be log of the amplitude/probability.  Is this necessary
% to mention anywhere?
%What is the meaning of locality for the discrete order?  Perhaps it
%will tell us something about how to choose the free parameters of the
%theory.  
%The idea then will be to arrive at an effective continuum action which is
%additive over disjoint, spacelike separated regions of the
%causal set.  
%Roughly speaking locality translates into the statement that the
%action is additive over disjoint, spacelike separated regions of the
%causal set.  
% Mention issue of why constant term does not dominate over R term?
%Once this is understood in detail then the appropriate form for the
%effective continuum action will follow.
% RDS -- Locality is difficult to obtain, but also powerful if we do
% obtain it.

In general it is difficult to define a discrete ``lattice'' which is
Lorentz invariant, as most regular lattices that one considers are not
invariant under Lorentz transformations.  
However, the set of points obtained from a random ``sprinkling'' into a
spacetime region is Lorentz invariant.
Taking advantage of this property of a random embedding, progress has been made
in understanding how an effectively local action may
arise in the context of causal sets \cite{daughton,salgado}.
%massless scalar field
%\be
%S=-\half\int(\nabla\phi)^2 d^2x \longrightarrow \oint + \half\int\phi\Box\phi d^2x
%\ee
%$G=$``$\Box^{-1}$''
Their work shows that Lorentz invariance can be made compatible with
locality on a lattice.
%  Shows how to
%compute something like the Dalambertian for a causet.
%2d causal metric
%4d links
%based on domain of support of Green's functions.
%[It would be good to provide some discussion of their work, but I am
%too lazy right now, so I'll just skip it.  Try to fit it in later, if
%there is time.]

This apparent success in %successfully 
combining locality, Lorentz
invariance, and discreteness demonstrates a great advantage of causal
sets.  Never before have all these three aspects been present in a
physical theory.

\chapter{Investigation of Transitive Percolation Dynamics}
\label{chaptranperc}

\section{Introduction}
\label{tranperc_intro}

% See also Daughton thesis, p. 43, he discusses some of the same issues.

%: domain in sum over histories
%As mentioned in section \ref{gencov}, [it isn't really mentioned]
The dynamics of causal sets will likely find its final expression as a
quantum measure defined over suitably chosen classes of ``histories'',
where in this case a history will be simply a causal set.\footnote{
Much of the text of this chapter is
taken directly from \cite{contlim}.}
One may
expect, in analog with the path integral formulation of quantum
mechanics, that the quantum measure will arise from a sum over
histories, which may have a form similar to
\bne
 \sum_{C,C'} A(C,C';\{q\})
 \label{soh}
\ene 
where $A$ is a complex amplitude for a pair of causal sets $C, C'$, possibly
depending on a set of parameters $\{q\}$.
A difficulty in defining the quantum measure in terms of a sum of this
nature is that the sum would likely have to be constrained to
``Schwinger histories'', which are pairs of histories that have the
same ``value'' at some time ``$T$'' which is to the future of any
constraint which is used to define the set of histories for which one
is seeking the measure.  Because there is no covariant notion of a
time $T$ in cosmology, and the notion of the ``value'' of a causal set
at a ``time $T$'' is also difficult to define, it is difficult to see
how to directly write down a measure of the form (\ref{soh}).
Instead, the quantum measure will probably arise via a construction
analogous to that which defines the classical measure (\ref{PC0}).
%In fact it appears to be quite difficult to define the quantum measure
%in terms of a sum of this nature, because the sum would have to be
%constrained to ``Schwinger histories'', i.e. pairs of histories that
%have the same ``value'' at some time ``$T$'' which is to the future of
%any constraint which is used to define the set of histories for which
%one is seeking the measure.  Because there is no covariant notion of a
%time $T$ in cosmology, and the notion of the ``value'' of a causal set
%at a ``time $T$'' is also difficult to define, it appears that the
%quantum measure for causal sets may in fact not have the form in
%(\ref{soh}), but instead arise by via a construction analogous to that
%which defines the classical measure (\ref{PC0}).
%where A is e^{i/hbar S(\gamma)}, guessing S, 
%hard to directly guess action for A, instead follow construction ...

Even though we do not know the exact form of the summand, a question
which presents itself is how to enumerate the causal sets which form
the domain of the sum itself.  This problem has been studied
extensively, often in the context partial orders as transitive,
acyclic, directed graphs.  In particular, Kleitman and Rothschild
\cite{KR} (see also \cite{3layer}) have shown that, in the asymptotic
limit $N\to\infty$, the number of distinct orders definable on $N$
elements is given by
\be
(1+O(1/N))\phi_p \frac{2^{3/4}}{\sqrt{\pi}} \, 2^{ N^2/4 + 3N/2 - \log_2N/2} \:,
\ee
where 
\be
\phi_p = \sum_{j=-\infty}^\infty 2^{-(j+1/2)^2} \approx 2.1289312 \:
\ee
for even $N$ and 
%A similar expression holds for odd $N$:
%\be
%(1+O(1/N))\frac{4\phi_2}{3\pi} \, 2^{N^2/4 - N\log_2N +
%(3/2+\log_2e)N - \log_2(N+1)} \:,
%\ee
%where
\be
\phi_p = \sum_{j=-\infty}^\infty 2^{-j^2} \approx 2.1289368
\ee
for odd $N$.
Thus, for any appreciable
value of $N$ (say $N>20$), in the absence of some special amplitude
$A(C,C';\{q\})$ which for example is zero %vanishes 
on all but a 
vanishingly small fraction
of the $N$-element causets, % $C$, 
it seems that, in practice, the
sum in (\ref{soh}) must performed by a simulation or other
approximation method.  An
important question then is how to sample the set of $N$-element causets.
%causal sets.
% I invoke KR to argue for Monte-carlo, then again to argue kill it.
% This is a little silly.  I should probably fix this.  Let me leave
% it for now, though, anyway.  Maybe even show it to Rafael...
% I think that it is not so bad, thinking deeper we see a flaw.
%
% Also one might complain about why not use metropolis algorithm here
% to do sum, but this is essentially Monte Carlo anyway.

%\section{The dynamics of transitive percolation}
%: \section{Transitive Percolation}
\label{perc}
There exists a simple ``model'' for generating partially ordered sets
at random, which is familiar in the field of random graph theory,
% [Why this last clause / ``parenthetic expression''?] I dunno.
which we call \emph{transitive percolation}.
The name, %which was
suggested by David Meyer \cite{tranperc}, arises from the fact that this model can be
regarded as a sort of one-dimensional directed percolation, where a
relation $i\prec j$ is thought of as a ``bond'' or ``channel'' between ``sites''
$i$ and $j$ in a one dimensional lattice
(c.f. e.g. \cite{schulman}).  
It is defined by a single real parameter $p$ (and a non-negative integer
$N$).
To generate an $N$-element poset at
random, start with a set of $N$ elements labeled $0, 1, 2, \ldots
N-1$, and introduce a relation between each of the ${N \choose 2}$
pairs of elements with a probability $p$ (with the element with the
smaller label preceding that with the greater), where $p$ is any real
number in $[0,1]$.  Since the resulting relation will not be
transitive in general, form its transitive closure (e.g. if $2\prec 3$
and $3 \prec 438$ then enforce that $2\prec 438$).  

%: form of KR posets -- (needs some editing if not also failure of MC)
If transitive percolation is to be used to sample the domain of
summation in (\ref{soh}), then we need to understand in detail the
resulting distribution on the set of $N$-element causal sets, so a
weight factor can be placed into the summand to correct for the bias
of the sampling technique.  Unfortunately, it is impossible to do
this, for the following reason.  The asymptotic enumeration of
$N$-element orders found by Kleitman and Rothschild mentioned above
was achieved by showing that almost all $N$-orders are ``3-layer''
orders.  (An ``$l$-layer order'' is one in which the set of elements is
partitioned into $l$ antichains $X_1$, $X_2$, $\ldots, X_l$, where each
element of $X_i$ precedes every element of $X_j$ for $j>i+1$, and no element of $X_i$ precedes any
element of $X_j$ for $i>j$.)  Furthermore, they found that almost all
3-layer orders have about $N/2$ elements in $X_2$ and about $N/4$ elements in the
other antichains.  Here ``almost all'' means that the fraction of
orders with this characteristic goes to 1 in the limit $N\to\infty$.
%Make some comment about the counter-intuitivity of this result?
This result tells us that essentially all posets sampled will be
3-layer, so that the weight factor will degenerate to zero for any
non-3-layer posets, %leading to a sort of ``entropy catastrophe''.
which bodes ill for the whole approach of doing a
Monte-Carlo sum.
%no weight factor will correct for the bias

%: Dhar type phase transitions
(In connection with generic, layered orders, Deepak Dhar \cite{dhar}
and Kleitman and Rothschild \cite{KRentropy} have studied the behavior
of an entropy function on these posets, $S(r)$, where $r$ is the ordering
fraction defined in \S \ref{MMdim}.
%, utilizing the asymptotic form of Kleitman and Rothschild.
They found an infinite number of first order phase transitions, at
each of
which $\partial S/\partial r$ vanishes over a finite interval of $r$.
The order parameter is the average height, which increases by one
across each transition.
%Regarding $r$ as a sort of energy, they found that there exist a
%number of phase transitions for multiple values of $r$, i.e. that
%$S(N,r)$ changes discontinuously with $r$ at these ``critical''
%values.  
In addition, they have found that, for a given $r$, most causets are highly
time-asymmetric.  The presence of the phase transitions suggests that
there may be a continuum limit.)
%the presence of a continuum limit.)
%that
%there may be a continuum limit there.
% This paragraph is crummy, but I am tired and want to go to bed.
% Please fix it someday, by studying phase transitions in Kittel,
% discussing the k-layer posets, etc.
%  Why is it here, anyway?  It is a feature of generic posets
%  (sort-of).  It seems misplaced in reading, though.

Obviously these 3-layer posets in no way resemble those which would
faithfully embed into a spacetime.  
%However, in the limit of large
%$N$, these are the only causal sets which will arise in a random
%sampling.  
%However, 
%given that they will
%dominate the sum with a super-exponental entropic weight factor, 
Since their number grows exponentially in $N^2$, one may
imagine that any dynamics for causal sets is doomed to failure, since
any Boltzmann-like weight which ``only'' grows exponentially in an
extensive quantity (e.g. energy) would be insufficient to overcome this
super-exponential entropic weight factor. % favoring the generic causets.
Thus we have a sort of entropy catastrophe, forcing generic causets
upon us regardless of our choice of dynamics.
However, the causal sets generated by the transitive percolation
algorithm look nothing like the generic 3-layer orders.  If this model
is to be regarded as a physical dynamics in itself, then this entropy catastrophe
is already forestalled with this quite naive dynamical model.  In
fact, %looking back at it, 
we can see that the dynamics of causal sets,
being inherently non-local, would be expected to have an action which
grows quadratically with an extensive quantity, rather than linearly.
Then this sort of non-local action is exactly what is needed to
overcome the entropic dominance of the generic orders.  (In fact, the
probability of arriving at a causal set with $R$ related pairs, via
the transitive percolation algorithm, grows like $e^{\beta R}$, where
$\beta=-\ln p$ acts as a sort of inverse temperature.  c.f. \cite{AChR})
Note that this situation is not so different from that of ordinary quantum
mechanics, where the smooth paths, which form a set of measure zero in
the space of all paths, are the ones which dominate the sum over
histories in the classical limit.
%
%seems doomed to
%failure, since the entropic weight factor favoring the generic causets
%grows faster than exponential, but one would naievely imagine that a
%Boltzman-like weighting of $e^beta E$, being ``only'' exponential in
%an extensive quantity, would be incapible of overcoming
%this favoring of generic causets at large N.
%This leads to a sort of entropy catastrophe.
%
%, but they will be the only ones
%which arise by randomly sampling causets for the domain of summation
%in (\ref{soh}). 
%Then this strategy of performing the sum by
%Monte-Carlo is doomed to failure, since the only causets which are
%sampled will be ones that we do not expect to 
%%contribute significantly to the overall measure.
%be physically significant.
%Thus there seems to be a sort of ``entropy catastrophe'' for causal
%sets, being that there is a super-exponential entropic weight factor favoring
%these unphysical generic 3-layer causets.
%However, it turns out that the causal sets generated by the
%transitive percolation algorithm 
%described above look nothing like the generic 3-layer orders.
%If this model is regarded as a physical
%causal set dynamics in itself, then
%the catastrophe is already forestalled with this naive model of
%random causal sets.

% [ I don't think that I will bother with this paragraph, as I am now
% somewhat convinced (but only somewhat!) that the genericity will emerge at a
% ``reasonable'' value of N. ] -- I changed my mind...
One important question which has not been addressed is at what value
of $N$ the Kleitman Rothschild result becomes valid.  Enumeration of
partial orders by computer shows no obvious tendency toward the
3-layer orders, for the meager values of $N$ which a computer allows.
%[[ Can I say this?  Can I say which N's have been examined? ]]
It is possible, however unlikely, that the result will be
of no consequence for causal sets, as it emerges only after $N$ is
much larger than will ever be needed for physically reasonable causal
sets, say $N\gg 10^{240}$.
%Should I put a footnote here explaining the significance of this number?
In any event it would be useful to have a feel
for the ``domain of validity'' of this asymptotic result.

We will see that in fact transitive percolation can be regarded as
much more than just an algorithm to generate causal sets at random to be
used in a Monte-Carlo sum over histories.  It is an important
special case of a generic class of ``sequential growth'' dynamics for
causal sets, which will be explained in detail in Chapter
\ref{dynamicschap}.
%Given the simplicity of this dynamical model, both conceptually and
%from an algorithmic standpoint, it offers a ``stepping stone''
%allowing us to look into some general features of causal set dynamics.
%[Express the importance of how it is a fixed point in cosmological
%renormalization here somehow?]
%From a physical point of view, 
%Transitive percolation, viewed as a dynamical model in itself, 
In particular, it has many appealing features, both as a model for a
relatively small region of spacetime and as a cosmological model for
spacetime as a whole.
%\section{popularity}
Incidentally, it has attracted the
interest of both mathematicians and physicists for reasons having
nothing to do with quantum gravity.  By physicists, it has been studied
as a problem in the statistical mechanical field of percolation.
%, as I have already alluded to.  
By mathematicians, it has been studied extensively as a branch of
random graph theory (a poset being the same thing as a transitive
acyclic directed graph).
% Omit just because it repeats a footnote in the intro?  Decide
% what do do here.
%
Conversely, random graph theory could be construed as the theory of
percolation on a complete graph.
Some physics references on transitive percolation %(viewed from whatever angle)
are 
\cite{schulman,AChR,contlim,scaling}.
%\cite{brightwell,posts,bb,schulman,AChR,contlim,scaling}.
In connection with random graph theory, there exist
%Another thing making it an attractive
%special case to work with is the availability in the mathematics
%literature of 
a large number of results governing the asymptotic behavior of
posets generated in this manner %\cite{mathrefs, AlonEtAl}.
\cite{brightwell,posts,bbdim,width,pt,cst,scc,simon,AlonEtAl}.
%[ Is this correct, what these references examine? ]
% I decided to put these later, in ``large scale effective theory''.

\section{Features}
% This is already mentioned above.
%Aside from its convenience, this percolation dynamics possesses a
%number of distinguishing features, including an underlying
%time-reversal invariance and a special relevance to causal set
%cosmology.
%%, as we describe briefly below.  
%%In addition, from the gravity viewpoint this model has
%%many ``physically appealing'' features:

\subsection{May resemble continuum spacetime}
\label{continuum}
In computer simulations, two independent coarse-graining invariant
dimension indicators, Myrheim-Meyer dimension and midpoint scaling
dimension, tend to agree with each other, which is encouraging if
these causal sets are to embed faithfully into spacetime with a well
defined dimension.\footnote{All these numerical calculations were
performed by R.~Sorkin.}  Another dimension indicator, which involves
counting small subcausets whose frequency provides an indicator of
dimension, behaves poorly.  However, this measure of dimension is not
invariant under coarse graining, so it only indicates that transitive
percolated causal sets themselves do not directly embed faithfully
into
%(an interval of?) 
Minkowski space, but some appropriately chosen subcauset
(e.g. coarse-grained) may still approximate a spacetime, which is what
one would expect for a dynamics of causal sets anyway.
%(Some other indicators of manifold-like behavior
%have tended to do much more poorly, but those are not invariant under
%coarse graining, whereas one would in any case expect to observe
%manifold like behavior only for a sufficiently coarse grained causal
%set.) 
% [ Exposition here depends on whether I put nugget dimension into
% the introduction.]

In the pure percolation model, however, these dimension
indicators vary with %scale, i.e. size of the interval considered,
time (i.e. with $N$, as the causal set ``grows'')
which suggests that one may wish to 
rescale $p$ in such a way as 
%if one wishes 
to hold the spacetime dimension constant.\footnote
{\singlesp This is only a suggestion because these estimators neglect
curvature.  Transitive percolation could of course produce something
resembling a region of a curved spacetime, such as de Sitter or
anti-de Sitter.}
One may ask, then, if the model can be generalized by having $p$ vary
with $N$ in an appropriate sense.  We will see in Chapter
\ref{dynamicschap} that something rather like this is in fact
possible.\footnote
{\singlesp It should be noted that the measure of dimension that is
varying with $N$ is that of the causal set in its entirety, not that
of a ``local region''.  In fact, it seems that, due to the homogeneity
of percolation, the dimension of a region will depend only on its
size.  Thus as $N$ increases, the dimension associated with that
$N$-element ``region'' (the entire causal set) changes
uniformly.  Then, more correctly, it is the \emph{scale} dependence of
dimension in transitive percolation which may suggest that $p$ should
vary somehow.}
%Granted, in the continuum GR description of cosmology, spacetime is
%described in terms of a manifold, whose dimension cannot change by
%definition.  Thus the notion of a time or scale dependent dimension is
%foreign to differential geometry/GR.  (There can be an effective scale
%dependent topology, but all the fine details of the manifold are
%retained at all length scales -- it is the coarse graining allowed by
%the discrete theory of causal sets that allows a precise notion of
%scale dependent topology.)  But one may intuitively see how even the
%Friedman universe, suitably extended, may become one dimensional after
%a suitably long time.  The argument is to extend the FRW universe by
%stringing an infinite number of copies together in a line, ``big bang
%to big crunch''.  Then at the largest scale this just looks like
%M^1.}
%
%  continuum limit is not Minkowski, but some weird cosmology, maybe
%Better would be Rafael's idea for smaller and smaller intervals in
%percolation.  This may be more fruitful!!
%As $N\to\infty$, for fixed $p$, $r\to\infty$.  This is reasonable with
%a cosmological interpretation.  So behavior is o.k., don't need
%varying $p$

\subsection{Homogeneous}
\label{homogeneous}

Consider the transitive percolation algorithm described in \S
\ref{perc}, and an arbitrary element of a causal set generated by this
model, say the one labeled 257.  Its future will be some causal set,
$\fut(257)$.  Because of the extreme symmetry of the transitive
percolation algorithm, the probability distribution of $\fut(257)$
will be completely independent of the structure of that portion of the
causal set which is to the past of element 257, %and also the portion 
or spacelike
to 257.  This is clear because, regardless of what is to the
past and unrelated to this element, each successive element will join to 257
with a fixed probability $p$.  Thus its future will behave the same as
that of any of the other elements.

%For transitive percolation, the future of any element of the
%causal set is completely independent of anything ``spacelike related'' to
%that element, or to its past.  %This is evident from the simple 
Therefore, the only spacetimes which a causal set generated by
transitive percolation could hope to resemble would be (space-time)
homogeneous, such as the Minkowski, de Sitter, or anti-de Sitter
spacetimes.  Likewise transitive percolation has no hope of resembling
a spacetime with
%can be no 
propagating degrees of freedom, such as gravitational
waves.
%but neither of these possibilities is compatible with the periodic
%re-collapses alluded to earlier.  At best, therefore, one could hope
%to reproduce a small portion of such a homogeneous spacetime.
% Whoops -- Goedel spacetime is spacetime homogeneous as well, so there
% must be more than just these two examples...
%[ I don't understand...  more than just spatially homogeneous I guess.  I see spacetime homogeneous ==> no singularity ==> entire universe is not described by transitive percolation ]

\subsection{Time reversal invariance}
Transitive percolation is independent of time orientation.  When
viewed from the perspective of a sequential growth dynamics, this may
not be so obvious, but it is clear when viewed from the more
static algorithm described above.

\subsection{Existence of a continuum limit}
Moreover, computer simulations suggest strongly that the model
possesses a continuum limit (see Section \ref{contlimsec}) and exhibits
scaling behavior in that limit with $p$ scaling roughly like
$c\log{n}/n$ \cite{scaling}.

\subsection{Originary transitive percolation}
\label{originarysec}
There exists another model which is very similar to transitive
percolation, called ``originary transitive percolation''.  It is most
clearly described in terms of a ``cosmological growth process'' which
%is discussed in detail 
is introduced in \S \ref{growth}.
For now, suffice it to say that the model is the same as
transitive percolation, except that every element (but one) must be a descendent
of at least one other element of the causet.
The net effect is
that the growing causal set is required to have an ``origin'' (=
unique minimum element).
%every causet which arises will
%have a unique minimal element, an ``origin''.
%The rule
%for randomly generating a causet is the same as for transitive
%percolation, except that each new 
%element is required to be related to
%at least one ``existing'' element (one with a smaller label).  
%(see section \ref{growth} for a more complete explanation of this
%process),
It turns out that originary transitive percolation is equivalent to
ordinary transitive percolation, if one ``discards'' all elements
which are not to the future of the first element.  That is, if one
generates a causet via transitive percolation with $N$ and $p$, and
then considers only the subcauset which contains
the (inclusive) future of ``element 0'', one obtains a model
equivalent to that of originary transitive percolation at the same $p$
(but of course smaller $N$).
%[ Is this correct?  Should I use it as an alternative description of
%originary percolation, and therefore avoid the ``suffice it to say'' above?]

\subsection{Suggestive large scale cosmology}
\begin{figure}[htbp]
\center
\scalebox{.7}{\includegraphics{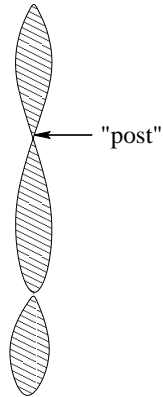}}
\caption{Transitive percolation cosmology}
\label{tranperccos}
\end{figure}

Consider a picture of causal set cosmology which involves %the causal
%set undergoing 
cosmological ``bounces'', where the causal set collapses
down to a single element, and then re-expands
%, so that this single
%element is related to every other element of the causal set, 
as illustrated in Figure \ref{tranperccos}.
Alon {\it et al.} \cite{AlonEtAl} call such an element a {\it post},
which is defined %formally 
as an element which is related to every
other in the causal set.  In the context of percolation dynamics, they
have proved rigorously that
%the causal set undergoes 
such cosmological bounces 
%(which they call {\it posts})
%where the causal set collapses down to a single element, and
%re-expands, so that this single element is related to every other
%element of the causal set.
occur with
probability 1 (if $p>0$), from which it follows that there are
infinitely many cosmological cycles, each cycle but the first having
the dynamics of originary percolation.
%This divides the causal set into ``cosmological eras'' as illustrated
%in Figure \ref{tranperccos}.
Then the ``cosmology'' of transitive percolation is quite suggestive,
consisting of a
universe which cycles endlessly through phases of expansion, stasis, and
contraction (via fluctuation) back down to a single element.
% The stasis doesn't follow, but let's leave it for now.
% Should I put a footnote explaining what width ~ constant?

Note that transitive percolation is only
homogeneous on average.
Thus, with this sto\-chas\-tic model, we see a phenomenon which does not arise
in deterministic theories --- a ``locally homogeneous spacetime''
which nevertheless 
possesses points where the universe contracts to
``volume 1'' and reexpands.
Also, this means that transitive percolation cannot produce
the entirety of an Einstein spacetime, it is only possible that its
continuum limit yield a portion of some homogeneous spacetime.
%Note that the appearance of posts is inconsistent with homogeneity for
%any Einstein spacetime.  
%However, transitive percolation is only
%homogeneous \emph{on average}.
%, so this result implies that, as a whole,
%transitive percolation will not reproduce any spacetime.  
%At
%best, it may reproduce a limited region of spacetime.  
%(This apparent
%contradiction, a homogeneous ``spacetime'' with posts, arises because we
%are more familiar with deterministic models for spacetime, wherein
%such an event never occurs.)

%Originary percolation might have an important role to play in
%cosmology, for the following reason.  If a given cosmological
%``cycle'' ends with the causet collapsing down to a single element,
%then the ensuing re-expansion is necessarily given by an originary
%causet.  

\subsection{Cosmological renormalization}
\label{cosrenormsec}

In Chapter \ref{dynamicschap}, the classical dynamics for causal sets will
be expressed in terms of a countable sequence of parameters, or
``coupling constants'', $t_n$.
Some work by Dou \cite{dou}, and more recently
\cite{causetcosmo,cosrenorm}, describes a ``cosmological
renormalization'' process, wherein at each cycle of expansion,
collapse to a single point, and re-expansion, the parameters
%``coupling constants'' 
describing the dynamics of the causal set are ``renormalized'', taking
new effective values in each subsequent cycle.  It is easy to show
that the percolation dynamics are the unique ``fixed points'' under
this renormalization flow, and furthermore that a large class of
dynamics (choices for the parameters $t_n$) converge to this fixed
point as the renormalization process extends to infinity.  Thus, if
our universe is in fact described by something resembling the dynamics
to be derived in Chapter \ref{dynamicschap}, and it does undergo
cycles of expansion and re-contraction, then after a long time it will
be increasingly described by this (originary) transitive percolation
dynamics.

Furthermore, if the parameters $t_n$ take the form given in
(\ref{lifelike}), which might be expected for a physically reasonable
dynamics, then the cosmological renormalization process makes an
interesting ``prediction'' regarding the early universe.
As discussed in \cite{causetcosmo}, for the dynamics of
(\ref{lifelike}), the early universe behaves as originary transitive
percolation for a period which grows longer and longer after successive cosmological
renormalizations.  
In addition,
%Under the action of the cosmological renormalization, 
the effective $p$ of this percolation phase
diminishes as $1/\sqrt{N}$, with $N$ being the number of elements to
the past of the current cosmological epoch.  Thus after a long time,
$p$ will be driven to an arbitrarily small value.
%The remarks of \cite{causetcosmo} suggest that 
%the early universe 
%% i.e. the universe immediately after one of these bounces, assuming
%% that they occur for a more general dynamics
%may be described by something like originary percolation, with an
%extremely small value of $p$.  
If this is the case, then it can provide some explanation to the
puzzle of why the universe is so large, homogeneous, and isotropic.
The reason is that, for extremely small $p$, originary percolation
will almost surely generate a ``Cayley tree'', which is a tree for
which each element has on average two immediate successors.  After
this rapidly expanding ``tree phase'' the universe
should make a transition into
%grow into
something resembling a spacetime
which is
%out of which will arise a spacetime
%which is 
spatially homogeneous and isotropic.
%some transition period
%This describes a universe which expands exponentially with
%time, and is homogeneous and spatially isotropic.
%, which is consistent
%with current observation.

\subsection{Phase transitions in the early universe}
For the (non-originary) transitive percolation dynamics, it is known
that there is a percolation phase transition at $p\sim{1/N}$, where
the causet transforms qualitatively from a large number of small
disconnected universes\footnote {\singlesp By the term ``universe'' I
mean simply a connected component of the order.}
%, that is, regarding the Hasse diagram as a (undirected) graph, 
%a set of vertices connected by some sequence of edges.}
%such that there exists a pa
for $p<p_{\mathrm{crit}}$ to a causet with one large
universe and a number of much smaller disconnected universes for
$p>p_{\mathrm{crit}}$.  
(This can be regarded as the ``early universe'' of transitive
percolation 
%in the sense that 
%the vales of $N$ considered in the simulations of \S
%\ref{contlimsec}, and discussed above, are generally much greater
%than $1/p$, and are qualitatively different in structure.  
because cosmological time can be measured by spacetime volume, so that
small $N$ corresponds to short or early times.)
A second phase transition gathers the
disconnected branches of the universe, leading to a single connected
universe.  This occurs near the percolation transition, at $p\sim
\ln N/N$ \cite{pt}.  In fact, this second phase transition occurs ``at the same
time'' as a third, at which the fraction of elements to the future of
element 0 becomes very close to 1.

There is, incidentally, still some hope of being able to reproduce the generic
3-layer posets by running the transitive percolation algorithm at the
percolation phase transition.
%which (might?) exists 
%at around $p \sim 1/N$.  
%I have not studied this possibility.  
%The hope would be 
It is possible that each piece (connected component) of the causet
would be a generic poset.

\subsection{Diffusion-like model}
\label{diffusion}

The expectation value of the ordering fraction $\E{r}$ of a causet generated
by transitive percolation can be computed exactly by writing a
recursion relation for the number of descendants of element
0 (we'll call this element $e_0$ for short).\footnote{
This recursion relation is due to R. Sorkin.}
For
the purposes of this discussion we switch to the reflexive convention for
defining a partial order, i.e. replace the irreflexivity condition
$x\not\prec x$ with reflexivity $x\prec x$ (and re-impose acyclicity with $x\prec y \prec x \implies x=y$).
Define
$f_n(k)$ to be the probability that, considering only those elements
whose labels are less than $n$, $e_0$ has exactly $k$ descendents.
Clearly $f_1(k)=\delta_1^k$.
The recursion relation can be defined %as follows
by noting that ``at stage $n$'', there are two ways that $e_0$ can
have $k$ descendents.  ``At stage $n-1$'' either $e_0$ had $k$
descendents, and element $n$ does not ``link to'' any element to the future
of $e_0$, or $e_0$ had $k-1$
descendents, and element $n$ does link to one of $e_0$'s descendents.
The former event occurs with probability $q^k$, while the latter with
probability $1-q^{k-1}$.  Thus
\bne
f_n(k) = q^k f_{n-1}(k) + (1-q^{k-1}) f_{n-1}(k-1) \:. \label{recursion}
\ene
The expected number of descendents of $e_0$ ``at stage $n$''
%(for an $N$-element causet) 
is then 
\be
x_n = \sum_{k=1}^n k f_n(k) \:.
\ee
Because of the symmetry of the percolation algorithm, the expected
number of descendents of element 1 at stage $n$ is equivalent to the
expected number of descendents of element 0 at stage $n-1$.  (To see
this, simply relabel the causet such that element $i$ is relabeled
$i-1$.)  Then the expected number of relations in an $N$-element
percolated causet, is $\sum_{n=1}^N x_n$, or switching back to the
irreflexive convention,
\be
\E{R(N)} = \sum_{n=1}^N (x_n-1) \:.
\ee
%This can now be solved ``exactly'' (on a
%computer), to get $<r>$ accurate to numerical rounding errors.
This recursion relation can be evaluated quite efficiently on a
computer, to yield values of $\E{r}$, so far for $N$ up
to $2^{21}$, accurate to numerical rounding errors.

If this Markov process is modeled by a differential equation, the
``field'' $f$ behaves as a wave moving at constant speed ``to the
right'', with a diffusive character.  It is possible that further
study along these lines will lead to an understanding of the
asymptotic behavior of this model, for example to understand the
scaling behavior discussed in \S \ref{scaling}.

A generalization of this recursion relation is possible for the
originary percolation model, but it is more expensive computationally
than the $O(N^2)$ algorithm described here.
%of (\ref{recursion}).
%can also write one for originary percolation (I could figure this out
%but I don't feel like it so much)

\subsection{Gibbsian distribution}
\label{two_temp}
%Also transitive percolation can be described within a model of a two
%temperature lattice gas.  See \cite{AChR}.
Transitive percolation is readily embedded in a
``two-temperature'' statistical mechanics model, and as such, happens
also to be ``exactly soluble'' in the sense that the partition function
can be computed exactly.  Details of this model will be described in
\cite{AChR}. %,scaling}. [[ Should we include this in the scaling
	     %paper? rds says no]]

\section{Continuum Limit}
\label{contlimsec}

%[ What do I have to do here to give credit to RDS for much of these ideas/text?]

Here we search for evidence of a continuum limit in the transitive
percolation dynamics.  One might question whether a continuum limit is
even desirable in a fundamentally discrete theory, but a continuum
{\it approximation} in a suitable regime is certainly necessary if the
theory is to reproduce known physics.  Given this, it seems only a
small step to a rigorous continuum limit, and conversely, the
existence of such a limit would encourage the belief that the theory
is capable of yielding continuum physics with sufficient accuracy.

% sound fast, diffusion slow, ratio between them goes to infinity in
% the continuum limit
Perhaps an analogy with kinetic theory can provide a useful illustration.
In quantum
gravity, the discreteness scale is set, presumably, by the Planck
length $l=(\kappa\hbar)^{1/2}$ (where $\kappa=8\pi{G}$), whose
vanishing therefore signals a continuum limit.  In kinetic theory, the
discreteness scales are set by the mean free path $\lambda$ and the
mean free time $\tau$, both of which must go to zero for a description
by partial differential equations to become exact.  Corresponding to
these two independent length and time scales are two ``coupling
constants'': the diffusion constant $D$ and the speed of sound
$c_{\mathrm{sound}}$.  Just as the value of the gravitational coupling
constant $G\hbar$ reflects (presumably) the magnitude of the
fundamental spacetime discreteness scale, so the values of $D$ and
$c_{\mathrm{sound}}$ reflect the magnitudes of the microscopic
parameters $\lambda$ and $\tau$ according to the relations
$$
D \sim {\lambda^2 \over \tau} , \quad c_{\mathrm{sound}} \sim
{\lambda \over \tau}
$$
or conversely
$$
   \lambda \sim {D \over c_{\mathrm{sound}}} , \quad 
   \tau \sim {D \over c_{\mathrm{sound}}^2}   \,.
$$
In a continuum limit of kinetic theory, therefore, we must have either
$D\to0$ or $c_{\mathrm{sound}}\to\infty$.  In the former case, we can
hold $c_{\mathrm{sound}}$ fixed, but we get a purely mechanical
macroscopic world, without diffusion or viscosity.  In the latter
case, we can hold $D$ fixed, but we get a ``purely diffusive'' world
with mechanical forces propagating at infinite speed.  In each case we
get a well defined --- but defective --- continuum physics, lacking
some features of the true, atomistic world.

If we can trust this analogy, then something very similar must hold in
quantum gravity.  To send $l$ to zero, we must make either $G$ or
$\hbar$ vanish.  In the former case, we would expect to obtain a
quantum world with the metric decoupled from non-gravitational matter;
that is, we would expect to get a theory of quantum field theory in a
purely classical background spacetime solving the source-free Einstein
equations.  
% i.e. QFT in Ricci flat background
In the latter case, we would expect to obtain classical
general relativity.  Thus, there might be two distinct continuum
limits of quantum gravity, each physically defective in its own way,
but nonetheless well defined.

For our purposes, the important point is that, although
we would not expect quantum gravity to exist as a continuum theory, it
could have limits which do, and one of these limits might be classical
general relativity.  It is thus sensible to inquire whether one of the
classical causal set dynamics we have defined describes classical
spacetimes.  In the following, we make a beginning on this question by
asking whether the special case of ``percolated causal sets'', as we
will call them, admits a continuum limit at all.%, i.e. whether the
%dynamics provides a non-trivial theory in the limit of $l\to 0$.  Here
%I examine a closely related limit, in which $N\to\infty$.
% [ Let's just leave it for now, maybe ask Rafael what he thinks.
% Also, above sentences are not correct, note ``In the following''.
% Also I think that this is explained below. ] ??

Of course, the physical content of any continuum limit we might find
will depend on what we hold fixed in passing to the limit, and this in
turn is intimately linked to how we choose the coarse-graining
procedure that defines the effective macroscopic theory whose
existence the continuum limit signifies.  Obviously, 
we will want to
send $N\to\infty$ for any continuum limit, but it is less evident
exactly how
we should coarse-grain and what coarse grained parameters we want to
hold fixed in taking the limit.  Indeed, the appropriate choices will
depend on whether the macroscopic spacetime region we have in mind is,
to take some naturally arising examples, ($i$) a fixed bounded portion
of Minkowski space of some dimension, ($ii$) an entire cycle of a
Friedmann universe from initial expansion to final recollapse, or
($iii$) an $N$-dependent portion of an unbounded spacetime $M$ that
expands to encompass all of $M$ as $N\to\infty$.  In the sequel, we
will have in mind primarily the first of the three examples just
listed.  Without attempting an definitive analysis of the
coarse-graining question, we will simply adopt the simplest
definitions that seem to us to be suited to this example.  
More
specifically, we will coarse-grain by the random selection procedure
of \S \ref{coarse_graining},
%randomly selecting a
%sub-causal-set of a fixed number of elements, 
and we will choose to
hold fixed some convenient invariants of that sub-causal-set,
including the ordering fraction, which, as mentioned in \S \ref{MMdim},
can be interpreted
as the dimension of the spacetime region it constitutes.\footnote
{\singlesp Strictly speaking this interpretation is correct only if the causal
 set forms an interval or Alexandrov neighborhood within the
 spacetime, but, as mentioned earlier, the notion of Myrheim-Meyer
 dimension remains useful in this wider context.}
%[[RDS: we hope!]]
As we will see, the resulting scheme has much in common with the kind
of coarse-graining that goes into the definition of continuum limit
in quantum field theory.  For this reason, we believe it can serve
also as an instructive ``laboratory'' in which this concept, and
related concepts like ``running coupling constant'' and ``non-trivial
fixed point'',
%% renormalization group scaling 
can be considered from a fresh perspective.
% Illustration of continuum limit and renormalization.  Lot's of
% people have thought about it.  Popular in this sense, people will be
% interested to see a study of this kind.

\subsection{The critical point at $p=0$, $N=\infty$}
Transitive percolation is a model of random causets which depends on
two parameters, $p\in [0,1]$ and $N\in\NaturalNumbers$.  For a given
$p$, the model defines a probability distribution on the set of
$N$-element causets.\footnote%
{\singlesp Strictly speaking this distribution has gauge-invariant meaning only in
 the limit $N\to\infty$ ($p$ fixed); for it is only insofar as the
 causal set ``runs to completion'' that generally covariant questions can
 be asked.  Notice that this limit is inherent in causal set dynamics
 itself, and has nothing to do with the continuum limit considered here,
 which sends $p$ to zero as $N\to\infty$.}
For $p=0$, the only causet with nonzero probability, obviously, is the
$N$-antichain.  Now let $p>0$.  With a little thought, one can
convince oneself that for $N\rightarrow\infty$, the causet will look
very much like a chain.  Indeed it has been proved \cite{bb} (see also
\cite{schulman}) that, as $N\rightarrow\infty$ with $p$ fixed at some
(arbitrarily small) positive number, $r\rightarrow1$ in probability,
where $r$ is the ordering fraction of the causal set. 
%defined in \S \ref{MMdim}.
%\be
%      r \equiv \frac{R}{N(N-1)/2} = { R \over \Nc2 } \,, 
%\ee
%$R$ being the number of relations in the causet,
%i.e. the number of pairs of causet elements $x$, $y$ 
%such that $x\prec{y}$ or $y\prec{x}$.
Note that the $N$-chain has the greatest possible number of
relations $\Nc2$, so $r\rightarrow1$ gives a precise meaning to ``looking
like a chain''.  

We see that for $N\rightarrow\infty$, there is a change in the
qualitative nature of the causet as $p$ varies away from zero, and the
point $p=0, N=\infty$ (or $p=1/N=0$) is in this sense a
critical point of the model.  It is the behavior of the model near this
critical point which we study in detail.
%will concern us in this paper.

The fact that a coarse grained causet is
automatically another causet will make it easy for us to formulate
precise notions of continuum limit, running of the coupling constant
$p$, etc.  
In this respect, we believe that % [[ Is ``we'' a problem here? ]]
this model combines
precision with novelty in such a manner as to furnish an instructive
illustration of concepts related to renormalizability, independently
of its application to quantum gravity.

\subsection{The large scale effective theory}

The transitive percolation model for generating random causal sets
is a ``microscopic'' dynamics,
% for causal sets (that of transitive percolation) 
and the procedure described in \S \ref{coarse_graining} provides 
a precise notion of coarse graining
(that of random selection of a sub-causal-set).
On this basis, we can
produce an effective ``macroscopic'' dynamics by imagining that a
causet $C$ is first percolated with $N$ elements and then
coarse-grained down to $m<N$ elements.  This two-step process
constitutes an effective random procedure for generating $m$ element
causets depending (in addition to $m$) on the parameters $N$ and $p$.
In causal set theory, number of elements corresponds to spacetime
volume, so we can interpret $N/m$ as the factor by which the
``observation scale'' has been increased by the coarse graining.  If,
then, $V_0$ is the macroscopic volume of the spacetime region
constituted by our causet, and if we take $V_0$ to be fixed as
$N\to\infty$, then our procedure for generating causets of $m$
elements provides the effective dynamics at volume-scale $V_0/m$
(i.e. length scale $(V_0/m)^{1/d}$ for a spacetime of dimension $d$).

What does it mean for our effective theory to have a continuum limit
in this context?  Our stochastic microscopic dynamics gives, for each
choice of $p$, a probability distribution on the set of causal sets
$C$ with $N$ elements, and by choosing $m$, we determine at which
scale to examine the corresponding effective theory.  This
effective theory is itself just a probability distribution $f_m$ on
the set of $m$-element causets, 
% So f_m already considers the dynamics, it is not just for a single
% fine-grained causet
so our dynamics will have a well defined continuum limit if there
exists, as $N\to\infty$, a trajectory $p~=~p(N)$ along which the
corresponding probability distributions $f_m$ on coarse grained
causets approach fixed limiting distributions $f_m^\infty$ for all
$m$.  The limiting theory in this sense is then a sequence of
effective theories, one for each $m$, all fitting together
consistently.  (Thanks to the associative (semi-group) character of
our coarse-graining procedure, the existence of a limiting
distribution for any given $m$ implies its existence for all smaller
$m$.  Thus it suffices that a limiting distribution $f_m$ exist for
$m$ arbitrarily large.)  In general there will exist not just a single
such trajectory $p=p(N)$, but a one-parameter family of them
(corresponding to the one real parameter $p$ that characterizes the
microscopic dynamics at any fixed $N$), and one may wonder whether all
the trajectories will take on the same asymptotic form as they
approach the critical point $p=1/N=0$.  The asymptotic form of this
trajectory has been studied extensively in the mathematics literature
\cite{pt,posts,bb988,bb,aav,scc,aal,AlonEtAl}, with a variety of motivations,
including for example the search for efficient sorting algorithms.

Consider first the simplest nontrivial case, $m=2$.  Since there are
only two causal sets of size two, the 2-chain and the 2-antichain, the
distribution $f_2$ that gives the ``large scale physics'' in this case
is described by a single number which we can take to be $f_2(\twoch)$,
the probability of obtaining a 2-chain rather than a 2-antichain.
(The other probability, $f_2(\twoach)$, is of course not independent,
since classical probabilities must add up to unity.)
Interestingly enough, the number $f_2(\twoch)$ has a direct physical
interpretation in terms of %as giving 
the Myrheim-Meyer dimension of the
fine-grained causet $C$. %\footnote
%%
%{\singlesp 
%The astute reader may detect a slight technical error in
%this sentence.  Above, $f_m$ was defined as the distribution on
%$m$-element orders which arises from coarse graining a microscopic
%dynamics, in this case transitive percolation on $N$-element orders.
%%can we define f_m_microscopicdynamics ?) Though 
%Here we refer to $f_m$ as being the distribution which arises from
%coarse graining a single causet $C$.  Pedantically, %Technically,
%here and in the sentence following $f_2(\twoch)$ should be replaced with
%another expression, the \emph{abundance} of 2-chains, where
%``abundance of 2-chains'' refers to the probability that two elements
%selected at random from $C$ will be a 2-chain.  The term abundance is
%introduced below in \S \ref{simulations}.} 
%%To circumvent this issue, regard
%%$f_m$ as being the distribution on $m$-orders which arises from coarse
%%graining any ``microscopic dynamics'', which should be deduced from
%%context.  In the case of the above sentence, this ``dynamics'' is a
%%distribution which has probability 1 for the $N$-order $C$, and 0 for
%%all other $N$-orders.}
%%
Indeed, it can be seen that $f_2(\twoch)$ is %just 
nothing but the expectation value of
what we called
above the ordering fraction $r$ of $C$ (an argument explaining why
this is so follows in the next section).
But the ordering fraction, in turn, %  As such, it 
determines %a fractal
the Myrheim-Meyer dimension
$d$ that indicates the dimension of the Minkowski
spacetime $\Minkowski^d$ (if any) in which $C$ would embed faithfully
as an interval \cite{meyer,myrheim}.  
%Moreover, the Myrheim-Meyer dimension is, as is not hard to prove,
%invariant under coarse-graining.  
Thus, by coarse graining %the causet $C$ 
down to two elements, we are effectively measuring 
a certain kind of spacetime dimensionality of $C$.
%its spacetime dimension, if any.
In practice, we would not expect $C$ to
embed faithfully %only after 
without some degree of coarse-graining, but the
original $r$ would still provide a good dimension estimate since it
is, on average, coarse-graining invariant.

As we begin to consider coarse-graining to sizes $m>2$, the degree of
complication grows rapidly, simply because the number of partial
orders defined on $m$ elements grows rapidly with $m$.  For $m=3$
there are five possible causal sets: \threech, \V, \Lcauset, \wedge,
and \threeach.  Thus the effective dynamics at this ``scale'' is given
by five probabilities (so four free parameters).  For $m=4$ there are
sixteen probabilities, for $m=5$ there are sixty-three, and for $m=6$,
7 and 8, the number of probabilities is respectively 318, 2045, and
16999.

\subsection{Evidence from simulations}
\label{simulations}
%In this section, we report on some computer simulations that address
%directly the question whether transitive percolation possesses a
%continuum limit in the sense defined above.
%
%In a subsequent paper, we
%will report on simulations addressing the subsidiary question of a
%possible scaling behavior in the continuum limit.
In order that a continuum limit exist, it must be possible to choose a
trajectory for $p$ as a function of $N$ so that the resulting
coarse-grained probability distributions, $f_1$, $f_2$, $f_3$,~\dots,
have well defined limits as $N\to\infty$.  
To study this question
numerically, one can simulate transitive percolation using the
algorithm described in Section \ref{perc}, while choosing $p$ so as to
hold constant (say) the $m=2$ distribution $f_2$ ($f_1$ being
trivial).  
Because of the way transitive percolation is defined, it is
intuitively obvious that $p$ can be chosen to achieve this, and that
in doing so, one leaves $p$ with no further freedom.
%This may be clear by 
(Observe that $\E{r}=f_2(\twoch)$ is 0 when $p=0$, 1 when $p=1$,
and increases smoothly and monotonically with $p$.  Thus for any
choice of $\E{r}\in[0,1]$ there
must a $p$ which yields that $\E{r}$, and since
$f_2(\twoach)=1-f_2(\twoch)$, the entire
distribution $f_2$.)
The decisive
question then is whether, along the trajectory thereby defined, the
higher distribution functions, $f_3$, $f_4$, etc. all approach
nontrivial limits.

% Maybe the following two paragraphs should be put above, but lets not
% bother for now.  Maybe later, if it still seems worthwhile.  I think
% that the reasoning is that in previous section we are discussing
% what the large scale effective theory means, f's mean physically,
% etc.  Here we are discussing how to do the computer simulations.
% Thus detail about how to compute f_2 comes here.
As mentioned above, holding $f_2$ fixed is the same thing as
holding fixed the expectation value $\E{r}$ of ordering fraction
$r=R/{{N}\choose{2}}$.  
To see in more detail why this is so, consider
the coarse-graining 
that takes us from the original causet
$C_N$ of $N$ elements to a causet $C_2$ of two elements.  
Since coarse-graining is just random selection, the probability
$f_2(\twoch)$ that $C_2$ turns out to be a 2-chain is just the
probability that two elements of $C_N$ selected at random form a
2-chain rather than a 2-antichain.  
In other words, it is just the
probability that two elements of $C_N$ selected at random are 
causally related.  
Plainly, this is the same as the {\it fraction} of pairs of
elements of $C_N$ such that the two members of the pair form a
relation $x\prec{y}$ or $y\prec{x}$.  Therefore, the ordering fraction
$r$ equals the probability of getting a 2-chain when coarse graining
$C_N$ down to two elements; and $f_2(\twoch)=\E{r}$, as claimed.

This reasoning illustrates, in fact, how one can in principle
determine any one of the distributions $f_m$ by answering the
question, ``What is the probability of getting this particular
$m$-element causet from this particular $N$-element causet if you
coarse grain down to $m$ elements?''  To compute the answer to such a
question starting with any given causet $C_N$, one examines every
possible combination of $m$ elements, counts the number of times that
the combination forms the particular causet being looked for, and
divides the total by ${N \choose m}$.  The ensemble mean of the
resulting {\it abundance}, as we will refer to it, is then $f_m(\xi)$,
% So abundance means for a particular causet : <a_x> = f_m(x)
where $\xi$ is the causet for which one is looking.
In practice, of course, we normally use
a more efficient counting algorithm than simply examining individually
all ${N \choose m}$ subsets of $C_N$.

\subsubsection{Histograms of 2-chain and 4-chain abundances}

\begin{figure}[htb]
\center
\scalebox{.71}{
\rotatebox{-90}{
\includegraphics{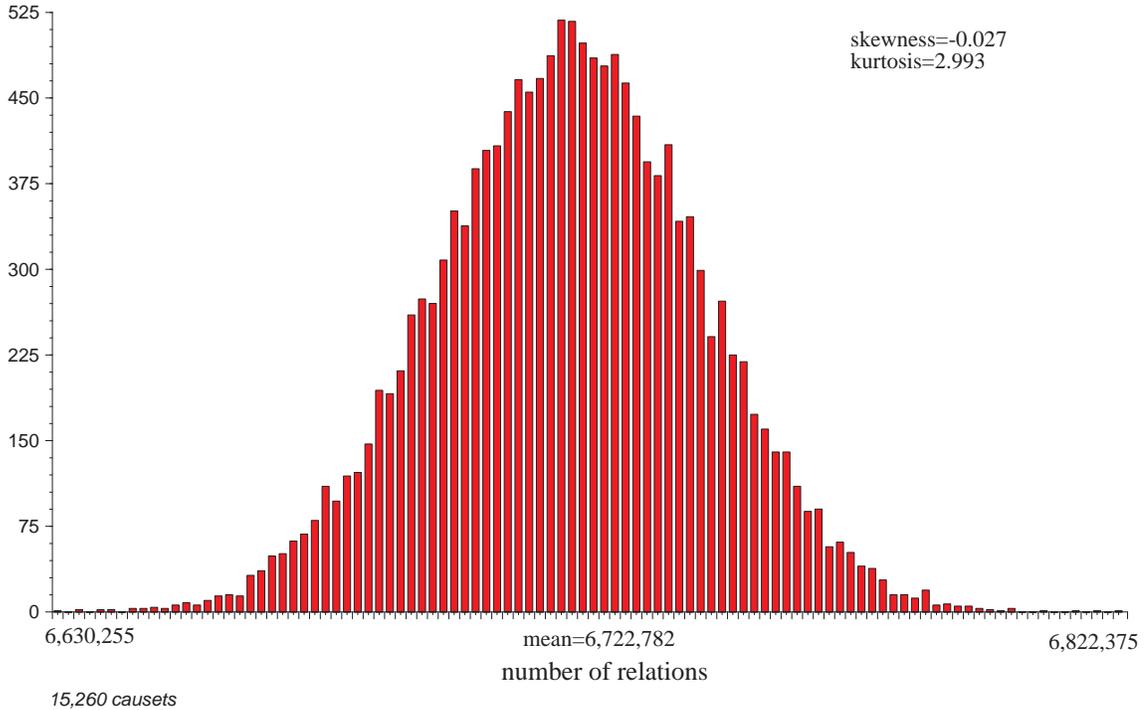} }}
\caption{Distribution of number of relations for $N=4096$, $p=0.01155$}
\label{hist2ch}
\end{figure}

\begin{figure}[htb]
\center \scalebox{.69}{ \rotatebox{-90}{
\includegraphics{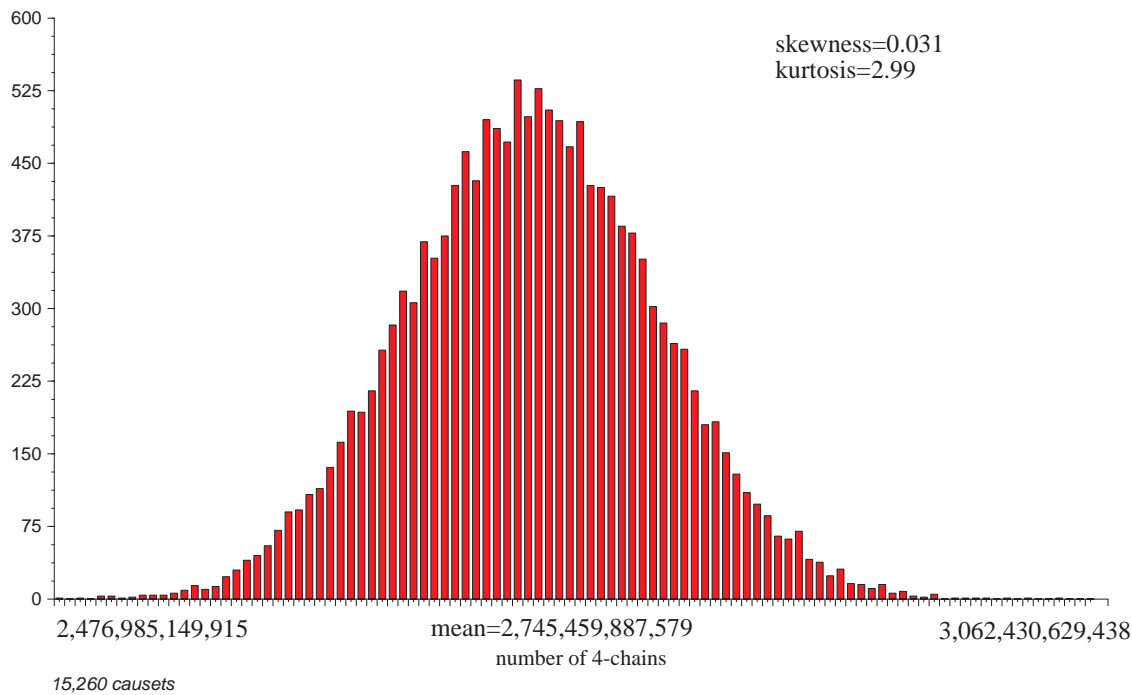} }}
\caption{Distribution of number of 4-chains for $N=4096$, $p=0.01155$}
\label{hist4ch}
\end{figure}

As explained in the previous subsection, the main computational
problem, once the random causet has been generated, is determining the
number of subcausets of different sizes and types.  To get a feel for
how some of the resulting ``abundances'' are distributed, we start by
presenting a couple of histograms.
Figure \ref{hist2ch} shows the number $R$ of relations
obtained from a simulation in which 15,260 causal sets were generated by
transitive percolation with $p=0.01155$, $N=4096$.
Visually, the distribution is Gaussian, in agreement with
the fact that its ``kurtosis''
\be
  \overline{ \left(x - \bar{x} \right)^4 }   \, \bigg/ \       
  {\overline{ \left(x-\bar{x}\right)^2 }\,}^2
\ee
of 2.993 is very nearly equal to its Gaussian value of 3 (the over-bar
denotes sample mean).  In these simulations, $p$ was chosen so that
the number of 3-chains was equal on average to half the total number
possible, i.e. the ``abundance of 3-chains'', $\mbox{(number of
3-chains)}/{N \choose 3}$,
was equal to $1/2$ on average.  The picture is qualitatively identical
if one counts 4-chains rather than 2-chains, as exhibited in
Fig. \ref{hist4ch}.
% These 4-chains (and 2-chains) were counted exactly, not sampled.  I don't think
% that it is worth mentioning, though.  Is it?  Naah.

(One may wonder whether it was to be expected that these distributions
would appear to be so normal.  If the variable in question, here the
number of 2-chains $R$ or the number of 4-chains ($C_4$, say), can be
expressed as a sum of independent random variables, then the central
limit theorem provides an explanation.  So consider the variables
$x_{ij}$ which are 1 if $i \prec j$ and zero otherwise.  Then $R$ is
easily expressed as a sum of these variables:
\be
        R = \sum_{i<j} x_{ij}
\ee
However, the $x_{ij}$ are not independent, due to transitivity.
Apparently, this dependence is not large enough to interfere much with
the normality of their sum.  The number of 4-chains $C_4$ can be
expressed in a similar manner
\be
    C_4 = \sum_{i<j<k<l} x_{ij} x_{jk} x_{kl}\,.
\ee
and similar remarks apply.)

Let us mention that for values of $p$ sufficiently close to 0 or 1,
these distributions will appear skew.  This occurs simply because the
numbers under consideration (e.g. the number of $m$-chains) are
bounded between zero and $N \choose m$ and must deviate from normality
if their mean gets too close to a boundary relative to the size of
their standard deviation.  Whenever we draw an error bar in the
following, we will ignore any deviation from normality in the
corresponding distribution.

Notice incidentally that the total number of 4-chains possible is
${4096\choose4}=11,710,951,848,960$.  Consequently, the mean 4-chain
abundance\footnote
{\singlesp Occasionally I will write simply ``abundance'', in
place of ``mean abundance'', assuming the average is obvious from
context.}
in our simulation is only
$\frac{2,745,459,887,579}{11,710,951,848,960}=0.234$, a considerably smaller
value than the mean 2-chain abundance of $\E{r}=\frac{6,722,782}{{4096 \choose
2}}=0.802$.  This was to be expected, considering that the 2-chain is
one of only two possible causets of its size, while the 4-chain is one
out 16 possibilities.  (Notice also that 4-chains are necessarily less
probable than 2-chains, because every coarse-graining of a 4-chain is
a 2-chain, whereas the 2-chain can come from every 4-element causet
save the 4-antichain.)

\subsubsection{Trajectories of $p$ versus $N$}

The question we are exploring is whether there exist, for
$N\to\infty$, trajectories $p=p(N)$ along which the mean abundances of
all finite causets tend to definite limits.  To seek such trajectories
numerically, we will select some finite ``reference causet'' and
determine, for a range of $N$, those values of $p$ which maintain its
mean abundance at some target value.  If a continuum limit does exist,
then it should not matter in the end which causet we select as our
reference, since any other choice (together with a matching choice of
target abundance) should produce the same trajectory asymptotically.
We would also anticipate that all the trajectories would behave
similarly for large $N$, and that, in particular, either all would
lead to continuum limits or all would not.  In principle it could
happen that only a certain subset led to continuum limits, but we know
of no reason to expect such an eventuality.  In the simulations
reported here, I have chosen as our reference causets the 2-, 3- and
5-chains.  I have computed six trajectories, holding the mean 2-chain
abundance fixed at 1/2, 1/3, and 1/10, the mean 3-chain abundance
fixed at 1/2 and .0814837, and the mean 5-chain abundance fixed at
1/2.  For $N$, I have used as large a range as available computing
resources allowed.  % Should I brag about how big?  Hard because can
		    % go lots further with <r>.

\begin{figure}[htb!]
\center
\scalebox{1.22}{
\includegraphics{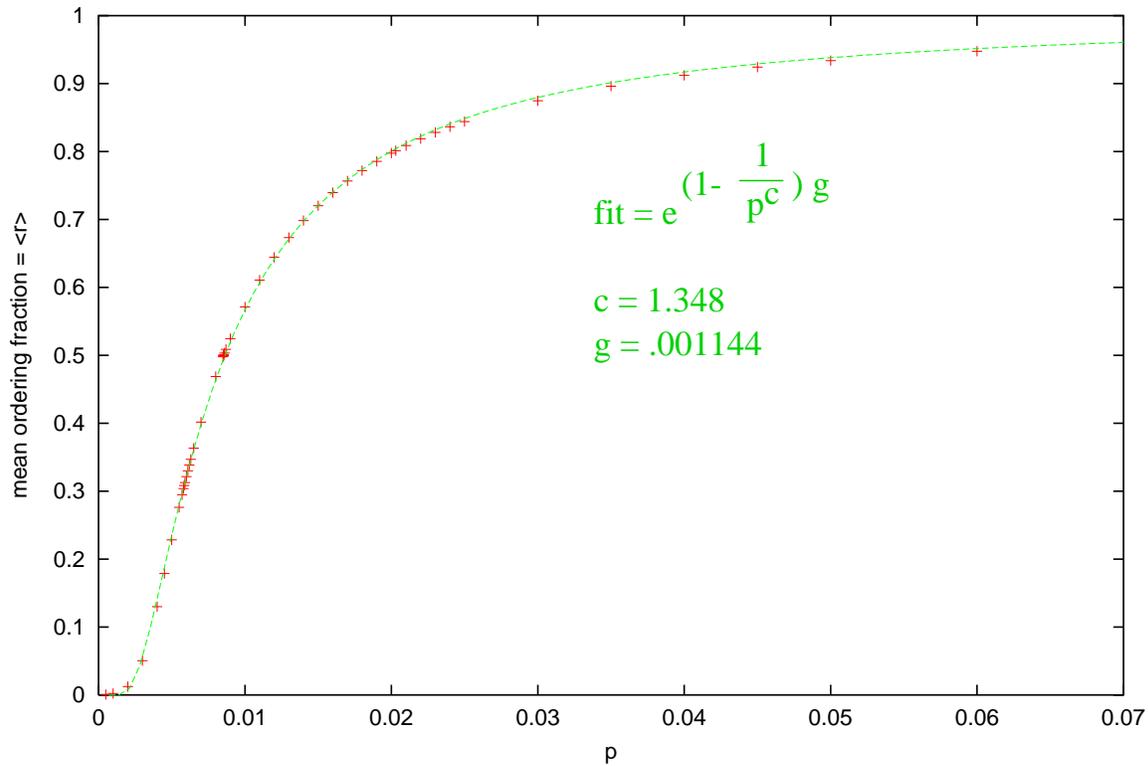} }
\caption{Ordering fractions as a function of $p$ for $N = 2048$}
\label{rvsp}
\end{figure}

Before discussing the trajectories as such, let us have a look at how
the mean 2-chain abundance $\E{r}$ (i.e. the mean ordering fraction)
varies with $p$ for a fixed $N$ of 2048, as exhibited in Figure
\ref{rvsp}.  (Vertical error bars are displayed in the figure but are
so small that they just look like horizontal lines.  The plotted
points were obtained from the exact calculation for the ensemble
average $\E{r}$ discussed in \S \ref{diffusion}, so the errors come
only from floating point roundoff.  The fitting function used in
Figure \ref{rvsp} will be discussed 
%briefly in \S \ref{pdrdp}, see also
in \cite{scaling}.
%will be discussed in a subsequent paper
%\cite{scaling}, where we examine scaling behavior; see also
%\cite{mathrefs}.)
%
As one can see, $\E{r}$ starts at 0 for $p=0$, rises rapidly to near 1
and then asymptotes to 1 at $p=1$ (not shown).  Of course, it was
evident a priori that $\E{r}$ would increase monotonically from 0 to 1
as $p$ varied between these same two values, but it is perhaps
noteworthy that its graph betrays no sign of discontinuity or
non-analyticity (no sign of a ``phase transition'').  To this extent,
it strengthens the expectation that the trajectories we find will all
share the same qualitative behavior as $N\to\infty$.

\begin{figure}[htbp]
\center
\scalebox{.78}{
\includegraphics{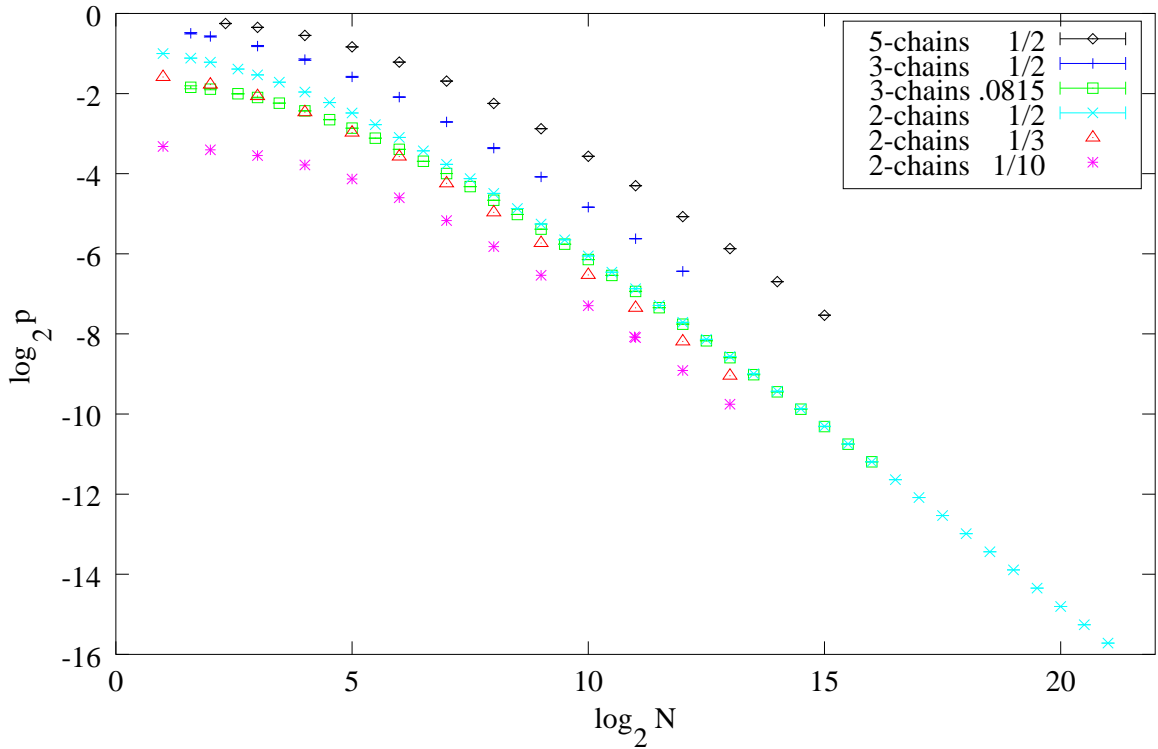} }
\caption%
{Flow of the ``coupling constant'' $p$ as $N\to\infty$ (six trajectories)}
\label{6traj}
\end{figure}

The six trajectories we have simulated are depicted in
Fig. \ref{6traj}.\footnote {\singlesp Notice that the error bars are
shown rotated in the legend.  This will be the case for all subsequent
legends as well.}  A higher abundance of $m$-chains for fixed $m$
leads to a trajectory with higher $p$.  Also note that, as observed
above, the longer chains require larger values of $p$ to attain the
same mean abundance, hence a choice of mean abundance = 1/2
corresponds in each case to a different trajectory.  The trajectories
with $\E{r}$ held to lower values are ``higher dimensional'' in the
sense that $\E{r}=1/2$ corresponds to a Myrheim-Meyer dimension of 2,
while $\E{r}=1/10$ corresponds to a Myrheim-Meyer dimension of 4.
Observe that the plots give the impression of becoming straight lines
with a common slope at large $N$.  This tends to corroborate the
expectation that they will exhibit some form of scaling with a common
exponent, a behavior reminiscent of that found with continuum limits
in many other contexts.  This is further suggested by the fact that
two distinct trajectories ($f_2(\twoch)=1/2$ and
$f_3(\threech)=.0814837$), obtained by holding different abundances
fixed, seem to converge for large $N$.

\begin{figure}[htbp]
\center
\scalebox{.80}{
\includegraphics{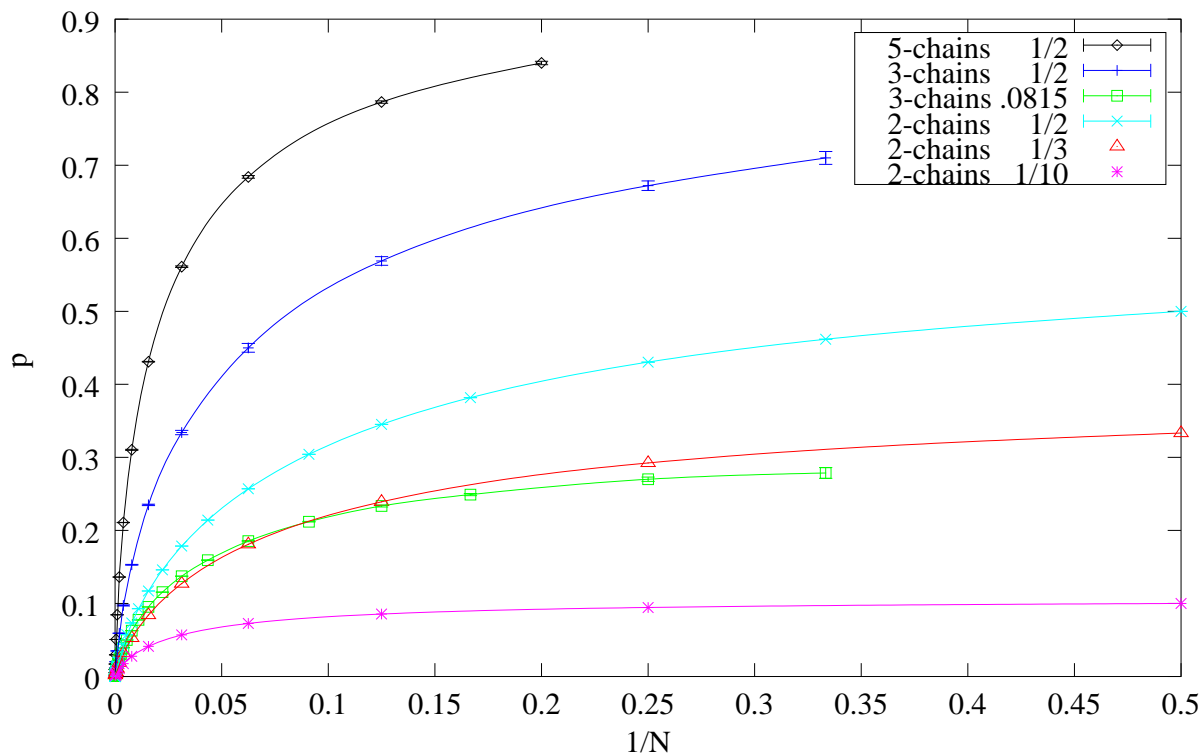} }
\caption{Six trajectories approaching the critical point at $p=0$, $N=\infty$}
\label{critpoint}
\end{figure}

By taking the abscissa to be $1/N$ rather than $\log_2 N$, we can
bring the critical point to the origin, as in Fig. \ref{critpoint}.
The lines which pass through the data points there are just splines
drawn to aid the eye in following the trajectories.  Note that the
curves tend to asymptote to the $p$-axis, suggesting that $p$ falls
off more slowly than $1/N$.  This suggestion is corroborated by more
detailed analysis of the scaling behavior of these trajectories, as
will be discussed in \cite{scaling}.
% logarithmic corrections?

\subsubsection{Flow of the coarse-grained theory along a trajectory}

We come finally to a direct test of whether the coarse-grained theory
converges to a limit as $N\to\infty$.  Independently of scaling or any
other indicator, this is by definition the criterion for a continuum
limit to exist.  I have examined this question by means of
simulations conducted for five of the six trajectories mentioned
above.  In each simulation I proceeded as follows.  For each chosen
$N$, I experimentally found a $p$ sufficiently close to the desired
trajectory.  Having determined $p$, I then generated a large number
of causets by the percolation algorithm described in
Section~\ref{perc}.  (The number generated varied from 64 to 40,000.)
For each such random causet, I computed the abundances of the
different $m$-element (sub)causets under consideration (2-chain, 3-chain,
3-antichain, etc), and combined the results to obtain the mean
abundances we have plotted here, together with their standard errors.
(The errors shown do not include any contribution from the slight
inaccuracy in the value of $p$ used.  Except for the 3- and 5-chain
trajectories these errors are negligibly small.)

To compute the abundances of the 2-, 3-, and 4-orders for a given
causet, I randomly sampled its four-element subcausets, counting the
number of times each of the sixteen possible 4-orders arose, and
dividing each of these counts by the number of samples taken to get
the corresponding abundance.  As an aid in identifying to which
4-order a sampled subcauset belonged I used the following invariant,
which distinguishes all of the sixteen 4-orders, save two pairs.
$$
      I(S) = \prod\limits_{x\in S} \left( 2 + |\past(x)| \right)
$$
Here, $\past(x) = \{y\in S | y \prec x\}$ is the exclusive past of the
element $x$ and $|\past(x)|$ is its cardinality.  Thus, we associate
to each element of the causet, a number which is two more than the
cardinality of its exclusive past, and we form the product of these
numbers (four, in this case) to get our invariant.  (For example, this
invariant is 90 for the ``diamond'' poset, \diamond.)

\begin{figure}[htb]
\center
\scalebox{.78}{
\includegraphics{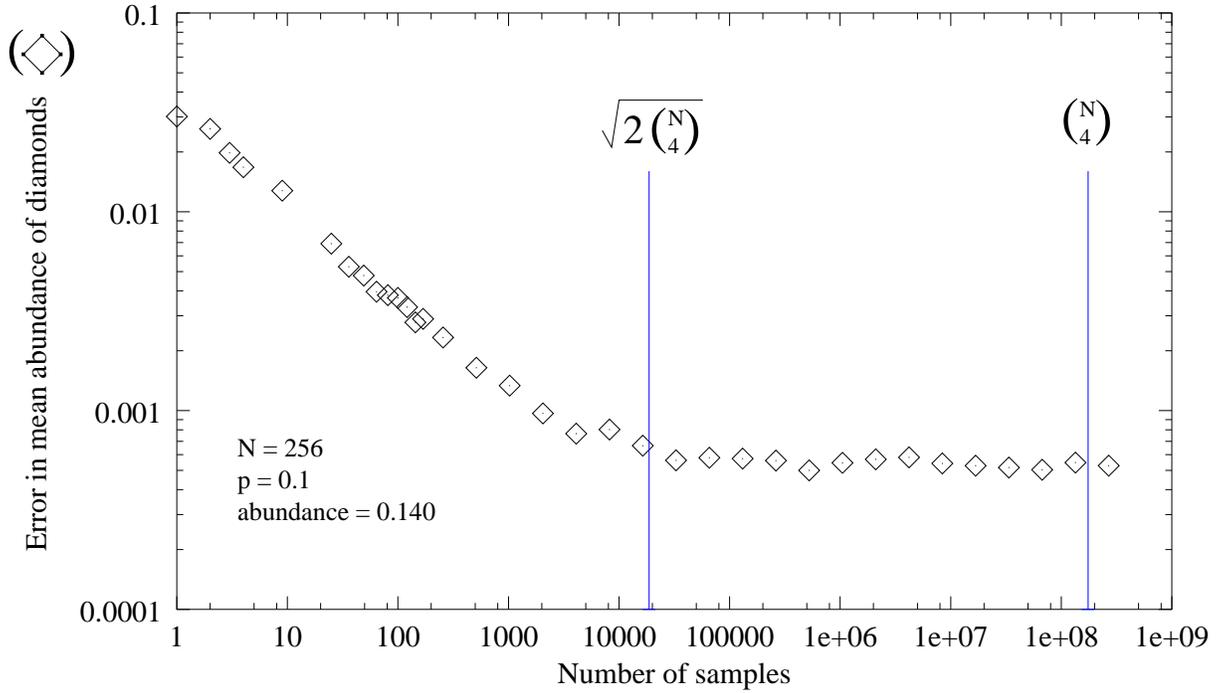} }
\caption{Reduction of error in estimated diamond abundance with
         increasing number of samples} 
\label{sampling_error}
\end{figure}

The number of samples taken from an $N$ element causet was chosen to
be $\sqrt{2{N \choose 4}}$, on the grounds that the probability to get
the same four element subset twice becomes appreciable with more than
this number of samples.  Numerical tests confirmed that this rule of thumb
tends to minimize the sampling error, as seen in Figure
\ref{sampling_error}.\footnote
{\singlesp The errors depicted in Fig. \ref{sampling_error} were found by
generating 100 causets by transitive percolation, and for each one
performing the indicated number of samples of 4 element subcausets
(with replacement),
counting the fraction of times that the diamond arose.  The errors
reported are the (square root of the) variance of the mean
of this quantity over the 100 causets.}
% Is this clear?  Should I just say standard error in the mean?  Will
% that be clear?

Once one has the mean abundances of all the 4-orders, the mean
abundances of the smaller causets can be found by further coarse
graining.  By explicitly carrying out this coarse graining, one easily
deduces the following relationships:
\bee
f_3(\threech) & = & f_4(\fourch) + \half\left(f_4(\Y)+f_4(\iY)\right) 
	+ \quart f_4(\threecho) + \quart\left(f_4(\new)+f_4(\inu)\right) 
	+ \half f_4(\diamond)\\
f_3(\V) & = & \half\, f_4(\Y) + \half f_4(\new) + \quart f_4(\diamond) +
	\thquart f_4(\flower) + \quart f_4(\Vo) + \quart f_4(\N) 
	+ \half f_4(\bowtie)\\
f_3(\Lcauset) & = & \thquart f_4(\threecho) +
	\quart\left(f_4(\new)+f_4(\inu)\right) +
	\half\left(f_4(\Vo)+f_4(\wedgeo)\right) + f_4(\pie) + \half
	f_4(\Lo)\\
&&+\half f_4(\N)\\
f_3(\wedge) & = & \half\, f_4(\iY) + \half f_4(\inu) + \quart f_4(\diamond) +
	\thquart f_4(\iflower) + \quart f_4(\wedgeo) + \quart f_4(\N) 
	+ \half f_4(\bowtie)\\
f_3(\threeach) & = & \quart\left(f_4(\flower)+f_4(\iflower)\right) 
	+ \quart\left(f_4(\Vo)+f_4(\wedgeo)\right) + \half f_4(\Lo) 
	+ f_4(\fourach)\\
f_2(\twoch) & = & f_3(\threech) +
	\frac{2}{3}\left(f_3(\V)+f_3(\wedge)\right) 
	+ \frac{1}{3} f_3(\Lcauset)\\
f_2(\twoach) & = & 1 - f_2(\twoch)
\eee
In the first six equations, the coefficient before each term on the
right is the fraction of coarse-grainings of that causet which yield
the causet on the left.

\begin{figure}[htbp]
\center
\scalebox{.65}{
\includegraphics{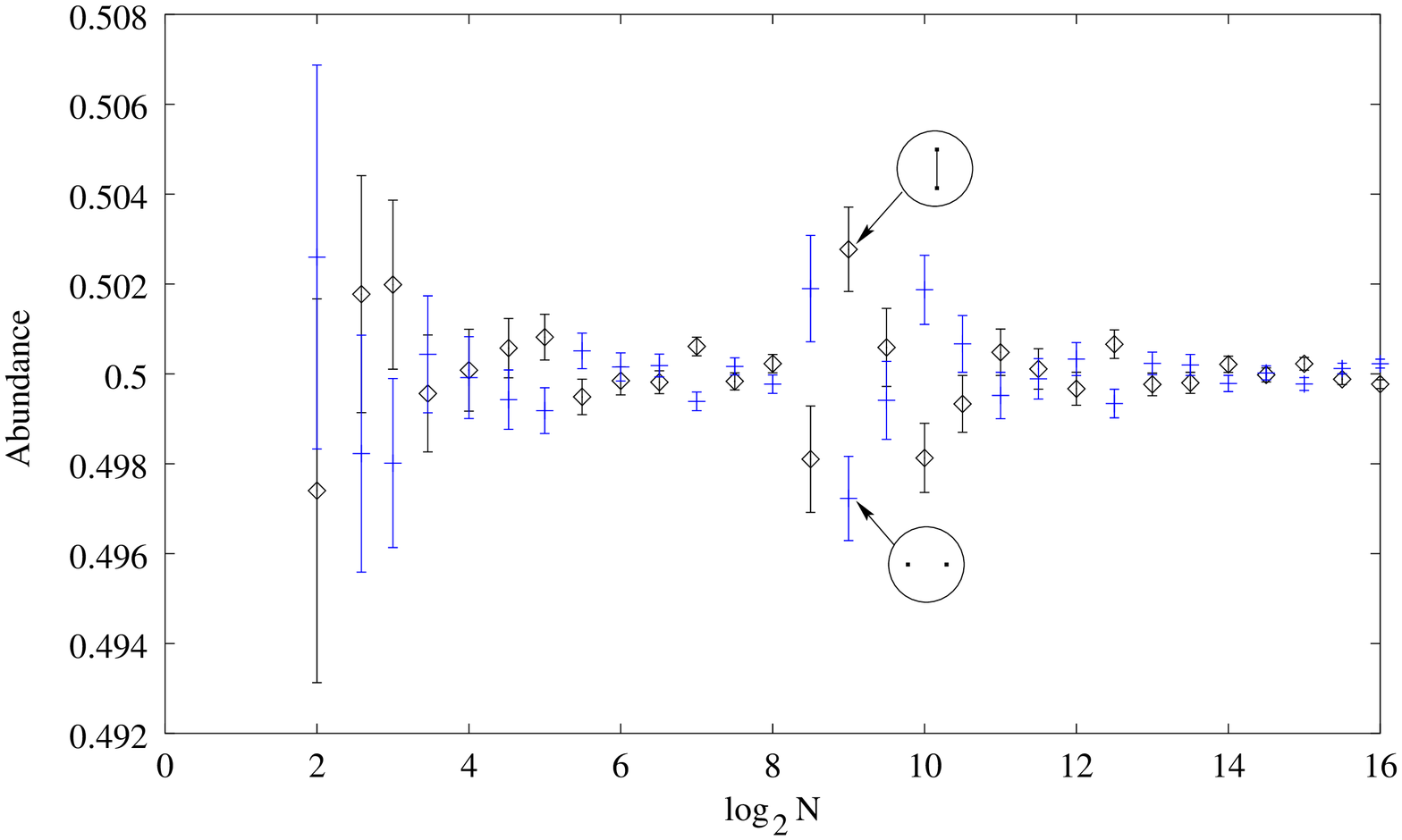} }
\caption{Flow of the coarse-grained probabilities $f_m$ for $m=2$.  
         The 2-chain probability is held at 1/2.}
\label{2ch-2}
\end{figure}

\begin{figure}[htb]
\center
\scalebox{.76}{ % .78
\includegraphics{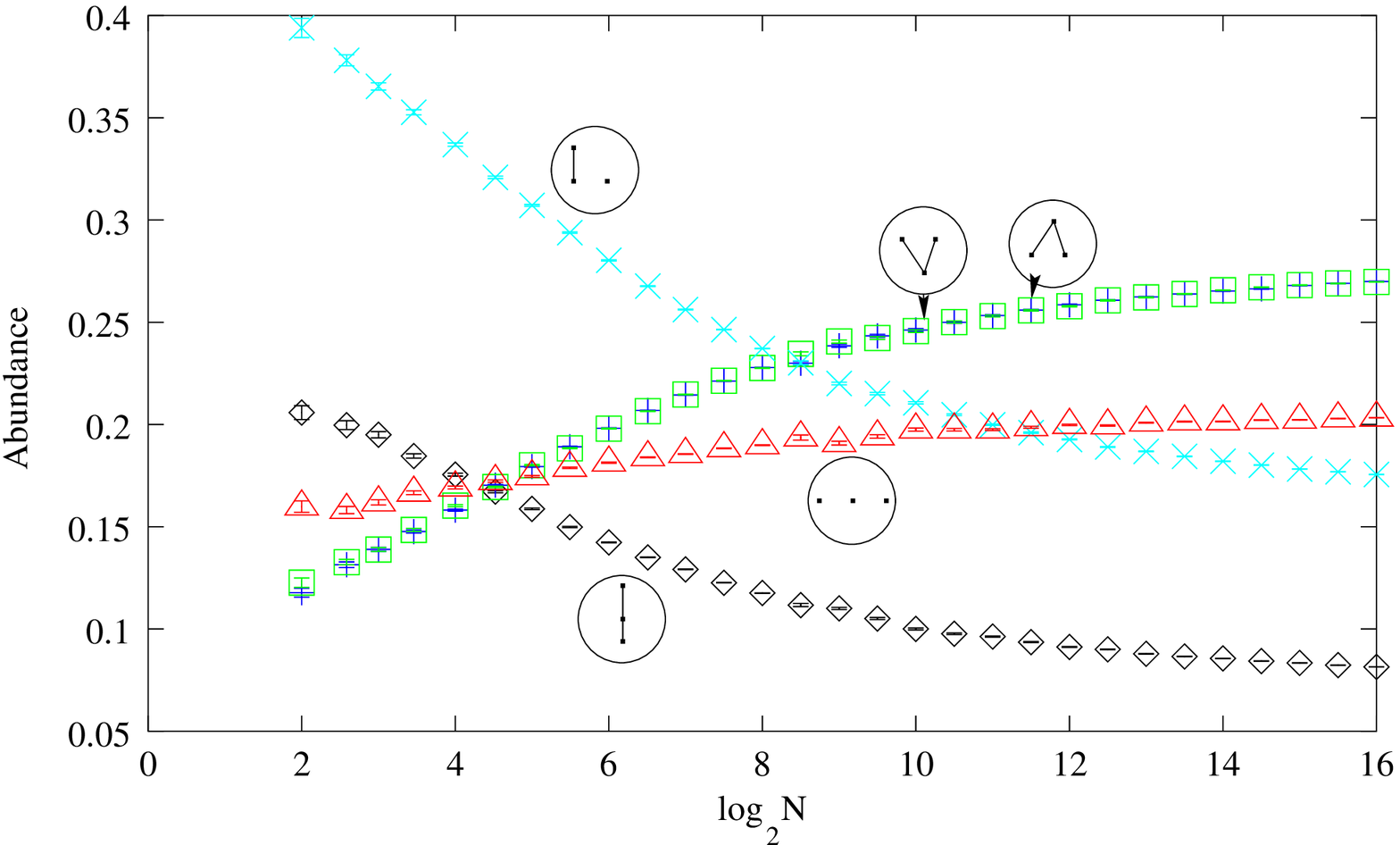} }
\caption{Flow of the coarse-grained probabilities $f_m$ for $m=3$.  
The 2-chain probability is held at 1/2.}
\label{2ch-3}
\end{figure}

\begin{figure}[htb]
\center
\scalebox{.61}{ % .63
\includegraphics{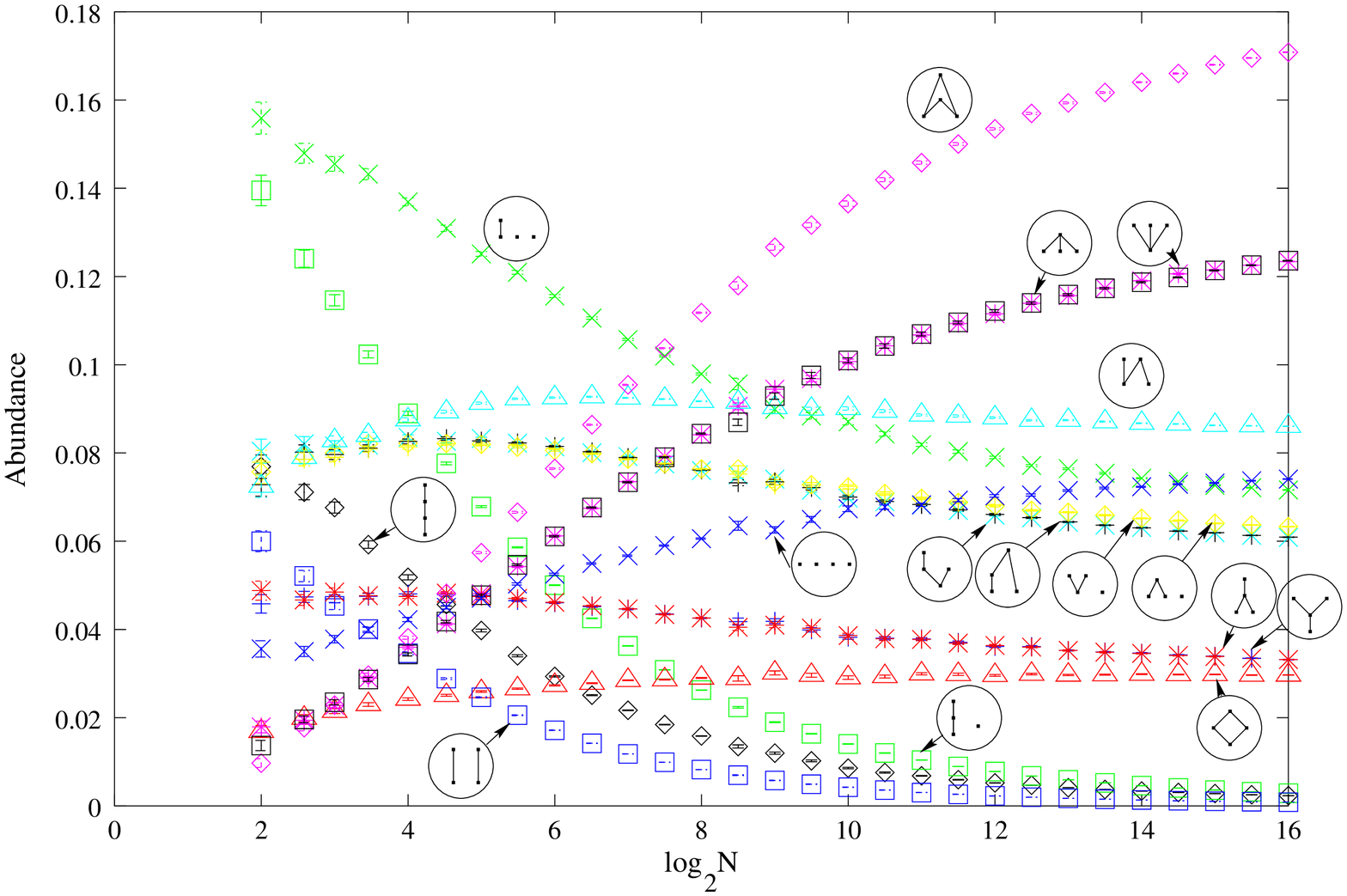} }
\caption{Flow of the coarse-grained probabilities $f_m$ for $m=4$.
         The 2-chain probability is held at 1/2.}
\label{2ch-4}
\end{figure}

In Figures \ref{2ch-2}, \ref{2ch-3}, and \ref{2ch-4}, we exhibit how
the coarse-grained probabilities of all possible 2, 3, and 4 element
causets vary as we follow the trajectory along which the
coarse-grained 2-chain probability $f_2(\twoch)=r$ is held at $1/2$.
By design, the coarse-grained probability for the 2-chain remains flat
at 50\%, so Figure \ref{2ch-2} simply shows the accuracy with which
this was achieved.  (Observe the scale on the vertical axis.)  Notice
that, since $f_2(\twoch)$ and $f_2(\twoach)$ must sum to 1, their
error bars are necessarily equal.  (The standard deviation in the
abundances decreases with increasing $N$.  The ``blip'' around
$\log_2N=9$ occurs simply because I generated fewer causets at that
and larger values of $N$ to reduce computational costs.)

The crucial question is whether the probabilities for the three and
four element causets tend to definite limits as $N$ tends to infinity.
Several features of the diagrams indicate that this is indeed
occurring.  Most obviously, all the curves, except possibly a couple
in Figure \ref{2ch-4}, appear to be leveling off at large $N$.  But we
can bolster this conclusion by observing in which direction the curves
are moving, and considering their interrelationships.

For the moment let us focus our attention on Figure \ref{2ch-3}.  A
priori there are five coarse-grained probabilities to be followed.
That they must add up to unity reduces the degrees of freedom to four.
This is reduced further to three by the observation that, due to the
time-reversal symmetry of the percolation dynamics, we must have
$f_3(\V)=f_3(\wedge)$, as duly manifested in their graphs.  Moreover,
all five of the curves appear to be monotonic, with the curves for
$\wedge$, $\V$ and $\threeach$ rising, and the curves for $\threech$
and $\Lcauset$ falling.  If we accept this indication of monotonicity
from the diagram, then first of all, every probability $f_3(\xi)$ must
converge to some limiting value, because monotonic bounded functions
always do; and some of these limits must be nonzero, because the
probabilities must add up to 1.  Indeed, since $f_3(\V)$ and
$f_3(\wedge)$ are rising, they must converge to some nonzero value,
and this value must lie below 1/2 in order that the total probability
not exceed unity.  In consequence, the rising curve $f_3(\threeach)$
must also converge to a nontrivial probability (one which is neither 0
nor 1).  Taken all in all, then, it looks very much like the $m=3$
coarse-grained theory has a nontrivial $N\to\infty$ limit, with at
least three out of its five probabilities converging to nontrivial
values.

Although the ``rearrangement'' of the coarse-grained probabilities
appears much more dramatic in Figure \ref{2ch-4}, similar arguments
can be made.  Excepting initial ``transients'', it seems reasonable to
conclude from the data that monotonicity will be maintained.  From
this, it would follow that the probabilities for $\flower$ and
$\iflower$ (which must be equal by time-reversal symmetry) and the
other rising probabilities, $\bowtie$, $\fourach$, and $\diamond$, all
approach nontrivial limits.  The coarse-graining to 4 elements,
therefore, would also admit a continuum limit with a minimum of 4 out
of the 11 independent probabilities being nontrivial.

To the extent that the $m=2$ and $m=3$ cases are indicative, then, it is
reasonable to conclude that percolation dynamics admits a continuum limit
which is non-trivial at all ``scales'' $m$.

\begin{figure}[htb]
\center
\scalebox{.67}{ % .69
\includegraphics{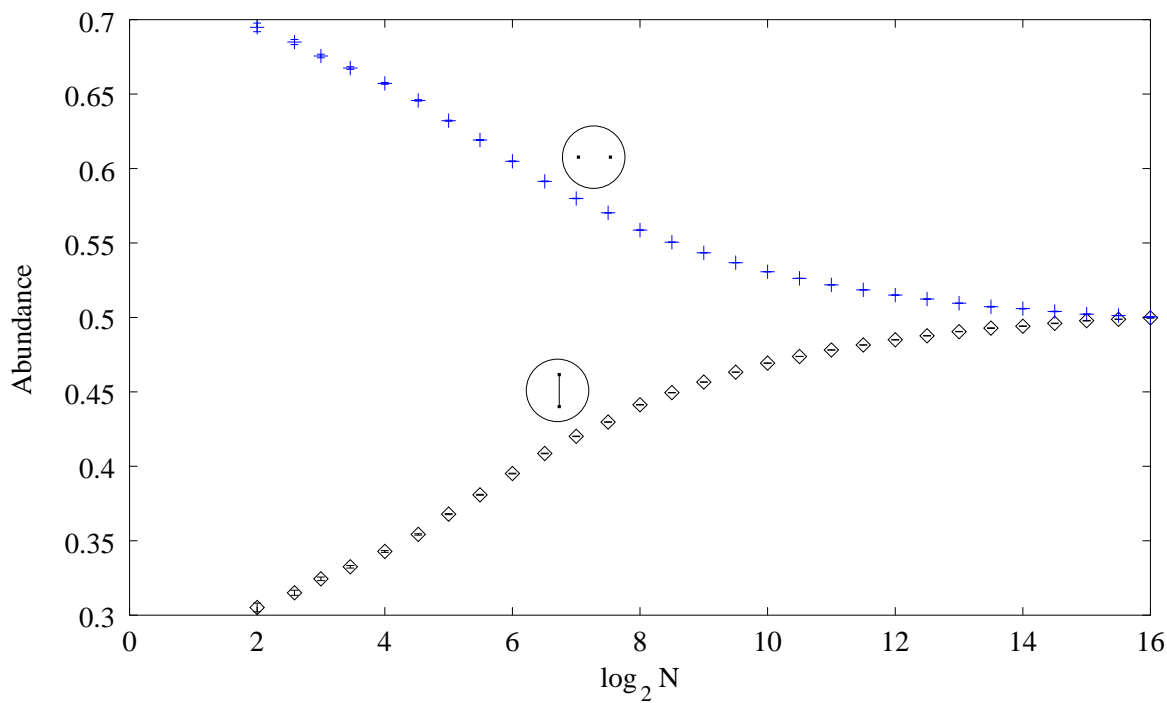} }
\caption{Flow of the coarse-grained probabilities $f_m$ for $m=2$.
         The 3-chain probability is held at 0.0814837.}
\label{3ch-2}
\end{figure}

\begin{figure}[htb]
\center
\scalebox{.67}{ % .7
\includegraphics{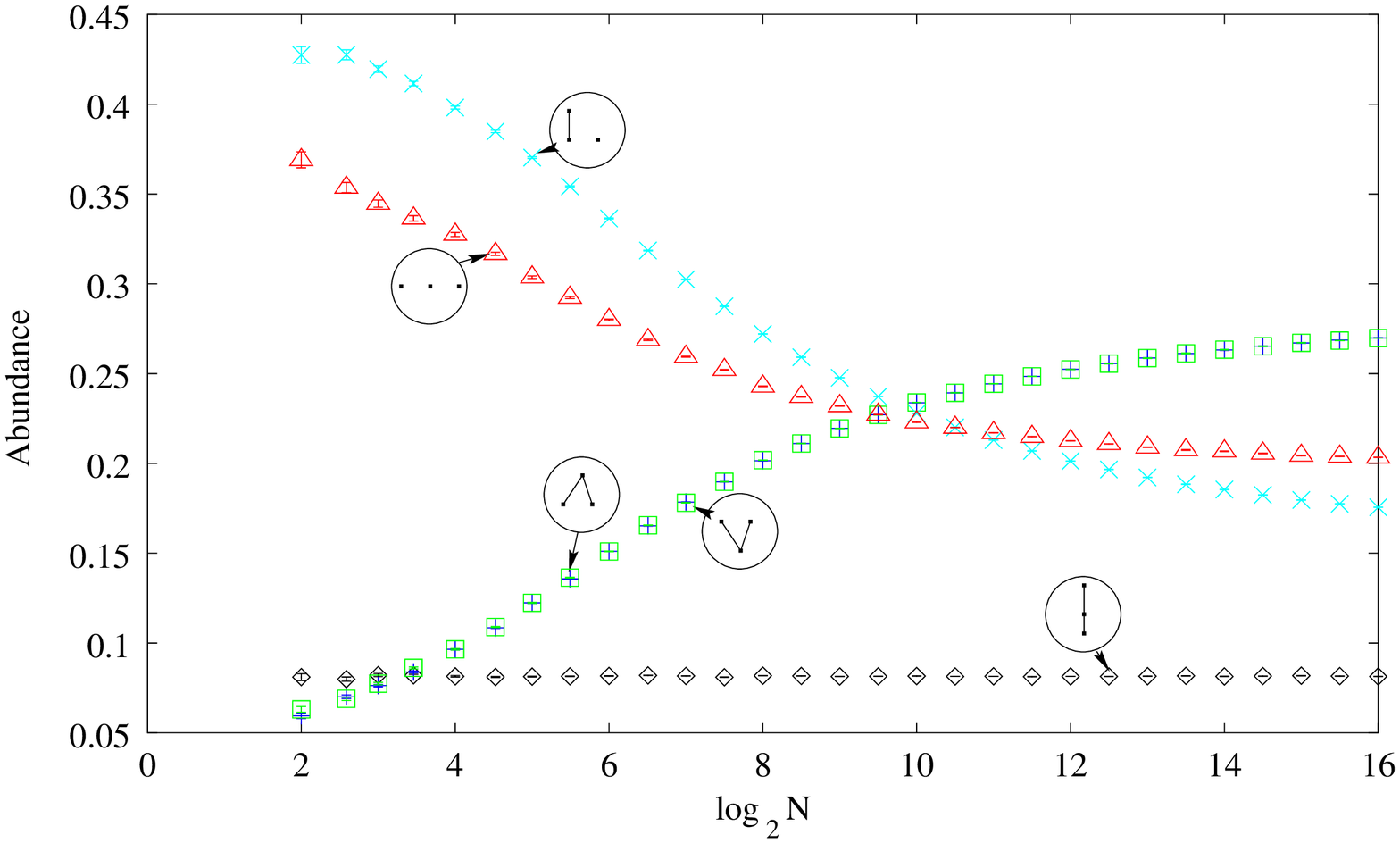} }
\caption{Flow of the coarse-grained probabilities $f_m$ for $m=3$.
        The 3-chain probability is held at 0.0814837.}
\label{3ch-3}
\end{figure}

\begin{figure}[htb]
\center
\scalebox{.64}{ % .68
\includegraphics{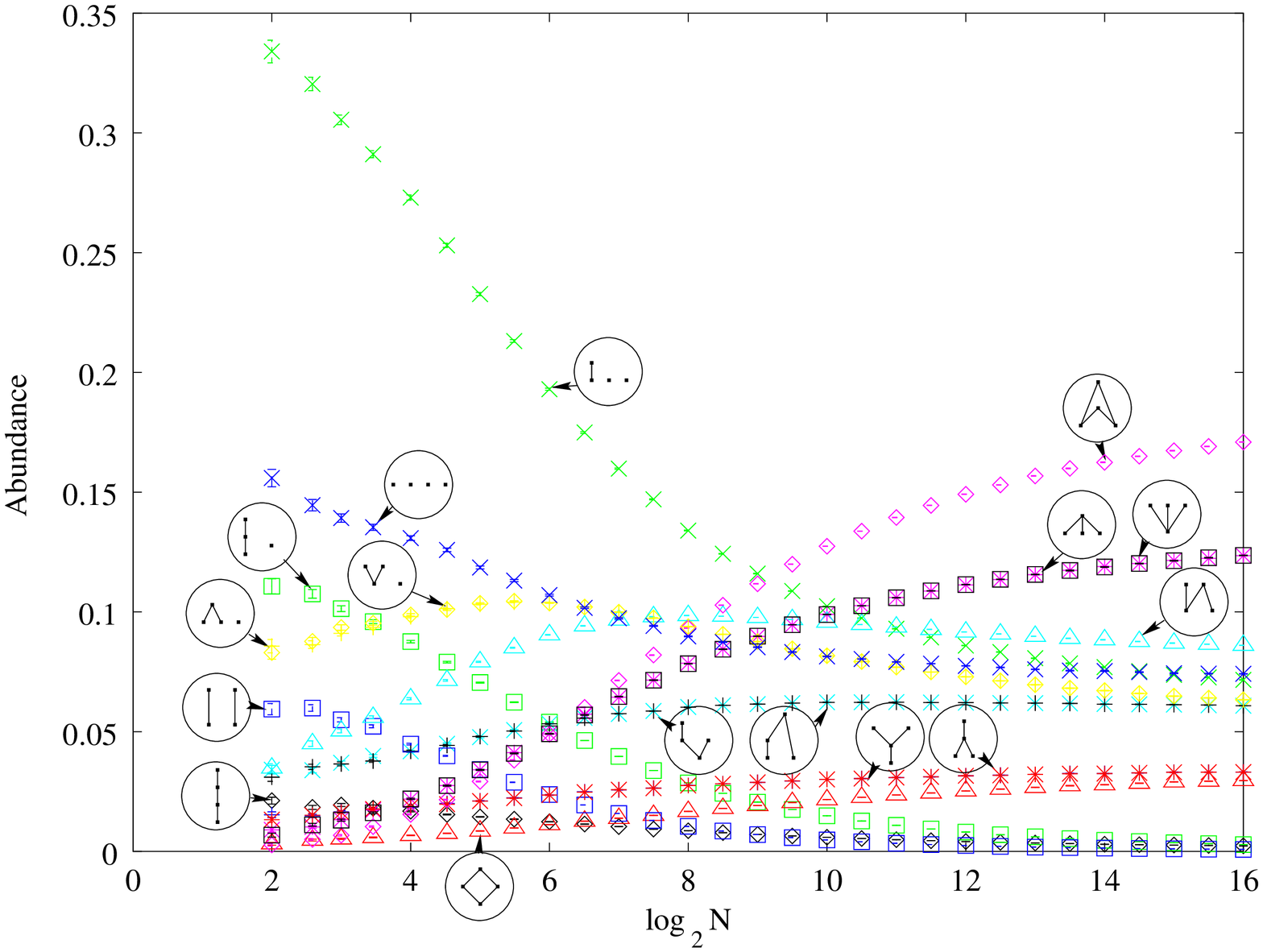} }
\caption{Flow of the coarse-grained probabilities $f_m$ for $m=4$.
         The 3-chain probability is held at 0.0814837.}
\label{3ch-4}
\end{figure}

The question suggests itself, whether the flow of the coarse-grained
probabilities would differ qualitatively if we held fixed some
mean abundance other than that of the 2-chain.  In Figures \ref{3ch-2},
\ref{3ch-3}, and \ref{3ch-4}, we display results obtained by fixing
the 3-chain abundance (its value having been chosen to make the
abundance of 2-chains be 1/2 when $N=2^{16}$).  Notice in Figure
\ref{3ch-2} that the abundance of 2-chains varies considerably along
this trajectory, whilst that of the 3-chain (in Figure \ref{3ch-3}) of
course remains constant.  Once again, the figures suggest strongly
that the trajectory is approaching a continuum limit with nontrivial
values for the coarse-grained probabilities of at least the 3-chain,
the ``V'' and the ``$\Lambda$'' (and in consequence, of the 2-chain and
2-antichain).

\begin{figure}[htb]
\center
\scalebox{.63}{ % .68
\includegraphics{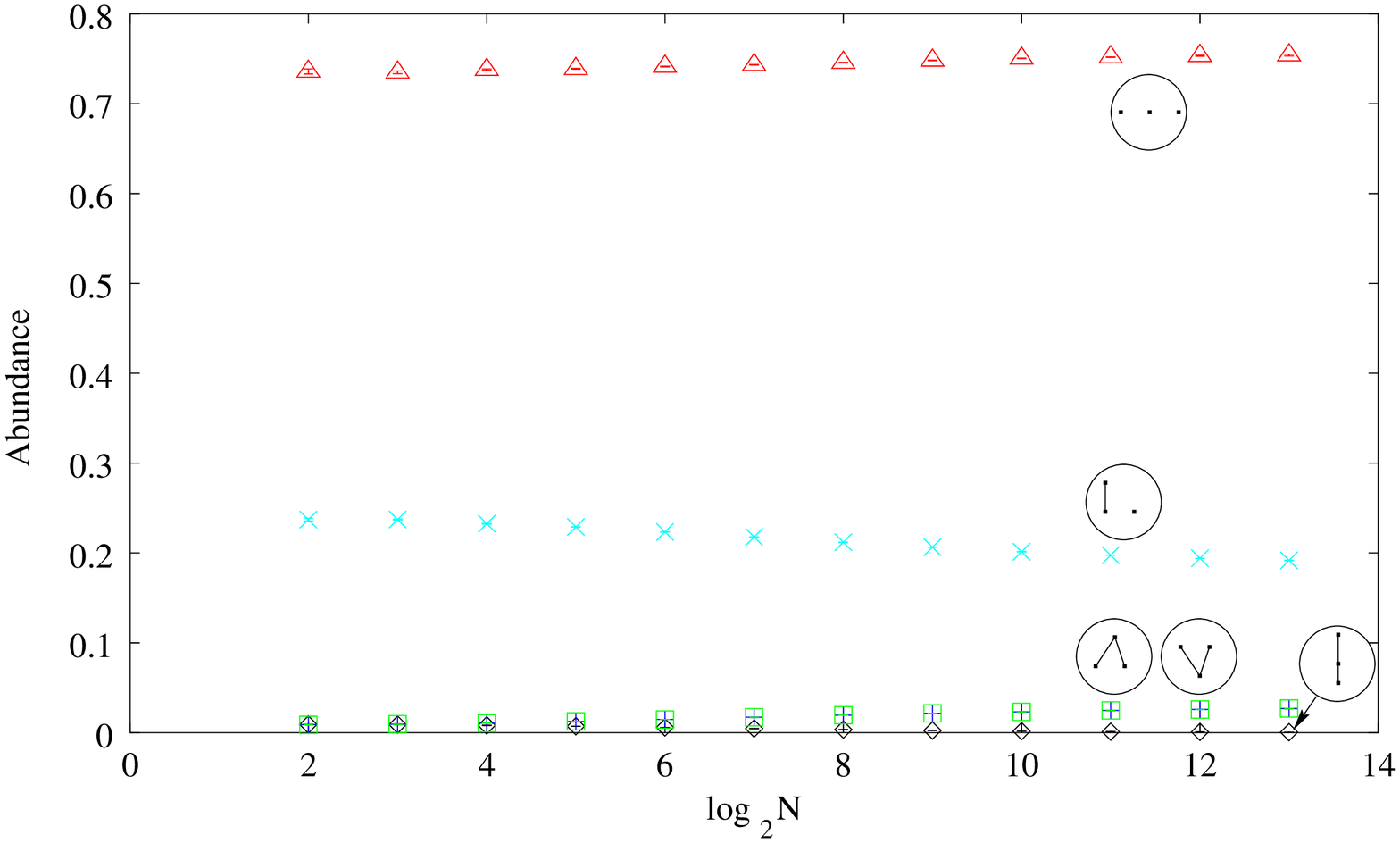} }
\caption{Flow of the coarse-grained probabilities $f_m$ for $m=3$.
         The 2-chain probability is held at 1/10.}
\label{4d-3}
\end{figure}

\begin{figure}[htb]
\center
\scalebox{.68}{ % .68
\includegraphics{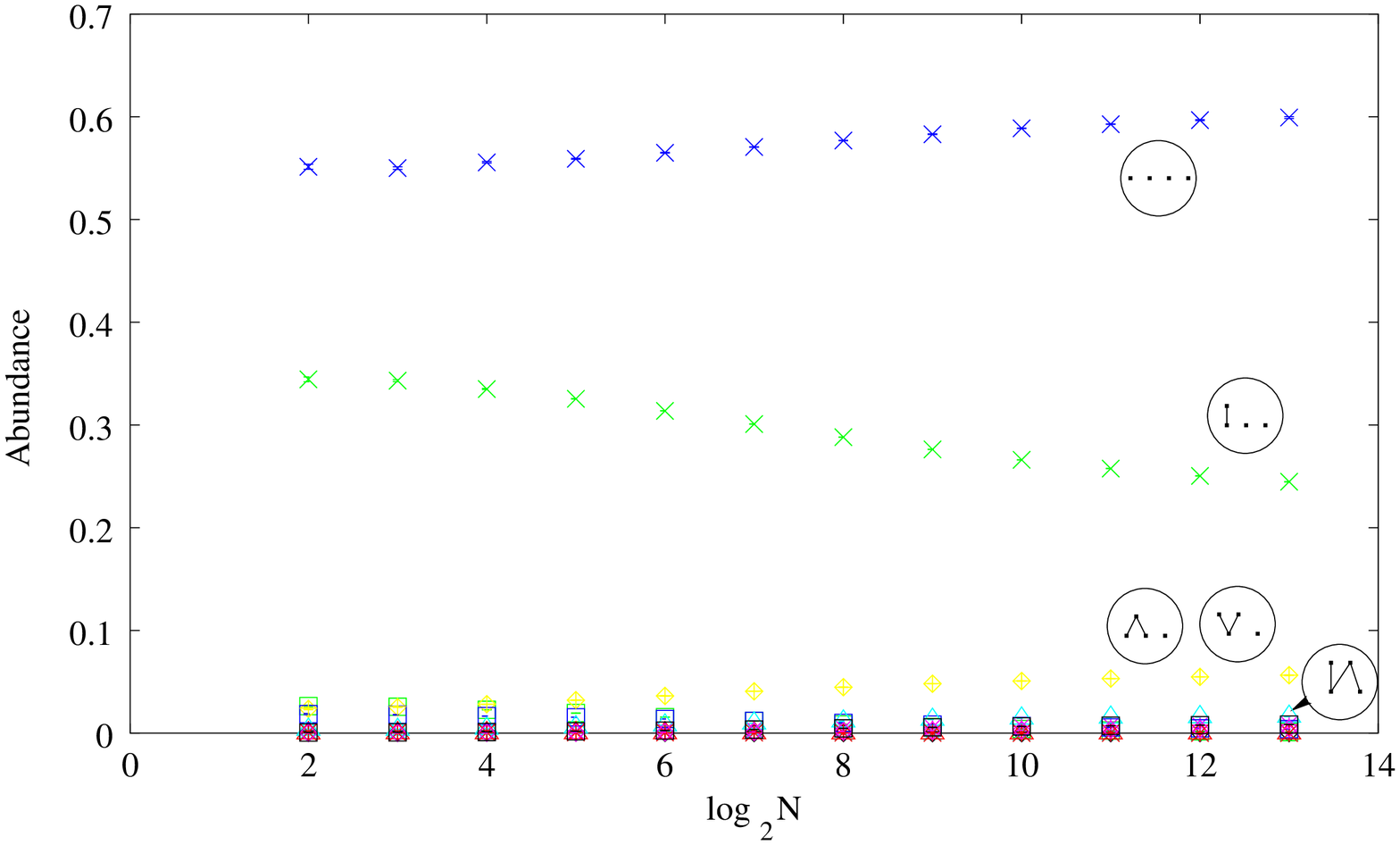} }
\caption{Flow of the coarse-grained probabilities $f_m$ for $m=4$.
      The 2-chain probability is held at 1/10.
      Only those curves lying high enough to be seen distinctly have
      been labeled.}
\label{4d-4}
\end{figure}

%that the more abundant sub-causets are those with fewer relations, for
%example the antichain and the $\Lo$.  The probability of
%coarse-graining to a chain is extremely small.
All the trajectories discussed so far produce causets with an ordering
fraction $r$ close to 1/2 for large $N$.  As mentioned earlier,
$r=1/2$ corresponds to a Myrheim-Meyer dimension of two.  Figures
\ref{4d-3} and \ref{4d-4} show the results of a simulation along the
``four dimensional'' trajectory defined by $r=1/10$.  (The value
$r=1/10$ corresponds to a Myrheim-Meyer dimension of 4.)  Here the
appearance of the flow is much less elaborate, with the curves arrayed
simply in order of increasing ordering fraction, $\threeach$ and
$\fourach$ being at the top and $\threech$ and (imperceptibly)
$\fourch$ at the bottom.  As before, all the curves are monotone as
far as can be seen.  Aside from the intrinsic interest of the case
$d=4$, these results indicate that our conclusions drawn for $d$ near
2 will hold good for all larger $d$ as well.

%% [RDS: david: do the numerical data confirm monotonicity for the
%% imperceptible curves in figure {4d-4}? ]
% Rafael agreed that they did.  I could show these curves as well, but this is extremely low priority...

\begin{figure}[htb]
\center
\scalebox{.53}{
\includegraphics{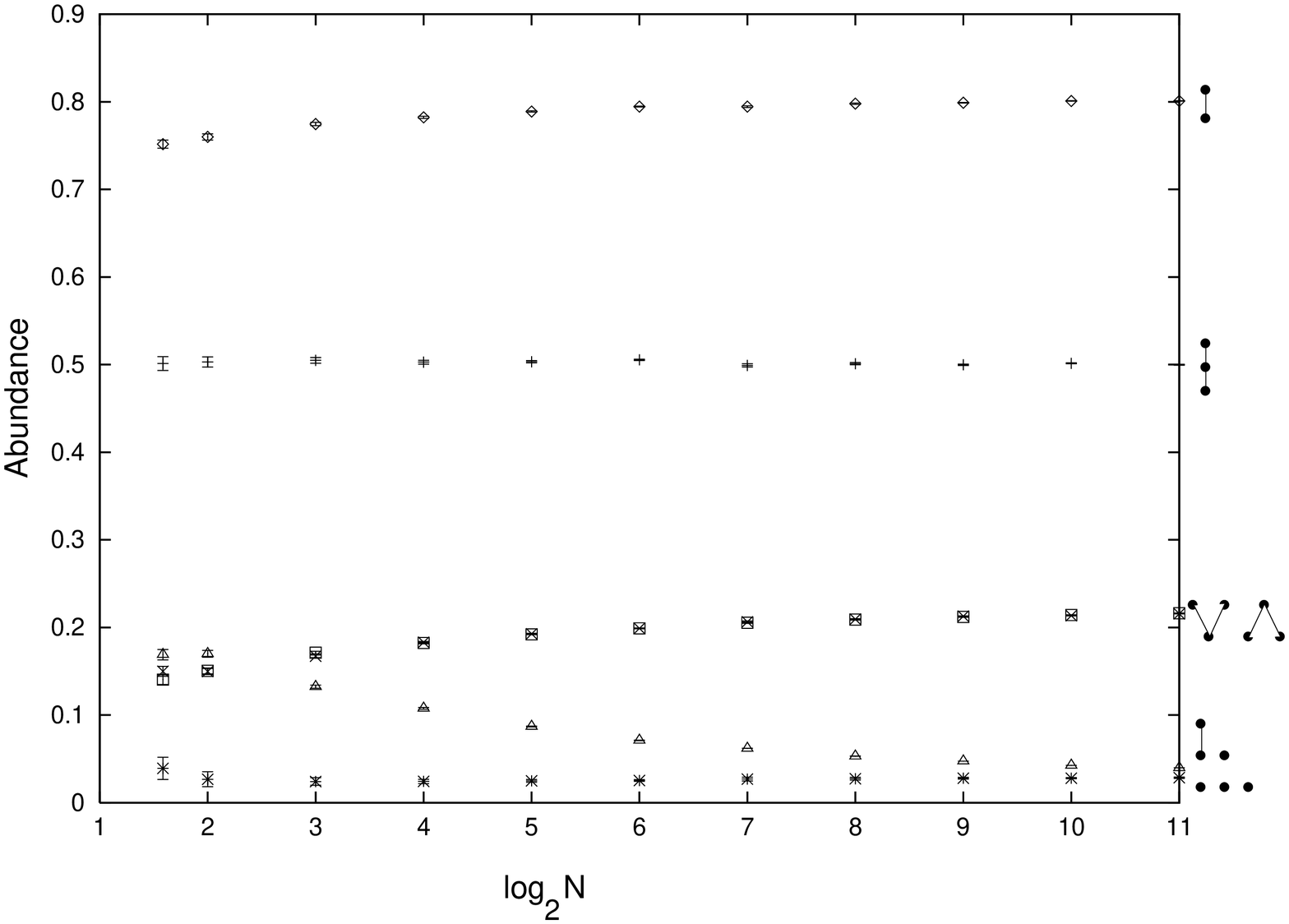} }
\caption{Flow of the coarse-grained probabilities $f_m$ for
$m=3$.  The 3-chain probability is held at 1/2.}
\label{3sub}
\end{figure}

Figure \ref{3sub} displays the flow of the coarse-grained
probabilities from a simulation in the opposite situation where the
ordering fraction is much greater than 1/2 (the Myrheim-Meyer
dimension is down near 1.)  Shown are the results of coarse-graining
to three element causets along the trajectory which holds the 3-chain
probability to 1/2.  Also shown is the 2-chain probability.  The
behavior is similar to that of Figure \ref{4d-3}, except that here the
coarse-grained probability rises with the ordering fraction instead of
falling.  This occurs because constraining $f_3(\threech)$ to be 1/2
generates rather chain-like causets whose Myrheim-Meyer dimension is
in the neighborhood of 1.34, as follows from the approximate limiting
value $f_2(\twoch)\approx0.8$.  The slow, monotonic, variation of the
probabilities at large $N$, along with the appearance of convergence
to non-zero values in each case, suggests the presence of a nontrivial
continuum limit for $r$ near unity as well.

\begin{figure}[htb]
\center
\scalebox{1.07}{ % 1.2
\includegraphics{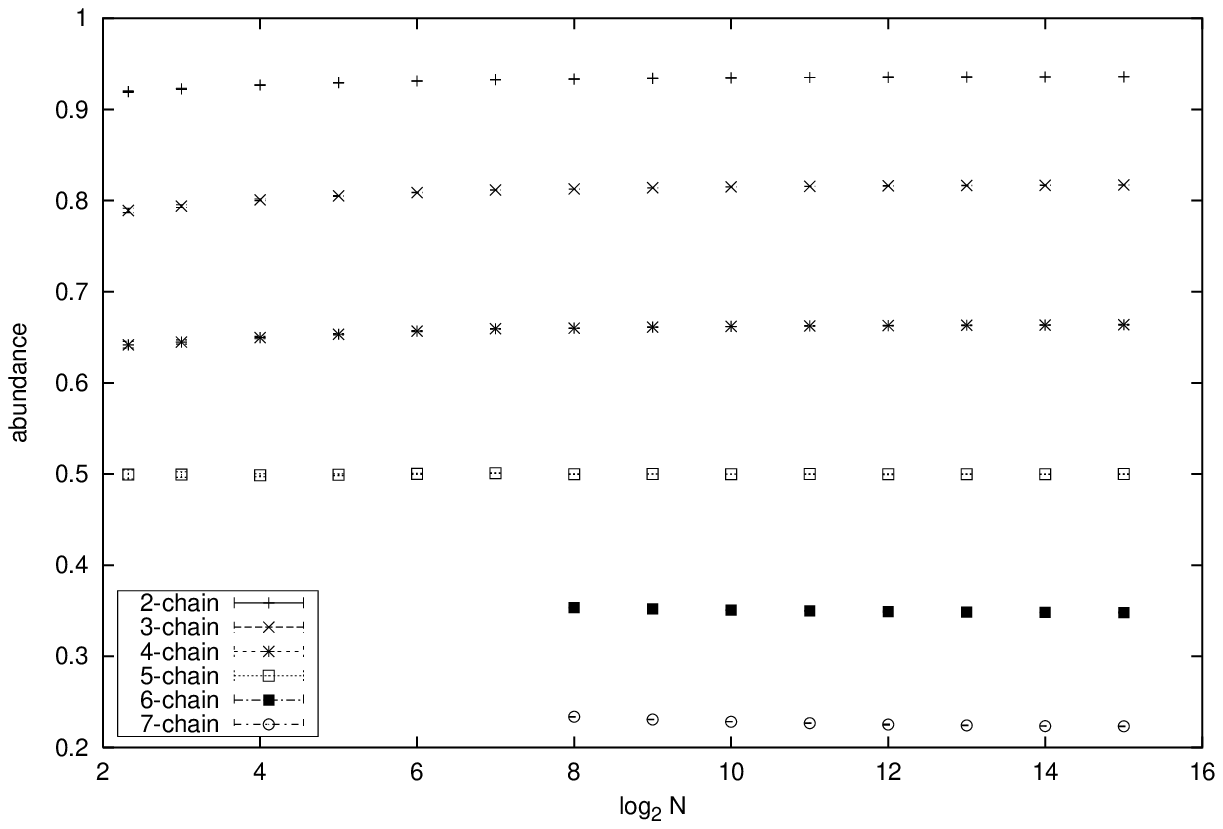} }
\caption{Flow of the coarse-grained probabilities $f_m$($m-$chain) for
   $m=2$ to $7$.  The 5-chain probability is held at 1/2.}
\label{5chtraj}
\end{figure}

\begin{figure}[htb]
\center
\scalebox{.68}{ % .7
\includegraphics{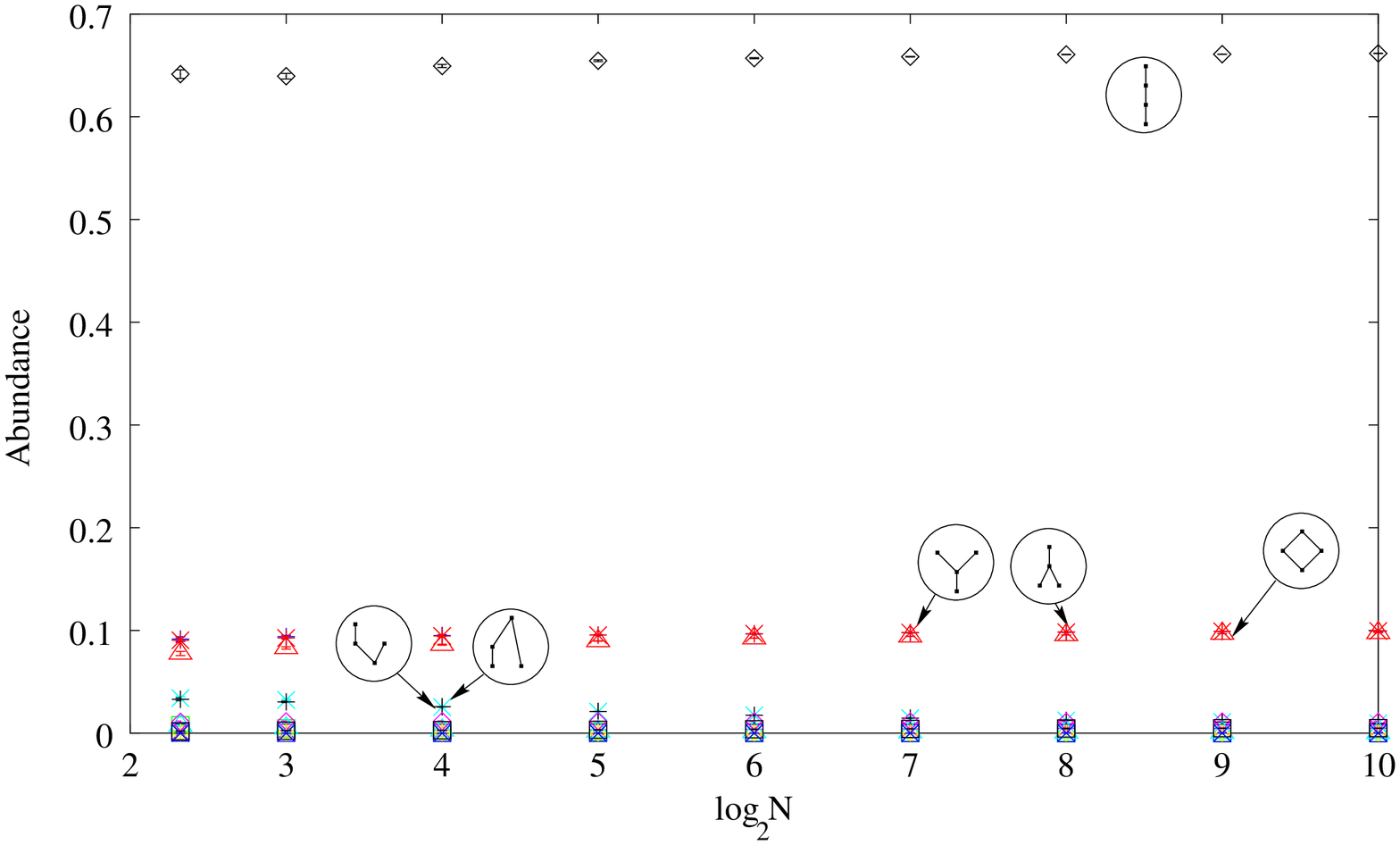} }
\caption{Flow of the coarse-grained probabilities $f_m$ for $m=4$.
         The 5-chain probability is held at 1/2.}
\label{5ch-4}
\end{figure}

% The results graphed in Figure \ref{5chtraj} arose from holding the
% mean 5-chain abundance $f_5(\hbox{\rm 5-chain})$ constant at 1/2, and
% tracking the abundances of the $k$-chains for $k$ between 2 and 7,
% inclusive.  
Figures \ref{5chtraj} and \ref{5ch-4} present the results of a final
set of simulations, the only ones we have carried out which examined
the abundances of causets containing more than four elements.  In
these simulations, the mean 5-chain abundance $f_5(\hbox{\rm
5-chain})$ was held at 1/2, producing causets that were even more
chain-like than before (Myrheim-Meyer dimension $\approx1.1$).  In
Figure \ref{5chtraj}, we track the resulting abundances of all
$k$-chains for $k$ between 2 and 7, inclusive.  (We limited ourselves
to chains, because their abundances are relatively easy to determine
computationally.)  As in Figure \ref{3sub}, all the coarse-grained
probabilities appear to be tending monotonically to limits at large
$N$.  In fact, they look amazingly constant over the whole range of
$N$, from 5 to $2^{15}$.  One may also observe that the coarse-grained
probability of a chain decreases markedly (and almost linearly over
the range examined!)  with its length, as one might expect.  It
appears also that the $k$-chain curves for $k\not=5$ are ``expanding
away'' from the 5-chain curve, but only very slightly.
Figure \ref{5ch-4} displays the flow of the probabilities for 
coarse-grainings to four elements.
%  along the $f_5(\hbox{\rm 5-chain})=1/2$ trajectory.
It is qualitatively similar to Figures \ref{4d-3}--\ref{3sub}, with
very flat probability curves, and here with a strong preference for
causets having many relations over those having few.  Comparing
Figures \ref{5ch-4} and \ref{4d-4} with Figures \ref{3ch-4} and
\ref{2ch-4}, we observe that trajectories which generate causets that
are rather chain-like or antichain-like seem to produce distributions
that converge more rapidly than those along which the ordering
fraction takes values close to 1/2.

In the way of further simulations, it would be extremely interesting
to look for continuum limits of some of the more general dynamical
laws discussed in \S \ref{cosmologies}.  In doing so,
however, one would no longer have available (as one does have for
transitive percolation) a very fast (yet easily coded) algorithm that
generates causets randomly in accord with the underlying dynamical
law.  Since the sequential growth dynamics of Chapter \ref{dynamicschap} is
produced by a stochastic process defined recursively on the causal
set, it is easily mimicked algorithmically; but the most obvious
algorithms that do so are too slow to generate efficiently causets of
the size we have discussed in this paper.  Hence, one would either
have to devise better algorithms for generating causets ``one off'',
or one would have to use an entirely different method to obtain the
mean abundances, like Monte Carlo simulation of the random causet.

\subsection{Concluding Comments} % on this Continuum Limit}

%: n0 to n1 region of spacetime --> gauge dependence
%Transitive percolation is a discrete dynamical theory characterized by
%a single parameter $p$ lying between $0$ and $1$, which gives 
%%one obtains 
%a probability distribution on causal sets.
%%It can be regarded
%%as a model for cosmology, or as representing a local region of spacetime.
%%Regarded as a
%%stochastic process, it describes the steady growth of a causal set by
%%the continual birth or ``accretion'' of new elements.  
%If the goal were to %we were to regard the dynamics as a
%model spacetime of cosmological scale, %however,
%it may be suitable
%to investigate the continuum limit of originary percolation.
%The non-originary dynamics, which we have investigated here,
%%What we have done here 
%may be more appropriate in the context of
%spacetime regions of sub-cosmological scale.
%%If we wish to regard
%As such we may imagine that we are limiting
%ourselves to that portion of the causet comprising the elements with
%labels 
%%born
%between $N_0$ and $N_1$, %of the stochastic process, 
%thus obtaining a model of random posets containing $N=N_1-N_0$ elements.  

Transitive percolation is a discrete dynamical theory characterized by
a single parameter $p$ lying between $0$ and $1$.  Regarded as a
stochastic process as described in \S \ref{growth}, it describes the
steady growth of a causal set by the continual birth or ``accretion''
of new elements.  If we limit ourselves to that portion of the causet
comprising the elements born between step $N_0$ and step $N_1$ of the
stochastic process, we obtain a model of random posets containing
$N=N_1-N_0$ elements.  
%This is the model we have studied in this paper.

Because the underlying process is homogeneous, this model does not
depend on $N_0$ or $N_1$ separately, but only on their difference.  It
is therefore characterized by just two parameters $p$ and $N$.  One
should be aware that this truncation to a finite model is not
consistent with discrete general covariance (c.f. \S \ref{gen_cov} and
\S \ref{gencov}), because it is the subset
of elements with certain {\it labels} that has been selected out of
the larger causet, rather than a subset characterized by any directly
physical condition.  Thus, we have introduced an ``element of gauge''
and we hope that we are justified in having neglected it.  That is, we
hope that the random causets produced by the model we have actually
studied are representative of the type of suborder that one would
obtain by percolating a much larger (eventually infinite) causet and
then using a label-invariant criterion to select a subset of $N$
elements.

%: how could it fail?
Leaving this question aside, let us imagine that our model
represents an interval (say) in a causet $C$ underlying some
macroscopic spacetime manifold.  With this image in mind, it is
natural to interpret a continuum limit as one in which $N\to\infty$
while the coarse-grained features of the interval in question remain
constant.  We have made this notion precise by defining
coarse-graining as random selection of a suborder whose cardinality
$m$ measures the ``coarseness'' of our approximation.  A continuum
limit then is defined to be one in which $N$ tends to $\infty$ such
that, for each finite $m$, the induced probability distribution $f_m$
on the set of $m$-element posets converges to a definite limit, the
physical meaning being that the dynamics at the corresponding
length-scale is well defined.  Now, how could our model {\it fail} to
admit such a limit?

% in $\lambda\phi^4$ theory, $\lambda>0$, $m>0$ l\to 0 forces \lambda
% \to \infty before l reaches 0
In a field-theoretic setting, failure of a continuum limit to exist
typically means that the coarse-grained theory loses parameters as the
cutoff length goes to zero.  For example, $\lambda\phi^4$ scalar field
theory in 4 dimensions depends on two parameters, the mass $\mu$ and
the coupling constant $\lambda$.  In the continuum limit, $\lambda$ is
lost (i.e. it necessarily vanishes, because it is driven to infinity
at finite cutoff if one tries to hold its renormalized value fixed),
although one can arrange for $\mu$ to survive.  (At least this is what
most workers believe occurs.)  Strictly speaking, one should not say
that a continuum limit fails to exist altogether, but only that the
limiting theory is poorer in coupling constants than it was before the
limit was taken.  Now in our case, we have only one parameter to start
with, and we have seen that it does survive as $N\to\infty$ since we
can, for example, choose freely the $m=2$ coarse-grained probability
distribution $f_2$.  Hence, we need not fear such a loss of parameters
in our case.

What about the opposite possibility?  Could the coarse-grained theory
{\it gain} parameters in the $N\to\infty$ limit, as might occur if the
distributions $f_m$ were sensitive to the fine details of the
trajectory along which $N$ and $p$ approached the ``critical point''
$p=0$, $N=\infty$?\footnote%
{\singlesp For example, at a water-ice phase transition, the state is
specified by the pressure, temperature, and density, while away from
the transition the pressure and temperature suffice.  Thus the theory
picks up a parameter in the limit of approaching the phase transition.}
%one is free to 
% consider an approach to the water-ice phase
%transition.  Away from the transition, for a given pressure and
%temperature, we have either ice or water.  However at the transition
%we are free to specify the density as well, so that by approaching the
%transition 
%
%Such an increase of the parameter set through a limiting process
% seems logically possible, although we know of no example of it from
% field theory or statistical mechanics, unless one counts the extra
% global parameters that come in with ``spontaneous symmetry
% breaking''.}
%  For example, a water-ice equilibrium is not characterized by the 
%  temperature and pressure alone, though it is by density and temperature.
% [[ What is Rafael thinking here?]]
Our simulations showed no sign of such sensitivity, although we did
not look for it specifically.  (Compare, for example, Figure
\ref{2ch-3} with Figure \ref{3ch-3} and \ref{2ch-4} with \ref{3ch-4}.)

A third way the continuum limit could fail might perhaps be viewed as
an extreme form of the second.  It might happen that, no matter how
one chose the trajectory $p=p(N)$, some of the coarse-grained
probabilities $f_m(\xi)$ oscillated indefinitely as $N\to\infty$,
without ever settling down to fixed values.  Our simulations leave
little room for this kind of breakdown, since they manifest the exact
opposite kind of behavior, namely monotone variation of all the
coarse-grained probabilities we ``measured''.

Finally, a continuum limit could exist in the technical sense, but it
still could be effectively trivial (once again reminiscent of the
$\lambda\phi^4$ case --- if you care to regard a free field theory as
trivial).  
% lambda is lost in the limit, hence the coupling vanishes, --> free
% field theory
Here triviality would mean that all --- or almost all --- of the
coarse-grained probabilities $f_m(\xi)$ converged either to 0 or to 1.
Plainly, we can avoid this for at least some of the $f_m(\xi)$.
%coarse-grained probabilities.  
For example, we could choose an $m$ and hold either
$f_m(m\hbox{-chain})$ or $f_m(m\hbox{-antichain})$ fixed at any
desired value.
(As $p\to1$, $f_m(m\hbox{-chain})\to1$ and
$f_m(m\hbox{-antichain})\to0$; as $p\to0$, the opposite occurs.)
However, in principle, it could still happen that all the other $f_m$
besides these two went to 0 in the limit.  (Clearly, they could not go to
1, the other trivial value.)  Once again, our simulations show the
opposite behavior.  For example, we saw that $f_3(\V)$ {\it increased}
monotonically along the trajectory of Figure \ref{2ch-3}.
%
%In fact it can easily be proved that all the other $f_m$ cannot
%vanish.  Consider the continuum limit chosen such that
%$f_m(m\hbox{-chain})$ is held fixed at $a$ not 0 or 1, for some finite
%$m$.  Then the degenerate case would insist that
%$f_m(m\hbox{-antichain})=1-a$, and $f_m(\chi)$ for all other $\chi$
%vanish.  Imagine choosing two elements at random from a causet $C$ (of
%cardinality $N$, for $N$ arbitrarily large) along this trajectory.
%They will be related with probability $a$ (since they are either part
%of an $m$-chain or an $m$-antichain).  If they are related then a
%third element chosen at random must also be related (since all
%$f_m(\chi)$ with $\chi$ having \Lcauset as a subcauset vanish),
%implying that $C$ is a chain.  But then $f_m(m\hbox{-antichain})=1-a$
%for $a\neq1$ is not possible.  If the first two element chosen at
%random had been unrelated, then a similar argument leads to the
%impossibility of $a\neq0$.  It seems reasonable that with more work a
%similar proof could be constructed demonstrating the impossibility of
%a less trivial continuum limit (e.g. one in which $f_m(\chi)\neq0$
%only if $\chi$ can be expressed as a disjoint union of chains and
%antichains).

Moreover, even without reference to the simulations, we can make this
hypothetical ``chain-antichain degeneracy'' appear very implausible by
considering a ``typical'' causet $C$ generated by percolation for
$N\gg 1$
with $p$ on the trajectory that, for some chosen $m$, holds
$f_m(m\hbox{-chain})$ fixed at a value $a$ strictly between 0 and 1.
Then our degeneracy would insist that $f_m(m\hbox{-antichain})=1-a$ and
$f_m(\chi)=0$ for all other $\chi$.  But this would mean that, in a
manner of speaking, ``every'' coarse-graining of $C$ to $m$ elements
would be either a chain or an antichain.  In particular the causet
\Lcauset could not occur as a subcauset of $C$; whence, since \Lcauset
is a subcauset of every $m$-element causet except the chain and the
antichain, $C$ itself would have to be either an antichain or a chain.
But it is absurd that percolation for any parameter value $p$ other than
0 and 1 would produce a ``bimodal'' distribution such that $C$ would
have to be either a chain or an antichain, but nothing in between.  It
seems likely that similar arguments could be devised against the
possibility of similar, but slightly less trivial trivial continuum
limits, for example a limit in which $f_m(\chi)$ would vanish unless
$\chi$ were a disjoint union of chains and antichains.

Putting all this together, we have persuasive evidence that the
percolation model does admit a continuum limit, with the limiting
model being nontrivial and described by a single ``renormalized''
parameter or ``coupling constant''.  
%In this sense, the percolation
%model may be said to be ``renormalizable''.  
Furthermore, the
associated scaling behavior one might think to find in such a case is
also present, as will be discussed further in \cite{scaling}.

However, the question remains as to whether this continuum resembles a
spacetime manifold, or is something more pathological.  Do the causal
sets yielded by the percolation dynamics resemble genuine spacetimes?
%But is the word ``continuum'' here just a metaphor, or can it be taken
%more literally?  
%This depends, of course, on the extent to which the
%causets yielded by percolation dynamics resemble genuine spacetimes.
Based on the meager evidence available at the present time, for
example that mentioned in \S \ref{continuum} and \S \ref{homogeneous}
we can only answer ``it is possible''.
%On one hand, we know that
%any spacetime produced by percolation would have to be homogeneous.
%We also know that two very different dimension estimators (Myrheim-Meyer
%and midpoint scaling) seem to agree on
%percolated causets, which one might not expect, were there no actual
%dimensions for them to be estimating.  Certain other indicators 
%%(those arising from counting ``nuggets'') 
%tend to
%behave poorly, on the other hand, but they are just the ones that are
%{\it not} invariant under coarse-graining (they are not ``RG
%invariants''), so their poor behavior is consistent with the expectation
%that the causal set will not be manifold-like at the smallest scales
%(``foam''), but only after some degree of coarse-graining.  
%%Could mention somewhere the failure (again, c.f. nugget dimension) of
%%transitive percolation to reproduce intervals in Minkowski space.
%%[I'll have to understand it better from Rafael first.]

Finally, there is the ubiquitous issue of ``fine tuning'' or ``large
numbers''.  In any continuum situation, a large number is being
manifested (an actual infinity in the case of a true continuum) and
one may wonder where it came from.  In our case, the large numbers
were $p^{-1}$ and $N$.  For $N$, there is no mystery: unless the 
%birth process ceases, 
causal set ``ends'' at some point, $N$ is guaranteed to be as large as
desired.  But why should $p$ be so small?  Here we can appeal to the
preliminary results of Dou mentioned in \S \ref{cosrenormsec}, which
state that under cosmological renormalization, certain physically
reasonable dynamics of causal sets are driven toward an effective
percolation-like phase with a value of $p$ that scales like
$N_0^{-1/2}$, where $N_0$ is the number of elements of the causet
preceding the most recent ``bounce''.  Since this is sure to be an
enormous number if one waits long enough, $p$ is sure to become
arbitrarily small if sufficiently many cycles occur.  The reason for
the near flatness of spacetime --- or if you like for the large
diameter of the contemporary universe --- would then be just that the
underlying causal set is very old --- old enough to have accumulated,
let us say, $10^{480}$ elements in earlier cycles of expansion,
contraction and re-expansion.
% It would be good to explain where this number comes from, but lets
% leave this as low priority.

%If ---
%cosmologically considered --- the causet that is our universe has
%cycled through one or more phases of expansion and recollapse, then
%its dynamics will have been filtered through a kind of ``temporal
%coarse-graining'' or ``RG transformation'' that tends to drive it
%toward transitive percolation.  But what we didn't mention earlier was
%that the parameter $p$ of this effective dynamics 

%Perhaps some comments should be made regarding the significance of
%these results (or lack thereof...?), given the apparent
%misunderstanding of our reviewer.  This is \emph{a} continuum limit,
%but not necessarily the continuum of spacetime.  We have made no claim
%of manifoldness in that sense, spacetimeness, only that some sort of
%non-trivial dynamics remains in the limit $N\to\infty$.  I should
%discuss this with Rafael...
%I wouldn't worry about it too much, though.  Let's make this concern
%low priority.

\section{Scaling}
\label{scaling}

The trajectories of Figure \ref{6traj} can be fit numerically to
explore the scaling behavior of this model.  Doing so seems to suggest
an asymptotic functional form of $\log N/N$, which agrees with the
prediction of \cite{pt}.  Not only does this evidence for scaling
support the conclusion of the existence of a continuum limit, but the
asymptotic form can also be used to suggest a choice for the
parameters $t_n$ of the classical dynamics (of (\ref{PolC}), say)
which may produce spacetimes with a dimension which is constant over
many length scales.  Details will appear in \cite{scaling}.

%Brightwell \& Bollob\'as, Pittel \& Tungol

%guide in choosing $t_n$

%\include{scaling}
\chapter{Classical Dynamics of Sequential Growth}
\label{dynamicschap}

\section{Introduction}
\label{dyn_intro}
As illustrated in Chapter \ref{introchap}, the causal set approach to
quantum gravity has experienced considerable progress in its kinematic
aspects.\footnote{Much of the text of this chapter is
taken directly from \cite{class_dyn}.}  For example, there exist
%one possesses 
natural extensions of the concepts of proper time and spacetime
dimensionality, which take us a significant way toward an answer
to the question, ``When does a causal set resemble a Lorentzian
manifold?''.
The dynamics of causal sets (the ``equations of
motion''), however, has not been very developed to date.  One of the
primary difficulties in formulating a dynamics for causal sets is the
sparseness of the fundamental mathematical structure.  When all one
has to work with is a discrete set and a partial order, even the
notion of what we should mean by a dynamics is not obvious.

%: dynamics --> sequential growth
Traditionally, one prescribes a dynamical law by specifying a
Hamiltonian to be the generator of the time evolution.  This practice
presupposes the existence of a continuous time variable, which we do
not have in the case of causal sets.  Thus, one must conceive of
dynamics in a more general sense.  
Considerable progress can be made
by envisaging evolution
as a process of stochastic growth to be described in terms of the
probabilities (in the classical case, or more generally the quantum
measures in the quantum case \cite{qmeasure}) of forming designated
causal sets.  
As mentioned in \S \ref{gencov}, %That is, 
the dynamical law will be a rule which assigns probabilities to
suitable classes of causal sets (a causal set being the ``history'' of
the theory in the sense of ``sum-over-histories'').  One can then use
this rule --- technically a probability measure --- to ask physically
meaningful questions of the theory.  For example one could ask ``What
is the probability that the universe possesses the diamond poset as a
partial stem?''.
% This repeats a lot of stuff from before, but I don't care anymore.

%: why classical dynamics
Why are we interested in a classical dynamics for causal sets, when our
ultimate aim is a quantum theory of gravity?  One obvious reason is that
the classical case, being much simpler, can help us to get used to a
relatively unfamiliar type of dynamical formulation, bringing out the
pertinent physical issues and guiding us toward physically suitable
conditions to place on the theory.  Is there, for example, an
appropriate form of causality that we can impose?  Should we attempt to
express the theory directly in terms of gauge invariant (labeling
independent) quantities, or should we follow precedent by enforcing
gauge invariance only at the end?  Some of these issues are well
illustrated with the theories we construct herein.

One of the best reasons to be interested in a classical dynamics for
causal sets is that quantum gravity must possess general relativity as
a classical limit.  Thus general relativity should be described as
some type of effective classical dynamics for causal sets, and one may
hope that the relevant dynamical law will be among the family
delineated here.  % o.k.?
(One can't be certain this will occur, because
general relativity, as a continuum theory, seems most likely to arise
as an effective theory for coarse-grained causal sets, rather than
directly as a limit of the microscopic discrete theory, and there is
no guarantee that this effective theory will have the same form as the
underlying exact one.)

%: matter
A question commonly asked of the causal set program is ``How could
nongravitational matter arise from only a partial order?''.  One
obvious answer is that matter can emerge as a higher level construct
via the Kaluza-Klein mechanism \cite{JFKK}, but this possibility has
nothing to do with causal sets as such.  The theory developed in this
chapter suggests a different mechanism, in that is possible to rewrite the
theory in such a way that the dynamics appears to arise from a kind of
``effective action'' for a field of Ising spins living on the
relations of the causal set.  A form of ``Ising matter'' is thus
implicit in what would seem at first sight to be a purely
``source-free'' theory.

\subsection{Sequential growth}
\label{growth}
%: lead sentence
%In order to discuss causal set dynamics in terms of a sequential
%growth process, it will be convenient to introduce some new
%terminology.

The dynamics which we will derive can be regarded as a process of
``cosmological accretion'' or ``growth''.  At each step of this
process an element of the causal set comes into being as the
``offspring'' of a definite set of the existing elements -- the
elements that form its past.  The phenomenological passage of time is
taken to be a manifestation of this continuing growth of the causet.
Thus, we do not think of the process as happening ``in time'' but
rather as ``constituting time'', which means in a practical sense that
there is no meaningful order of birth of the elements other than that
implied by the relation $\prec$.

In order to define the dynamics, however, we will treat the births as
if they happened in a definite order with respect to some fictitious
``external time''.  In this way, we introduce an element of ``gauge''
into the description of the growth process which we will have to
compensate by imposing appropriate conditions of ``gauge invariance''.
This fictitious order of birth can be represented as a natural
labeling of the elements, that is, a labeling by integers
$0,1,2,3,\ldots$ which are compatible with the causal order (recall
definition of natural labeling in \S \ref{gencov}).
% $\prec$ in
%the sense that $x\prec{y}\implies{label}(x)<label(y)$.
The relevant notion of gauge invariance (which we will call ``discrete
general covariance'') is then captured by the statement that the
labels carry no physical meaning.  We discuss this more extensively
later on.

%%: sequential growth and general covariance
%[ Should I include this somewhere?  It is Rafael's words anyway, so
% maybe don't bother.] [ fine point, low priority]
%A sequential description of this sort is advantageous in representing
%the future as developing out of the past, but on the other hand it
%could seem to rely on an external parameter time (the ``time'' in
%which the growth occurs), thereby violating the principle that
%physical time is encoded in the intrinsic order-relation of the causal
%set and nothing else.  
%If physically real, such a parameter time would
%yield a distinguished labeling of the elements and thereby a notion of
%``absolute simultaneity'', in contradiction to the lessons of both
%special and general relativity.  
%To avoid such a consequence, we
%postulated a principle of {\it discrete general covariance}, according
%to which no probability of the theory can depend on --- and no
%physically meaningful question can refer to --- the imputed order of
%births, except insofar as that order reflects the intrinsic precedence
%relation of the causal set itself.

It is helpful to visualize the growth of the causal set in terms of
paths in a poset $\poscau$ of finite causal sets.  (Thus viewed, the
growth process will be a sort of Markov process taking place in
$\poscau$.)  Each finite causet (or rather each isomorphism
equivalence class of finite causets) is one element of this poset.  If
a causet can be formed by accreting a single element to a second
causet, then the former (the ``child'') follows the latter (the
``parent'') in $\poscau$ and the relation between them is a link.
Drawing $\poscau$ as a Hasse diagram of Hasse diagrams, we get
Fig. \ref{poscau}.  (Of course this is only a portion of the infinite
diagram; it includes all the causal sets of fewer than five elements
and 8 of the 63 five element causets.
\begin{figure}[htb]
\center
\scalebox{.61}{\includegraphics{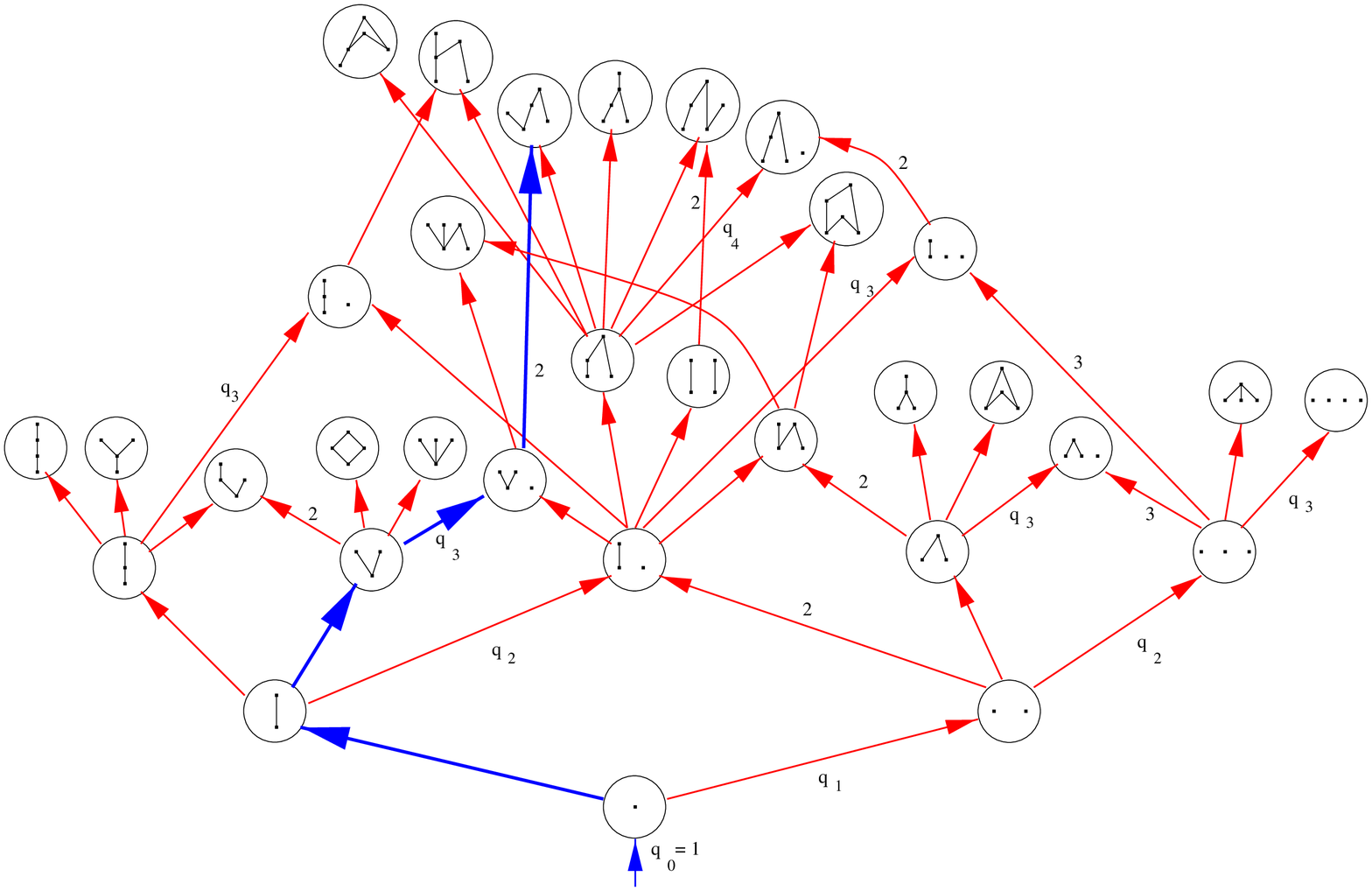}} 
\caption{The poset of finite causets}
\label{poscau}
\end{figure}
The ``decorations'' on some of the transitions in Fig. \ref{poscau}
are for later use.)  Any natural labeling of a causet $C\in\poscau$
determines uniquely a path in $\poscau$ beginning at the empty causet
and ending at $C$.  Conversely, any choice of upward path through this
diagram determines a naturally labeled causet, or rather a set of
them, since inequivalent labelings can sometimes give rise to the same
path in $\poscau$.\footnote
{\singlesp We could restore uniqueness by ``resolving'' each link
 $C_1\prec{C_2}$ of $\poscau$ into the set of distinct embeddings
 $i:C_1\to{C_2}$ that it represents.  Here, two embeddings count as
 distinct iff no automorphism of the child relates them (cf. the
 discussion of the Markov sum rule below).}
We want the physics to be independent of labeling, so different paths
in $\poscau$ leading to the same causet should be regarded as
representing the same (partial) universe, the distinction between them
being ``pure gauge''.

%: terminology
The causal sets which can be formed by adjoining a single maximal
element to a given causet will be called collectively a \emph{family}.
The causet from which they come is their \emph{parent}, and they are
\emph{siblings} of each other.  Each one is a \emph{child} of the
parent.  The child formed by adjoining an element which is to the
future of every element of the parent will be called the \emph{timid
child}.  The child formed by adjoining an element which is spacelike
to every other element will be called the \emph{gregarious child}.  
A child which is not the timid child will be called a \emph{bold}
child.  
(At times I may not be careful
to distinguish
between a child and the transition probability leading to that child,
assuming that the intended meaning is obvious from context.)
%
%In places where the intended meaning is obvious, I may not be careful
%to distinguish
%between a child and the transition probability leading to that child.
%How much do we actually use these definitions??  A lot for each
%except bold, which isn't used too much.  Should it be omitted?
%Maybe consider this later.
%never used:
%The set of parents of a given
%causal set $C$ form an \emph{antifamily}, each such parent of $C$ being
%an \emph{antichild} and $C$ itself being the \emph{antiparent}.

Each parent-child relationship in $\poscau$ describes a `transition'
$C\to C'$, from one causal set to another induced by the birth of a
new element.  The past of the new element (a subset of $C$) will be
referred to as the \emph{precursor set} of the transition (or
sometimes just the ``precursor of the transition'').  Normally, this
precursor set is uniquely determined up to automorphism of the parent
by the (isomorphism equivalence class of the) child, but (rather
remarkably) this is not always the case.\footnote
{\singlesp An example of the latter situation is
shown in Figure \ref{diffprecursor}, in which a parent causal set $C$
(whose elements are the 10 squares) can
undergo a transition to a child causal set $C'$ by adjoining a new
element, which is shown as one of the two circles.  
%Two different choices for the precursor for the new element
%are depicted by the two circles.
%as either of the circles.
Observe that a new element 
%in the place of 
at either circle will lead to
the same 11 element causet, but there is no automorphism of the parent
which maps one precursor into the other.}
%shows two possible
%choices of precursor for a new (11th) element (
%, consider the
%causal set of Figure \ref{diffprecursor}.
% [ This example may fit better in the examples section.  Let's
% decide on this later.  There is can be promoted to () or regular text.]
\begin{figure}[htb]
\center
\scalebox{.9}{\includegraphics{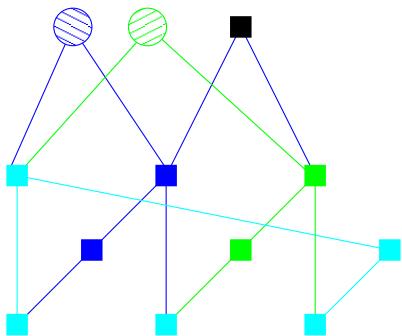}} 
\caption{Inequivalent precursors lead to same child}
\label{diffprecursor}
\end{figure}
The symbol $\C_n$ will denote the set of causets with $n$ elements,
and the set of all transitions from $\C_n$ to $\C_{n+1}$ will be
called \emph{stage $n$}.

As just remarked, each parent-child transition corresponds to a
choice of partial stem in the parent (the precursor of the transition).
Since there is a one-to-one correspondence between partial stems and
antichains, a choice of child also corresponds to a choice of (possibly
empty) antichain in the parent, the antichain in question being the set
of maximal elements of the 
precursor. % of the transition.
%past of the new element.  
Note also that the
new element will be \emph{linked} to each element of this antichain.

% Finally, notice that each parent-child transition determines a link in
% the poset of causets $\poscau$; hence a succession of such transitions
% determines a path in $\poscau$.  However the converse is not quite true
% because a given link in $\poscau$ determines parent and child only up to
% separate isomorphisms for both.  Thus given a concrete parent and a
% concrete child one does not yet know unambiguously the corresponding
% precursor set in the parent.  (In general the precursor set is not even
% determined up to automorphism of the parent!)  Modulo these
% reservations, one can think of a labeled causal set $C$ as
% equivalent to a path in $\poscau$ beginning with the one-element causal
% set and ending with $C$; but the more precise statement is that each
% labeled causal set determines a path, but not vice-versa.
%
% Why is this paragraph deleted?  I guess it doesn't add anything?
%% RDS [I think it only repeats things that we've now said earlier.  I
%% wouldn't strongly oppose keeping most of it as a footnote, if you
%% think it will add clarity] [ fine point, low priority]

\subsection{Some examples}
To help clarify the terminology introduced in the previous section, we
give some examples.  The 20 element causet of Figure \ref{20elts} was
generated by the stochastic dynamics described herein, with the choice
of parameters given by Equation (\ref{lifelike}) below (with $t=1$).
\begin{figure}[htbp]
\center
\scalebox{.9}{\includegraphics{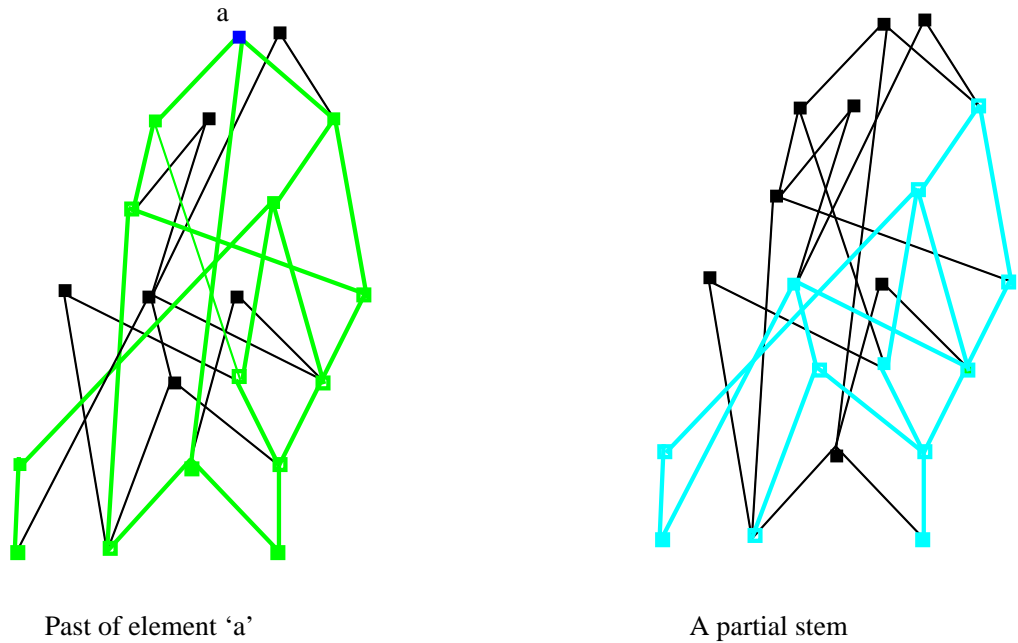}}
\caption{An example of a (`typical'?) 20 element causal set}
\label{20elts}
\end{figure}
In the copy of this causet on the left, the past of element $a$ is
highlighted.  Notice that since we use the irreflexive convention for
the order, $a$ is not included in its own past.  In the the copy on the
right, a partial stem of the causet is highlighted.

Figure \ref{family} shows \No and its children.
\begin{figure}[htbp]
\center
\scalebox{.64}{\includegraphics{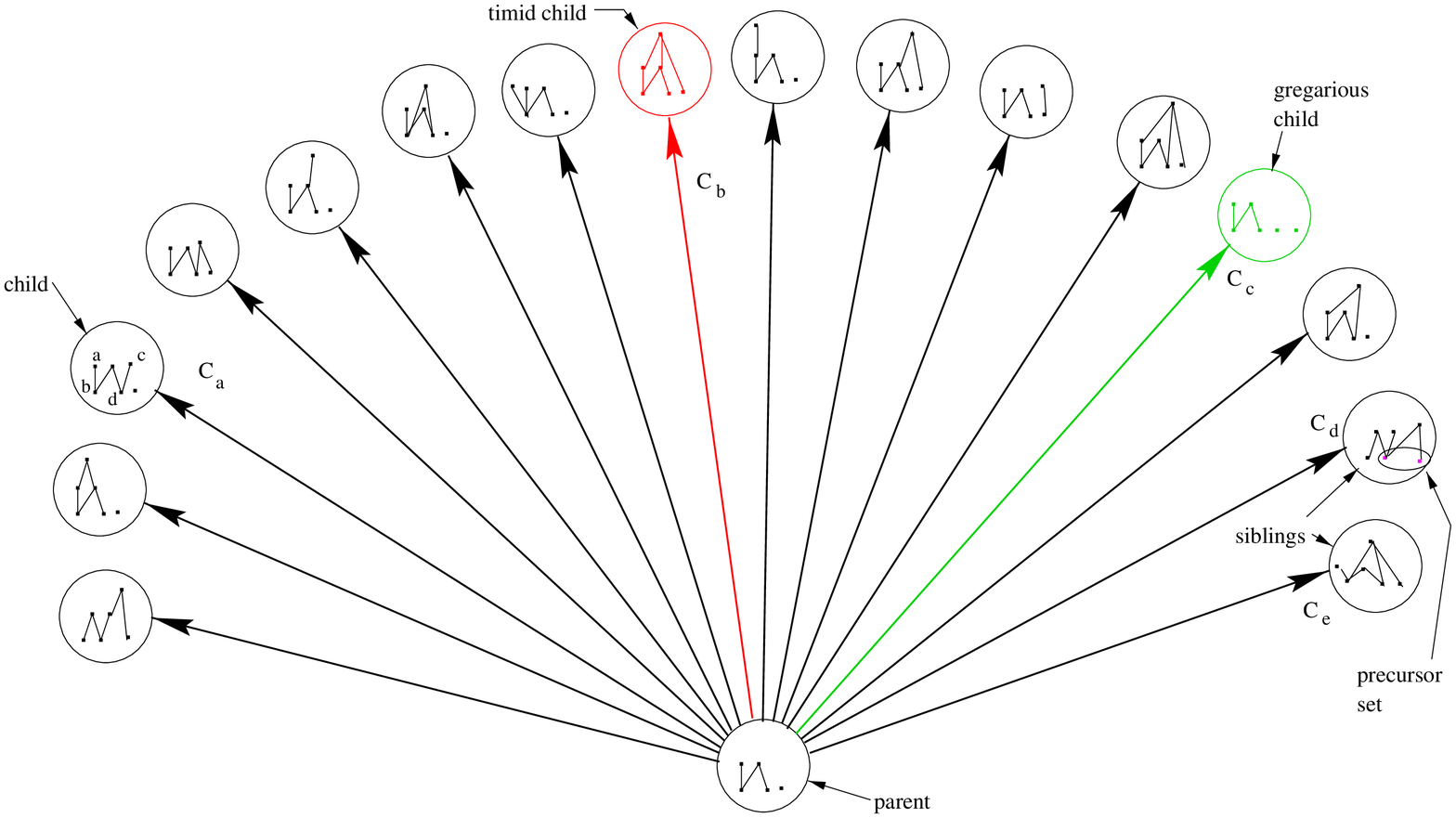}}
\caption{A family}
\label{family}
\end{figure}
The timid child is $C_b$ and the gregarious child is $C_c$.  The
precursor set leading to the transition to $C_d$ is shown in the
ellipse.  An example of an automorphism of $C_a$ is the map
$a\leftrightarrow c, b\leftrightarrow d$ (the other elements remaining
unchanged).

\section{Physical requirements on the dynamics}
\label{requirements}

The dynamics of transitive percolation, which was introduced in \S
\ref{perc}, can be expressed as a growth dynamics of the sort
presented in the previous section, by stating that each new
element forges a causal bond independently with each existing element
with probability $p\in[0,1]$.  (Any causal relation implied by
transitivity must then be added in as well.)
%As explained in the previous section, 
%As discussed in the previous section, one can think of transitive
%percolation as a sort of ``birth process'', but as such, 
However, this is only
one special case drawn from a much larger universe of possibilities.
As preparation for describing these more general possible dynamical
rules, let us consider the growth-sequence of a causal set universe.

First element `0' appears (say with probability one, since the
universe exists).  Then element `1' appears, either related to `0' or
not. Then element `2' appears, either related to `0' or `1', or both,
or neither.  Of course if $1\succ0$ and $2\succ1$ then $2\succ0$ by
transitivity.  Then element `3' appears with some consistent set of
ancestors, and so on and so forth.  Because of transitivity, each new
element ends up with a partial stem of the previous causet as its
precursor set.  
% and each such partial stem will occur with some probability.  Taken
% collectively, these {\it transition probabilities} determine the
% dynamics in question. ( this is explained in next+1 paragraph )
The result of this process, obviously, is a naturally
labeled causet (finite if we stop at some finite stage, or infinite if
we do not) whose labels record the order of succession of the
individual births.  For illustration, consider the path in
fig. \ref{poscau} delineated by the heavy arrows.  Along this path,
element `0' appears initially, then element `1' appears to the future
of element `0', then element `2' appears to the future of element `0',
but not to the future of `1', then element `3' appears unrelated to
any existing element, then element `4' appears to the future of
elements `0', `1' (say, or `2', it doesn't matter) and `3', then
element `5' appears (not shown in the diagram), etc.

Let us emphasize once more that the labels 0, 1, 2, etc. are not
supposed to be physically significant.  Rather, the ``external time''
that they record is just a way to conceptualize the process, and any
two birth sequences related to each other by a permutation of their
labels are to be regarded as physically identical.

So far, we have been describing the kinematics of sequential growth.
% Except in mentioning the tranperc dynamics...  Should I fix this?  Naah.
In order to define a dynamics for it, we may give, for each
$n$-element causet $C$, the {\it transition probability} from it to
each of its possible children.  Equivalently, we give a transition
probability for each partial stem within $C$.  We wish to construct a
general theory for these transition probabilities by subjecting them
to certain natural conditions.  In other words, we want to construct
the most general (classically stochastic) ``sequential growth
dynamics'' for causal sets.\footnote
{By choosing to specify our stochastic process in terms of transition
 probabilities, we have assumed in effect that the process is Markovian.
 Although this might seem to entail a loss of generality, the loss is
 only apparent, because the condition of discrete general covariance
 introduced below would have forced the Markov assumption on us,
 even if we had not already adopted it.}
In stating the following conditions, we will employ the terminology
introduced above. %in the Introduction of \S \ref{dyn_intro}.

\subsection{The condition of internal temporality}

By this imposing sounding phrase, we mean simply that each element is
born either to the future of, or unrelated to, all existing elements;
that is, no element can arise to the past of an existing element.

We have already assumed this tacitly in describing what we mean by a
sequential growth dynamics.  An equivalent formulation is that the
labeling induced by the order of birth must be {\it natural}, as
defined above.  The logic behind the requirement of internal
temporality is that all physical time is that of the intrinsic order
defining the causal set itself.  For an element to be born to the past
of another would be contradictory: it would mean that an event
occurred ``before'' another which intrinsically preceded it.

\subsection{The condition of discrete general covariance}
\label{gen_cov}
As we have been emphasizing, the ``external time'' in which the causal
set grows (equivalently the induced labeling of the resulting poset)
is not meant to carry any physical information.  We interpret this in
the present context as being the condition that the net probability of
forming any particular $n$-element causet $C$ is independent of the
order of birth we attribute to its elements.  
%\footnote % [again, repeats what is already here -- low priority ]
%{This last condition guarantees that the ``parameter time'' of our
% stochastic process {\it is compatible with} physical temporality, as
% recorded in the order relation $\prec$ that gives the causal set its
% structure.  In a broader sense, general covariance itself is also an
% aspect of internal temporality, since it guarantees that the
% parameter time {\it adds nothing to} the relation $\prec$.}
Thus, if
$\gamma$ is any path through the poset $\poscau$ of finite causal sets
that originates at the empty causet and terminates at $C$, then the
product of the transition probabilities along the links of $\gamma$
must be the same as for any other path arriving at $C$.  (So general
covariance in this setting is a type of path independence).
We should recall here, however, that, as observed earlier, a link in
$\poscau$ can sometimes represent more than one possible transition.
Thus our statement of path-independence, to be technically correct,
should say that the answer is the same no matter which transition
(partial stem) we select to represent the link.  
Obviously, this immediately entails that all
such representatives share the same transition probability.

We might with justice have required here conditions that are
apparently much stronger, including the condition that {\it any} two
paths through $\poscau$ with the same initial and final endpoints have
the same product of transition probabilities.  However, it is easy to
see that this already follows from the condition stated.\footnote
{\singlesp If $\gamma$ does not start with the empty causet $C_0$, but at $C_s$,
we can extend it to start at $C_0$ by choosing any fixed path from $C_0$
to $C_s$.  Then different paths from $C_s$ to the end-point $C_e$ correspond to
different paths between $C_0$ and $C_e$, and the equality of net
probabilities for the latter implies the same thing for the former.}
We therefore do not make it part of our definition of discrete general
covariance, although we will be using it crucially.

% See \S \ref{gencov} for discussion on this paragraph.
Finally, it is well to remark here that just because the ``arrival
probability at $C$'' is independent of path/labeling, that does not
necessarily mean that it carries an invariant meaning.  
On the
contrary a statement like ``when the causet had 8 elements it was a
chain'' is itself meaningless before a certain birth order is chosen.
This, also, is an aspect of the gauge problem, but not one that
functions as a constraint on the transition probabilities that define
our dynamics.  Rather it limits the physically meaningful {\it
questions} that we can ask of the dynamics.  Technically, we expect
that our dynamics (like any stochastic process) can be interpreted as
a probability measure on a certain $\sigma$-algebra, and the
requirement of general covariance will, in addition providing a
constraint on the transition probabilities of the growth process, 
% then 
serve to select the subalgebra of sets whose measures have direct
physical meaning.

% [ This may repeat some things already stated, but fooie, it will
% work.  low priority concern. ]

\subsection{The Bell causality condition}

The condition of ``internal temporality'' may be viewed as a very weak type
of causality condition.  The further causality condition we introduce now
is a discrete analog of the statement that no influence can propagate
faster than light.  This condition is 
quite strong, being similar to that from which one derives Bell's
inequalities.  
We believe that such a condition is appropriate for a
classical theory, and we expect that some analog will be valid in the
quantum case as well.  (On the other hand, we would have to abandon Bell
causality if our aim were to reproduce quantum effects from a classical
stochastic dynamics, as is sometimes advocated in the context of ``hidden
variable theories''.  Given the inherent non-locality of causal sets, 
there is no logical reason why such an attempt would have to fail.)

The physical idea behind our condition is that events occurring in some
part of a causal set $C$ should be influenced only by the portion of $C$
lying to their past.  In this way, the order relation constituting $C$
will be causal in the dynamical sense, and not only in name.  In terms of
our sequential growth dynamics, we make this precise as the requirement
that the ratio of the transition probabilities leading to two possible
children of a given causet depend only on the triad consisting of the
two corresponding precursor sets and their union.

Thus, let $C\rightarrow{}C_1$ designate a transition from $C\in\C_n$ to
$C_1\in\C_{n+1}$, and similarly for $C\rightarrow{}C_2$.  Then, the Bell
causality condition can be expressed as the equality of two 
ratios\footnote%
{In writing (\ref{BCr}), we assume for simplicity that both numerators
 and both denominators are nonzero.}%, this being the only case we will
% have occasion to treat in the present paper.}
\bne
  \frac {prob(C \rightarrow C_1)} {prob(C \rightarrow C_2)} 
   =
  \frac {prob(B \rightarrow B_1)} {prob(B \rightarrow B_2)} 
\label{BCr}
\ene
where $B\in\C_m$, ${m}\le{n}$, is the union of the precursor set of
$C\rightarrow C_1$ with the precursor set of $C\rightarrow C_2$,
$B_1\in\C_{m+1}$ is $B$ with an element added in the same manner as in
the transition $C\rightarrow{C_1}$, and $B_2\in\C_{m+1}$ is $B$ with
an element added in the same manner as in the transition $C\rightarrow
C_2$.\footnote%
{\singlesp Recall that the precursor set of the transition $C\to{}C_1$ is the
 subposet of $C$ that lies to the past of the new element that forms
 $C_1$.}
(Notice that if the union of the precursor sets is the entire parent
causet, then the Bell causality condition reduces to a trivial
identity.)

To clarify the relationships among the causets involved, it may help
to characterize the latter in yet another way. % latter?  
Let $e_1$ be the element born in the transition $C\to{}C_1$ and let
$e_2$ be the element born in the transition $C\to{}C_2$.  Then
$C_i=C\union\{e_i\}$ ($i=1,2$), and we have
$B=(\past\,e_1)\union(\past\,e_2)$ and $B_i=B\union\{e_i\}$ ($i=1,2$).

By its definition, Bell causality relates ratios of transition
probabilities belonging to one ``stage'' of the growth process to
ratios of transition probabilities belonging to previous stages.  For
illustration, consider the case depicted in fig. \ref{bc_ex}.
\begin{figure}[htbp]
\center
\scalebox{.6}{\includegraphics{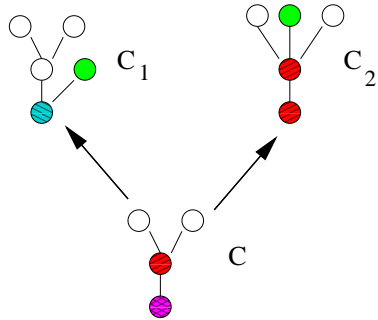}}
\caption{Illustrating Bell causality}
\label{bc_ex}
\end{figure}
The precursor $P_1$ of the transition $C\rightarrow{}C_1$ contains
only the earliest (minimum) element of $C$, shown in the figure as a
pattern-filled dot.  The precursor $P_2$ of $C\rightarrow{}C_2$
contains as well the next earliest element, shown as a (different
pattern)-filled dot.  The union of the two precursors is thus
$B={P_1}\union{P_2}=P_2$.  The elements of $C$ depicted as open dots
belong to neither precursor.  Such elements will be called
\emph{spectators}.  Bell causality says that the spectators can be
deleted without affecting relative probabilities.  Thus the ratio of
the transition probabilities of Figure \ref{bc_ex} is equal to that of
Figure \ref{bc_ex2}.
\begin{figure}[htbp]
\center
\scalebox{.6}{\includegraphics{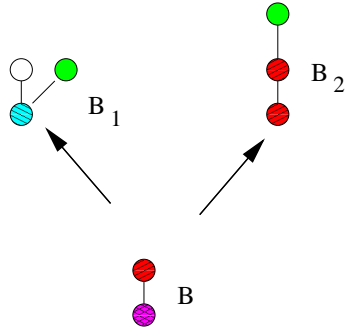}}
\caption{Illustrating Bell causality - spectators do not affect relative
probability}
\label{bc_ex2}
\end{figure}

% I could say more about how this definition of Bell causality
% captures the causality requirements of Bell, c.f. discussion with
% Sorkin & Malament 10/16/000 !
%[ Yes, it would be nice to do this, but I just don't have time.
%Other things are higher priority, such as scaling and grant apps!!! ]

\subsection{The Markov sum rule}
%We place a combinatorial factor in front of each term in the sum rule
%equation
%\be
%\sum_i \chi_i \alpha_i = 1 ,
%\ee
%where the sum is to be taken over all possible transitions in each
%family. 
As with any Markov process, we must require that the sum of the full
set of transition probabilities issuing from a given causet be unity.
However, the set we have to sum over depends in a subtle manner on the
extent to which we regard causal set elements as ``distinguishable''.
Heretofore we have identified distinct transitions with distinct
precursor sets of the parent.  In doing so, we have in effect been
treating causet elements as distinguishable (by not identifying with
each other, precursor sets related by automorphisms of the parent),
and this is what we shall continue to do.  Indeed, this is the
counting of children used implicitly by transitive percolation, so we
keep it here for consistency.  With respect to the diagram of Figure
\ref{poscau}, this method of counting has the effect of introducing
coefficients into the sum rule, equal to the number of partial stems
of the parent which could be the precursor set of the transition.  For
the transitions depicted there, these coefficients (when not one) are
shown next to the corresponding arrow.\footnote%
{\singlesp One might describe the result of setting these coefficients
 to unity as the case of ``indistinguishable causet elements''.  A
 preliminary investigation suggests that in this case a dynamics with
 a richer structure obtains: instead of the transition probability
 depending only on the size of the precursor set and the number of its
 maximal elements, it is sensitive to more details of the precursor
 set's structure.  However, since the dynamics of transitive
 percolation does not satisfy this modified sum rule, it appears
 difficult to derive a closed form expression for the transition
 probabilities in this case.}
%\subsubsection{``Bosonic'' Sum Rule}
%One can set all the $\chi_i=1$.  This is somewhat like regarding all
%causal set elements as ``indistinguishable''.  Details about finding
%the corresponding solution to the dynamical equations can be found on
%page u of the ``Classical Growth Dynamics Calculations'' folder.)
% In fact any coefficients could be used in the sum rule, leading to a
% different dynamics.  There is no obvious way to get a closed form
% expression for the dynamics for these generalizations, and
% furthermore no physical motivation, save for the ``bosonic'' rule.
% [ Should some of this stuff be promoted from a footnote, and
% discussed in a little more depth?]

These sum-rule coefficients admit an alternative
description in terms of embeddings of the parent into the child (as a
partial stem).  Instead of saying ``the number of partial stems of the
parent which could be the past of the new element'', we could say ``the
number of order preserving injective maps from the parent onto partial
stems of the child, divided by the number of automorphisms of the
child''.
To see this, let $e$ be such an embedding of parent $P$ into child $C$.
\be
   P \stackrel{e}{\longrightarrow} C
\ee
Given that the child has precisely one more element than the parent,
the injective
map $e$ singles out an element as the unique member of $C$ not
belonging to the image of $e$.  This element can be regarded as the
new element that arises from the transition $P\to C$.
%, i.e. the single element of $C\less(P)$.
%Call 
The past of this element (in $P$), $\Pi(e)$, will then be a possible
precursor of this transition
%Associate the precursor $\Pi(e)$ of the transition with the past of this element the precursor $\Pi(e)$
\bne
\Pi(e) = e^{-1}(\past(C\setminus e(P))) \label{etoprecursor} \:.
\ene
However, the maps $e$ overcount the number of precursors by the number of
automorphisms of the child, since each $e$ composed with an automorphism
$\alpha\in \mathrm{Aut}(C)$ yields a new map $f=\alpha\circ e$ 
%which in general is not equal to $e$, but nevertheless 
which corresponds to the same precursor.  
%Also two maps
%which lead to the same precursor must be related by an automorphism.
% Any pair of maps which give the same
%precursor are related by an automorphism, and an automorphism in $C$
%which leaves invariant the image of the map $e$ gives the same precursor.
To prove this, we must show that
\be
\Pi(e) = \Pi(f) \iff f = \alpha \circ e
\ee
%for some $\alpha \in \mathrm{Aut}(C)$.  To show $\Leftarrow$ 
To prove $\since$, insert (\ref{etoprecursor}) for $f$
\bee
\Pi(f)&=&(\alpha\circ e)^{-1} (\past(C\setminus \alpha(e(P)))) \\%\:.
%\ee
&=&(\alpha e)^{-1} (\past(\alpha(C\setminus e(P)))) \footnotemark  \\
%\eee
%(The second equality holds because 
&=&(\alpha e)^{-1} \alpha(\past(C\setminus e(P)))\\
&=&e^{-1}\alpha^{-1} \alpha(\past(C\setminus e(P)))\\
&=&\Pi(e) \:.
\eee
\footnotetext{\singlesp $C\setminus\alpha(X)=\alpha(C\setminus X)$ as sets, for
any set $X\subseteq C$.  An element 
%$y{}\in{}\alpha(C\setminus{}X) \iff
%\alpha^{-1}y\in{}C\setminus{}X \iff \alpha^{-1}y\notin{}X \iff y\notin
%\alpha(X)$
\bee
\lefteqn{y{}\in{}\alpha(C\setminus{}X)}\\
&&\iff \alpha^{-1}y\in{}C\setminus{}X \\
&&\iff \alpha^{-1}y\notin{}X \\
&&\iff y\notin \alpha(X) \:,
\eee
which is exactly the condition that $y\in C\setminus\alpha(X)$.}
The third equality holds because $\alpha$ is an automorphism.
To prove $\implies$, write $f$ as $\beta\circ e$, for some map
$\beta:C\to C$.  We then have
\bee
e^{-1}(\past(C\setminus e(P))) &=& f^{-1}(\past(C\setminus f(P)))\\
&=& e^{-1}\beta^{-1} (\past(C\setminus \beta(e(P))))\\
&=& e^{-1}\beta^{-1} (\past(\beta(C\setminus e(P)))) \:.
\eee
For this last equality to hold in general, $\beta$ must be an
automorphism.  %Thus each automorphism of $C$ 
% [[ Is this clear enough now?  Do I need to explain again how the
% counting works? ]]

\section{The general form of the transition probabilities}

We seek to derive a general prescription which gives, consistent with
our requirements, the transition probability from an element of
$\C_n$ to an element of $\C_{n+1}$.  To avoid having to deal with
special cases, we will assume throughout that no transition
probability vanishes.  Thus the solution we find may be termed
``generic'', but not absolutely general.

% [ stuff here cut to put into originary dynamics section ]

\subsection{Counting the free parameters}
A theory of the sort we are seeking provides a probability for each
transition, so without further restriction, it would contain a free
parameter for every possible antichain of every possible (finite)
causet.  We will see, however, that the requirements described above in
Section \ref{requirements} drastically limit this freedom.

\begin{lemma}
There is at most one free parameter per family.
\label{1param_family}
\end{lemma}

\noindent 
\textbf{Proof:} 
Consider a parent and its children (the set of possible transitions
from the parent).
A Bell causality equation will relate any pair of transitions whose
union of precursors is not the entire parent causet, since the
remaining elements will then be spectators, whose removal provides
such a relation.
%which can be removed to
%give a non-trivial Bell causality equation.
Since the precursor for the gregarious child is
empty, the complement of its union with any other partial stem (save that of the timid child,
whose precursor is the entire parent)
will be non-empty, resulting in a Bell causality equation relating the
pair of transition probabilities.
%will leave room for spectators
%the complement of its union with any partial stem (save the
%entire parent causet) will be a non-empty set,
Thus every child, except
the timid child, participates in a Bell causality equation with the
gregarious child.  
(See also the proof of Lemma \ref{bc_consis} in Appendix
\ref{consistency}.)
Since Bell causality equates ratios, all 
these transitions are determined up to an overall factor.
This leaves two free parameters for the family.  The Markov sum
rule gives another equation, which exhausts itself in determining
the probability of the timid child.  Hence precisely one
free parameter per family remains after Bell causality
and the sum rule are imposed.  $\Box$

\begin{lemma}
The probability to add a completely disconnected element (the
``gregarious child transition'') depends only on the cardinality of the
parent causal set.
\label{q's}
\end{lemma}

\noindent \textbf{Proof:}
\begin{figure}[htbp]
\center
\scalebox{.9}{\includegraphics{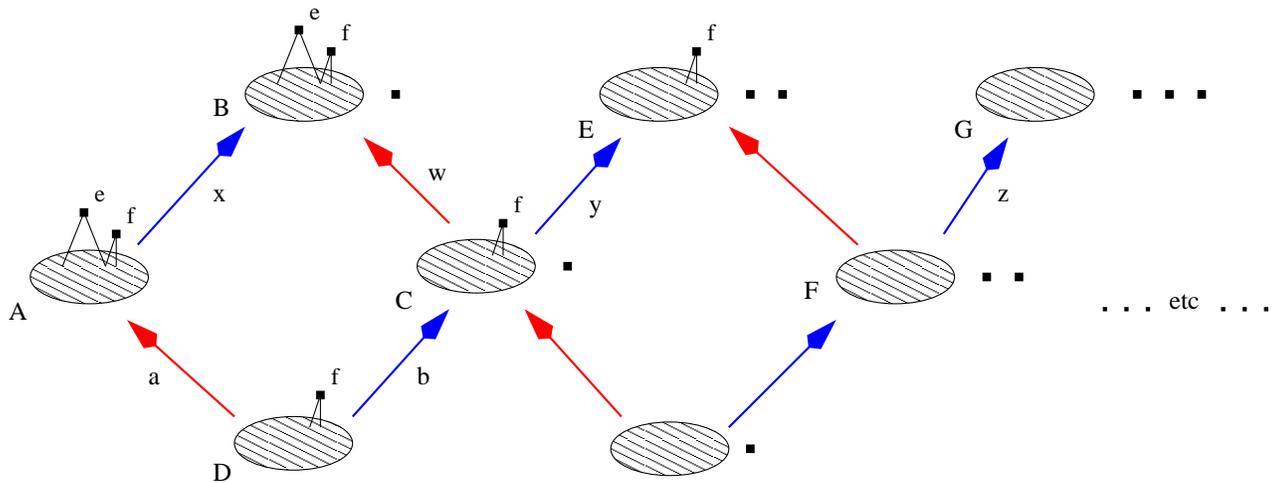}}
\caption{Equality of ``gregarious child'' transitions}
\label{GC}
\end{figure}
Consider an arbitrary causet $A$, with a maximal element $e$, as
indicated in Figure \ref{GC}.  Adjoining a disconnected element to $A$
produces the causet $B$.  Then, removing $e$ from $B$ leads to the
causet $C$, which can be looked upon as the gregarious child of the
causet $D=A\backslash\{e\}$.  Adding another disconnected element to
$C$ leads to a causet $E$ with (at least) two completely disconnected
elements.  The lower case letter attached to each arrow represents the
corresponding transition probability.  Now, by general covariance,
\be
      ax = bw 
\ee 
and by Bell causality,
\be 
    \frac{y}{w} = \frac{b}{a}
\ee 
(the disconnected element in $C$ acts as the spectator here).  Thus
\be 
    ax = bw = ay \Longrightarrow x=y
\ee 
(Recall that we have assumed that no transition probability vanishes.)
%Can repeat, regarding $C$ as ``new $A$'' (in effect moving maximal
%element $e$ to a disconnected element), thus eventually equating $x$
%to the transition probability from $n$-antichain to
%$n+1$-antichain $q_n$.
Repeating our deductions with $C$ in the place of $A$ in the above
argument (and another maximal element $f$ in the place of $e$), we see
that $y=z$, where $z$ is the probability for the transition from $F$
to $G$ (which has at least three completely disconnected elements) as shown.  Continuing in this way until we reach the antichain
$A_n$ shows finally that $x=q_n$, where we define $q_n$ as the
transition probability from the $n$-antichain to the
($n+1$)-antichain.  Since our starting causet $A$ was not chosen
specially, this completes the proof.  $\qed$

% Since each family has a gregarious child, choose this $q_n$ to be the
%free parameter at stage $n$.

If our causal sets are regarded as entire universes, then a
gregarious child transition corresponds to the spawning of a new,
completely disconnected universe (which is not to say that this new
universe will not connect up with the existing universe in the
future).  Lemma \ref{q's} proves that the probability for this to
occur does not depend on the internal structure of the existing
universe, but only on its size, which seems eminently reasonable.
In the sequel, we will call this probability $q_n$.

With Lemmas \ref{1param_family} and \ref{q's}, we have reduced the
number of free parameters (since every family has a gregarious child)
to 1 per stage, or what is the same thing, to one per causal set
element.  In the next sections we will see that no further reduction
is possible based on our stated conditions.  Thus, the transition
probabilities $q_n$ can be identified as the free parameters or
``coupling constants'' of the theory.  They are, however, restricted
further by inequalities that we will derive below.

%We have shown that Bell causality is self-consistent and that all the
%$q_n$'s (gregarious child transitions) on a level are equal.  General
%covariance follows from these two statements.  The sum rule is
%obviously consistent with everything, since it just gives the
%probabilities for the timid child transitions.  Thus all our
%conditions are mutually are self consistent.

\subsection{The general transition probability in closed form}

Given the $q_n$, the remaining transition probabilities (for the
non-gregarious children) are determined by Bell causality and the sum
rule, as we have seen.  Here we derive an expression in closed form
for an arbitrary transition probability in terms of causet invariants
and the parameters $q_n$.

\subsubsection{Mathematical form of transition probabilities}
We use the following notation:\\
%[This table is somewhat silly, but I'm too lazy to change it right
%now.  Maybe in the next run through?  Also it looks even more silly
%double spaced...]
\smallskip
\begin{tabular}{|c|c|}
\hline
$\alpha_n$ & an arbitrary transition probability from $\C_n$ to $\C_{n+1}$\\
$\beta_n$  & a transition whose precursor set is not the entire parent
             (`bold' transition)\\
$\gamma_n$ & a transition whose precursor set \emph{is} the entire parent
             (`timid' transition)\\
\hline
\end{tabular}\\

\smallskip
\noindent
Notice that the subscript $n$ here refers only to the number of elements
of the parent causet; it does not exhibit which particular transition
of stage $n$ is intended.  A more complete notation might provide $\alpha$,
$\beta$ and $\gamma$ with further indices to specify both the parent
causet and the precursor set within the parent.
%[[ How much of a faux pas is this?]]

We also set $q_0\ideq{}1$ by convention.

\begin{lemma}
Each transition probability $\alpha_n$ of stage $n$ has the form
\bne
           q_n \sum_{i=0}^n {\xi_i \frac{1}{q_i}}
\label{FN}
\ene
where the $\xi_i$ are integers depending on the individual transition
in question.
\label{integer_form}
\end{lemma}

\noindent \textbf{Proof:} 
This is easily seen to be true for stage %s 0 and  [[Rafael complained
					 %about \alpha_0=1 not
					 %declared yet.  It seems to
					 %be stated in $q_0\ideq{}1$
					 %by convention , but maybe
					 %its not clear to the reader.
1.  
Assume it is true for stage $n-1$.  
Consider a non-timid transition probability $\beta_n$ of stage $n$.  Bell
causality gives 
\bne
     \frac{\beta_n}{q_n} = \frac{\alpha_{n-1}}{q_{n-1}} \label{bc}
\ene
where $\alpha_{n-1}$ is an appropriate transition probability from
stage $n-1$.  (Bell causality has the property that not all spectators
have to be removed in writing the right
hand side of the equation.  Any partial stem of the subcauset of
spectators may be ``kept'' in selecting which transition probabilities
to place on the right hand side.)
%Instead of removing all spectators to form the right
%hand side of the Bell causality condition, any one of the (maximal)
%spectators can be omitted in selecting which transition probabilities
%to place on the right hand side.)
%The equality of (\ref{bc}) hold even 
So by induction
\bne 
  \beta_n = \alpha_{n-1} \frac{q_n}{q_{n-1}} 
  = \sum_{i=0}^{n-1} \xi_i \frac{q_{n-1}}{q_i} \frac{q_n}{q_{n-1}} 
  = \sum_{i=0}^{n-1} \xi_i \frac{q_n}{q_i} .
\label{BT}
\ene 
For a timid transition probability $\gamma_n$, we use the Markov sum rule: 
\bne
 \label{TT}
 \gamma_n = 1 - \sum_j  \beta_{nj}
\ene
where $j$ labels the possible bold transitions (i.e. the set of proper
partial stems of the parent).\footnote%
{Of course, more than one stem will in general correspond to the same
link in $\poscau$.  If we redefined $j$ to run over links in $\poscau$, then
(\ref{TT}) would read
$\gamma_n=1-\sum_j\chi_j\beta_{nj}$, where $\chi_j$ is the
``multiplicity'' of the $j$th link.}
But then, substituting (\ref{BT}) yields immediately
$$
  \gamma_n 
   = 1 - \sum\limits_j \sum\limits_{i=0}^{n-1} {\xi_{ji}\over q_i} q_n
   = 1 - \sum\limits_{i=0}^{n-1} {\sum_j \xi_{ji} \over q_i} q_n \,,
$$
which we clearly can put into the form (\ref{FN}) by taking
$\xi_i=-\sum_j\xi_{ji}$ for $i<n$ and
$\xi_n=1$.  $\Box$
% where the $\chi_j\in\Integers$ (actually the positive integers) 
% define the sum rule.  (are
% ?arbitrary coefficients in the sum rule?  Should I discuss the fact
% that \emph{any} choice of sum rule will give a viable theory?)  
%% RDS I'd vote against it, at this point.
% \be
% \gamma_n = 1 - \sum_j \chi_j \sum_{i=0}^{n-1} \xi_{ij} 
%	 \frac{q_n}{q_i}
% = 1 - \sum_{i=0}^{n-1} \left( \sum_j \chi_j \xi_{ij} \right) 
%	 \frac{q_n}{q_i}
% = \sum_{i=0}^n \left( \sum_j \chi_j \xi_{ij} \right)
%	 \frac{q_n}{q_i}
% \ee
% The sums in the parentheses are of course integers.  (Choose the
% $\xi_{jn}$ such that $\sum_j \chi_j \xi_{jn} = 1$.  (Do I even
% need to state this?)) 

\subsubsection{Another look at transitive percolation}
The transitive percolation model introduced in Chapter
\ref{chaptranperc} is consistent with the four conditions of Section
\ref{requirements}.  To see this, consider an arbitrary causal set
$C_n$ of size $n$.  Recall that, expressed in terms of a sequential
growth process, transitive percolation states that at each stage the
new element joins to each pre-existing element with probability $p$,
with extra relations added to insure transitivity.
Then the transition probability $\alpha_n$ from $C_n$
to a specified causet $C_{n+1}$ of size $n+1$ is given by
\bne
	\alpha_n = p^m (1-p)^{n-\pp}  \label{pp}
\ene
where $m$ is the number of maximal elements in the precursor set and
$\pp$ is the size of the entire precursor set.
(This becomes clear if one recalls how the precursor set of a newborn
element is generated in transitive percolation: first a set of ancestors
is selected at random, and then the ancestors implied by transitivity
are added.  From this, it follows immediately that a given stem
$S\subseteq{}C_n$ results from the procedure iff (i) every maximal
element of $S$ is selected in the first step, and (ii) no element of
$C_n\less{}S$  is selected  in the first step.)  
In particular, we see that the
``gregarious transition'' will occur with probability $q_n=q^n$, where
$q=1-p$. 

% \begin{figure}[htbp]
% \center
% \scalebox{.9}{\includegraphics{perc_gc.eps}}
% \caption{Demonstration that transitive percolation is generally covariant}
% \label{perc_gc}
% \end{figure}
% To see that transitive percolation obeys general covariance, refer to
% figure \ref{perc_gc}.  The ellipse at the bottom represents an
% arbitrary causet, with two of its children shown.  One child arises
% from a precursor set described by $m_1$ and $\pp_1$ (precursor set
% ($m_1$,$\pp_1$)), and the other child arises from a precursor set
% ($m_2$,$\pp_2$).  Those two children have a common child, obtained by
% adding the `other' element.  General covariance demands that the
% product of the probabilities along one path should equal the product
% along the other path, i.e. 
% \be 
% p^{m_1}q^{n-\pp_1}
% p^{m_2}q^{n+1-\pp_2} = p^{m_2}q^{n-\pp_2} p^{m_1}q^{n+1-\pp_1} . \qed
% \ee 
Now consider our four conditions.  Internal temporality was built in
from the outset, as we know.  Discrete general covariance is seen to
hold upon writing the net probability of a given $C_n$ explicitly in
terms of causet invariants (writing it in ``manifestly covariant
form'', c.f. the discussion in \S \ref{class_measure}) as
% Perhaps this should be explained more?  At least a reference to
% covariant forms of writing dynamics below!
% [[Is this clear enough?  Maybe worry more...  Maybe ask Rafael?  I
% guess he didn't like my demonstration above?  -- He probably thought
% this was more concise and thus clearer:]]
\be
   \Pr(C_n) = W(C_n) \, p^L \, q^{{n\choose 2}-R}
\ee
where $L$ is the number of links in $C_n$, $R$ the number of
relations, and $W$ the number of (natural) labelings of $C_n$.  (To
see how this arises, note that each of the links in the causal set
have to be ``put in by hand'', i.e. a link will never enter during the
transitive closure stage of the algorithm, so the $L$ links occur with probability $p^L$.
Furthermore, each of the ${n\choose 2}-R$ non-relations of the causet
must have not been ``selected'' during pre-transitive closure stage of
the algorithm, each of which occurs with probability $q=1-p$.
Finally, the transitive percolation algorithm generates labeled
posets, so any given unlabeled causet $C_n$ can arise in $W(C_n)$
different ways.)

To see that transitive percolation obeys Bell causality, consider an
arbitrary parent causet.  The transition probability to a given child
is 
exhibited
in eq. (\ref{pp}).  Consider two different children, one with
$(m,\pp)$=($m_1$,$\pp_1$) and the other with
$(m,\pp)$=($m_2$,$\pp_2$).  
Bell causality requires that the ratio of their
transition probabilities be the same as if the parent were 
reduced to the
union of the precursor sets of the two transitions, i.e.
it requires
\be
\frac{p^{m_1} q^{n-\pp_1}}{p^{m_2} q^{n-\pp_2}} = 
	\frac{p^{m_1} q^{n'-\pp_1}}{p^{m_2} q^{n'-\pp_2}}
\ee
where $n'$ is the cardinality of the union of the precursor sets of the two
transitions.  
Thus, Bell causality is satisfied by inspection.

Finally, the Markov sum rule is essentially trivial.  At each stage of
the growth process, a preliminary choice of ancestors is made by a
well-defined probabilistic procedure, and each such choice is mapped
uniquely onto a choice of partial stem.  Thus the induced probabilities
of the partial stems sum automatically to unity.
% For transitive percolation, a transition from an $n$ element causet to
% an $n+1$ element causet is made by choosing which elements will be to
% the past of the new element (each with probability $p$).  This
% enumerates the children, with the required weighting.  If there are
% multiple possibilities for the (labeled) precursor set of a
% transition, but they are all equivalent causally, then each of these
% labeled precursor sets will be `considered' once for the single
% (unlabeled) transition.  Thus transitive percolation obeys the sum
% rule.

\subsubsection{The general transition probability}

In the previous section we have shown that transitive percolation
produces transition probabilities 
(\ref{pp})
consistent with all our conditions.
By equating the right hand side of (\ref{pp}) to the general form
(\ref{FN}) of Lemma \ref{integer_form}, we can solve for the
$\xi_i$ and thus obtain the general solution of our conditions:
\be
\alpha_n = \sum_{i=0}^n{\xi_i \frac{1}{q_i}} \, q_n
	= p^m (1-p)^{n-\pp} 
	= (1-q)^m q^{n-\pp}
\ee
Expanding the factor $(1-q)^m$, and using the fact that $q_n=q^n$ for
transitive percolation, we get
\be
\xi_i = (-)^{\pp-i} {m \choose \pp-i} .
\ee
So an arbitrary transition probability in the general dynamics is,
according to (\ref{FN})
\be
\alpha_n = \sum_{i=0}^n (-)^{\pp-i} {m \choose \pp-i} \frac{q_n}{q_i} .
\ee
Noting that the binomial coefficients are zero for $\pp-i \notin
\{0..m\}$, and rearranging the indices, 
we obtain
\bne
\label{gen_trans_prob}
\fbox{
$\displaystyle \alpha_n = 
	\sum_{k=0}^m (-)^k {m \choose k} \frac{q_n}{q_{\pp-k}}$ .}
\ene
This form for the transition probability exhibits its causal nature
particularly clearly: except for the overall normalization factor $q_n$,
$\alpha_n$ depends only on invariants of the associated precursor set.

% Should we come up with a better notation for general transition
% probabilities than $\alpha_n$?  Maybe something like $\alpha_{n\pp
% m}$?  $\alpha_{n,\pp,m}$? $\alpha_{\C_{parent}\C_{precursor}}$?
%% RDS we could, but I'm too lazy

\subsection{Inequalities}

Since the $\alpha_n$ are classical probabilities, each must lie between
0 and 1, and this in turn restricts the possible values of the $q_n$.
Here we show that it suffices to impose only one inequality per stage;
all the others (two per child) then follow.  More precisely, what we
show is that, if $q_n>0$ for all $n$, and if $\alpha_n\ge{0}$ for the
``timid'' transition from the $n$-antichain, then all the 
$\alpha_n$ lie in $[0,1]$.  This we establish in the following two
``Claims''.
\\[3mm]
\noindent \textbf{Claim} \,
\emph{In order that all the transition probabilities $\alpha_n$ fall between 0
and 1, it suffices that each timid transition probability be $\ge 0$.}
\\[3mm]
\noindent \textbf{Proof:}  
As described in the proofs of lemmas \ref{1param_family} and \ref{bc_consis},
% put lemma 3 here?
each bold transition (of stage $n$) is given (via Bell causality) by
\be
\alpha_n = \alpha_m \frac{q_n}{q_m}
\ee
where $m$ is some natural number less than $n$.  The $q$'s are
positive.  So if the probabilities of the previous stages are
positive, then the bold probabilities of stage $n$ are also positive.
It follows by induction that all but the timid transition probabilities
are positive (since $\alpha_0=q_0=1$ obviously is).  But for the timid
transition of each family, we have
\bne
  \gamma_n = 1 - \sum_i \beta_i
  \label{GF}
\ene
where each $\beta_i$ is positive.  If any of the $\beta_i$ is greater
than one, $\gamma_n$ will obviously be negative.  Also (\ref{GF})
plainly cannot be greater than one.  Consequently, if we
require that $\gamma_n$ be positive, then all transition probabilities
in the family will be in $[0,1]$.  $\Box$

In a timid transition, the entire parent is the precursor set, so $\pp=n$.
The inequalities constraining each probability 
of a given family
to be in $[0,1]$ therefore
reduce to the sole condition 
\bne
   \sum_{k=0}^m (-)^k {m \choose k} \frac{1}{q_{n-k}} \geq 0 \,.
\label{tnm}
\ene

\vspace{2mm}
\noindent \textbf{Claim} \,
\emph{The most restrictive inequality of stage $n$ is the one arising from the
$n$-antichain, i.e. the one for which $m=n$.  All other inequalities 
of stage $n$ follow from this inequality and the inequalities for
smaller $n$.}
\\[2mm]

\noindent \textbf{Proof:}
Assume that we have, for $m=n$, 
\be
\sum_{k=0}^n (-)^k {n \choose k} \frac{1}{q_{n-k}} \geq 0 .
\ee
Add to this the inequality from stage $n-1$,
\be
\sum_{k=0}^{n-1} (-)^k {n-1 \choose k} \frac{1}{q_{n-k-1}} =
\sum_{k=0}^{n} (-)^{k-1} {n-1 \choose k-1} \frac{1}{q_{n-k}} \geq 0
\ee
to get
\be
\sum_{k=0}^{n-1} (-)^k {n-1 \choose k} \frac{1}{q_{n-k}} \geq 0 .
\ee
This is the inequality of stage $n$ for $m=n-1$.  (We have used the
identity ${n \choose k} = {n-1 \choose k} + {n-1 \choose k-1}$.)
Adding to it the inequality of stage $n-1$ with $m=n-2$ yields the
inequality of stage $n$ for $m=n-2$.  Repeating this process will give
all the inequalities of stage $n$. $\qed$

It is helpful to introduce the quantities
\bne
\fbox{ $ \displaystyle
  t_n = \sum_{k=0}^n (-)^{n-k} {n \choose k} \frac{1}{q_k}  $} 
\label{TD}
\ene
Obviously, we have $t_0=1$ (since $q_0=1$), and we have seen that the
full set of inequalities restricting the $q_n$ will be satisfied iff
$t_n\ge0$ for all $n$.  (Recall we are assuming $q_n>0, \, \forall n$.)
Moreover, given the $t_n$, we can recover the $q_n$ by inverting
(\ref{TD}): 
\begin{lemma}
\bne
\label{q_of_t}
  \fbox{ $ \displaystyle
  \frac{1}{q_n} = \sum_{k=0}^n {n \choose k} t_k$}
\ene
\end{lemma}
\noindent \textbf{Proof:}
This follows immediately from the identity
\be
  \sum\limits_{k=0}^n {n \choose k} (-)^{n-k} {k \choose m} = \delta^n_m \:,
\ee
which itself follows from Equations (27) and (28) on the top of pg. 37
of \cite{stanley}.  $\qed$\\
Thus, the $t_n$ may be treated as free parameters (subject only to
$t_n\ge{}0$ and $t_0=1$), and the $q_n$ can then be derived from 
(\ref{q_of_t}).
If this is done, 
the remaining transition probabilities $\alpha_n$ can be
re-expressed more simply in terms of the $t_n$ by inserting
(\ref{q_of_t}) into (\ref{gen_trans_prob}) to get
\be
\frac{\alpha_n}{q_n}  
     =  \sum_{l} t_l \sum_{k} (-)^k {m \choose k} {\pp-k \choose l}
     =  \sum_{l} t_l {\pp-m \choose \pp-l}
\ee
whence
\bne
%\fbox{ 
% $ \displaystyle
  \alpha_n = \frac
             {\sum_{l=m}^{\pp} {\pp-m \choose \pp-l} t_l}
	     {\sum_{j=0}^n {n \choose j} t_j} \:. \label{aot}
%  $} 
\ene
Here, we have used an identity for binomial coefficients
that can be found on page 63 of \cite{feller63}.
To express (\ref{aot}) more concisely, define 
\be
\fbox{
 $ \displaystyle
\lambda(\pp,m)=\sum_{l=m}^{\pp} {\pp-m \choose \pp-l} t_l %\:.
 $}
\ee
Then $q_n=\lambda(n,0)^{-1}$ and %we have
%so that %it can be written as
\bne
\fbox{ 
 $ \displaystyle
  \alpha_n = \frac{\lambda(\pp,m)}{\lambda(n,0)}
  $}  \label{alpha_of_t}
\ene

In this way, we arrive at the general solution of our inequalities.
(Actually, we go slightly beyond our ``genericity'' assumption that
$\alpha_n\not=0$ if we allow some of the $t_n$ to vanish; but no harm is
done thereby.)

% [[ something funny is happening here.  See notes from discussion with
% Joohan on my paper copy of the paper. Unfortunately I left it in the
% relativity lounge, so let's defer this issue 'till later. ]]

Let us conclude this section by noting that (\ref{q_of_t}) implies
\bne
  q_0 \equiv 1 \geq q_1 \geq q_2 \geq q_3 \geq \cdots
 \label{qin}
\ene
If we think of the $q_n$ as the basic parameters or ``coupling
constants'' of our sequential growth dynamics, then it is as if the
universe had a free choice of one parameter at each stage of the
process.  
% Say this differently?  -- why?
We thus get an ``evolving dynamical law'', but the evolution
is not absolutely free, since the allowable values of $q_n$ at every
stage are limited by the choices already made.  On the other hand, if
we think of the $t_n$ as the basic parameters, then the free choice is
unencumbered at each stage.  However, unlike the $q_n$, the $t_n$
cannot be identified with any dynamical transition probability.
Rather, they can be realized as ratios of two such probabilities,
namely as the ratio $x_n/q_n$, where $x_n$ is the transition
probability from an antichain of $n$ elements to the timid child of
that antichain.  (Thus, if we suppose that the evolving causet at the
beginning of stage $n$ is an antichain, then $t_n$ is the probability
that the next element will be born to the future of {\it every}
element, divided by the probability that the next element will be born
to the future of {\it no} element.)

\subsection{Proof that this dynamics obeys the physical requirements}
\label{obeys}

To complete our derivation, we must show that the sequential growth
dynamics given by (\ref{gen_trans_prob}) or (\ref{alpha_of_t}) obeys the
four conditions set out in Section \ref{requirements}.

\subsubsection{Internal temporality}
This condition is built into our definition of the growth process.

\subsubsection{Discrete general covariance}
We have to show that the product of the transition probabilities
$\alpha_n$ associated with a labeling of a fixed finite causet $C$ is
independent of the labeling.  But this follows immediately from 
(\ref{gen_trans_prob}) [or (\ref{alpha_of_t})]
once we notice that what remains after the overall product
$$
  \prod\limits_{j=0}^{|C|-1}q_j
$$
is factored out, is a product over all elements $x\in{C}$ of poset
invariants depending only on the structure of $\past(x)$.

\subsubsection{Bell causality}
Bell causality states that the ratio of the transition probabilities
for two siblings depends only on the union of their precursors.
Looking at (\ref{gen_trans_prob}), consider the ratio of two
such probabilities
$\alpha_{n1}$ and $\alpha_{n2}$.  The $q_n$ factors will cancel,
leading to an expression which depends only on $\pp_1$, $\pp_2$, $m_1$,
and $m_2$.  Since these are all determined by the structure of the
precursor sets, Bell causality is satisfied.

\subsubsection{Markov sum rule}
The sum rule states that sum of all transition probabilities
$\alpha_n$ from a given parent $C$ (of cardinality $|C|=n$) is unity.  
Since a child can
be identified with a partial stem of the parent,
we can write this condition, in view of (\ref{alpha_of_t}), as
\bne
\sum_S \sum_l t_l {|S|-m(S) \choose l-m(S)} = \sum_j t_j {n \choose j}
\label{Msr}
\ene
where $S$ ranges over the partial stems of $C$.
This must hold for any $t_l$, since they may be chosen
freely.  Reordering the sums and equating like terms yields
\bne
   \forall l, \ \sum_S {|S|-m(S) \choose l-m(S)} = {n \choose l} \,,
\label{identities}
\ene
an infinite set of identities which must hold if the sum rule
is to be satisfied by our dynamics.

The simplest way to see that (\ref{identities}) is true is to resort
to transitive percolation, for which $t_l=t^l$ where $t=p/q=p/(1-p)$
(c.f. (\ref{ttranperc}) below).
%(Set $q_n=q^n$ in
%(\ref{q_of_t}) and solve for $t$.)
In that case we know that the sum rule is satisfied, so by inspection of
(\ref{Msr}), we see that the identity (\ref{identities}) must be true.

A more intuitive proof is illustrated well by
the case of $l=3$.  Group the terms on the left side 
according to the number of maximal elements:
\def\clt#1#2{#1\,|\,#2}
\be
\begin{array}{ccccccccr}
\sum\limits_{\clt{S}{m(S)=0}} {|S|-0 \choose 3-0} & + &
\sum\limits_{\clt{S}{m(S)=1}} {|S|-1 \choose 3-1} & + & 
\sum\limits_{\clt{S}{m(S)=2}} {|S|-2 \choose 3-2} & + & 
\sum\limits_{\clt{S}{m(S)=3}} {|S|-3 \choose 3-3} & = & {n \choose 3}\\
0 & + & 
\sum\limits_{\clt{S}{m(S)=1}} {|S|-1 \choose 2} & + &
\sum\limits_{\clt{S}{m(S)=2}} (|S|-2) & + &
\sum\limits_{\clt{S}{m(S)=3}} 1 & = & {n \choose 3}
\end{array}
\ee
The first term is zero because the only partial stem with zero maximal
elements is empty (i.e. $|S|=0$).  
The second term is a sum over all partial stems with one maximal
element.  This is equivalent to a sum over elements, with the element's
inclusive past forming the partial stem.  The summand chooses every
possible pair of elements to the past of the maximal element.  Thus the
second term overall counts the 3-element subcausets of $C$ with a
single maximal element.  There are two possibilities
here, the three-chain \threech and the ``lambda'' \wedge.  
The third term sums over partial stems with two maximal elements, 
which is equivalent to summing over 2 element antichains, 
the inclusive past of the antichain being the partial stem.  
The summand then counts the
number of elements to the past of the two maximal ones.  Thus the third
term overall 
counts the number of three element subcausets with
precisely two maximal elements.  Again there are two possibilities, the
``V'' \V, and the ``L'', \Lcauset.
Finally, the fourth term is a sum over partial stems with three maximal
elements, and this can be interpreted as a sum over all three element
antichains \threeach.
As this example illustrates, then,
the left hand side of (\ref{identities}) counts the number of $l$ element
subcausets of $C$, placing them into ``bins'' according to the number
of maximal elements of the subcauset.  
Adding together the bin sizes yields the total number of $l$ element
subsets of $C$, which of course equals ${n \choose l}$.

\subsection{Sample cosmologies}
\label{cosmologies}
% single out a few
%simple choices of the free parameters and exhibit some properties of
%the resulting ``cosmologies''; 

The physical consequences of differing choices of the $t_n$ remain to
be explored.  To get an initial feel for this question, we list some
simple examples.  (Recall our convention that $t_0=1$, or
equivalently, $q_0=1$, where $q_0$ is the probability that the
universe is born at all.)
%\footnote% too silly...
%Could say something like ``At the level of purely logical discussion,
%...''  Maybe this might suggest how silly logic becomes when
%attempting to address questions such as these...
%{So, is the answer to the old question why something exists rather than
% nothing, simply that it is notationally more convenient for it to be so?})
\begin{itemize}

\item{``Dust universe''}
\be
     t_0=1, \ t_i = 0, \; i \geq 1
\ee
This universe is simply an antichain, since, according to (\ref{q_of_t}), 
$q_n=1$ for all $n$.  An ``example'' is shown in Figure \ref{dust}.
\vspace{2mm}
\begin{figure}[htbp]
\center
\scalebox{.6}{\includegraphics{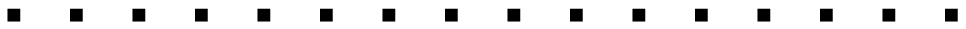}}
\caption{16 element dust universe}
\label{dust}
\end{figure}

\item{``Forest universe''}
\be
       t_0 = t_1 = 1; \; t_i = 0, \; i \geq 2
\ee
This yields a universe consisting wholly of trees,
since (see the next example) $t_2=t_3=t_4=\cdots=0$ implies that no
element of the causet can have more than one past link.  
The particular choice of $t_1=1$ has in addition the remarkable property
that, as follows easily from (\ref{alpha_of_t}), every allowed transition
of stage $n$ has the same probability $1/(n+1)$.  Figure
\ref{forest} displays an example.
\begin{figure}[htbp]
\center
\scalebox{1}{\includegraphics{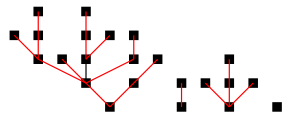}}
\caption{23 element forest universe}
\label{forest}
\end{figure}

\item{Case of limited number of past links}
\be
        t_i=0, \; i > n_0
\ee
Referring to expression 
(\ref{alpha_of_t})
one sees at once that $\alpha_n$ vanishes if $m>n_0$.  Hence, no element
can be born with more than $n_0$ past links or ``parents''.  This means
in particular that any realistic choice of parameters will have $t_n>0$
for all $n$, since an element of a causal set
faithfully embeddable in Minkowski space
would have an infinite number of past links.  See Figure \ref{sorest}
for an example.
\begin{figure}[htbp]
\center
\scalebox{.6}{\includegraphics{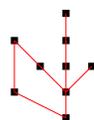}}
\caption{9 element universe with $t_1=t_2=1$}
\label{sorest}
\end{figure}

\item{Transitive percolation}
\bne
             t_n = t^n \label{ttranperc}
\ene
\begin{figure}[htbp]
\center
\scalebox{.45}{\includegraphics{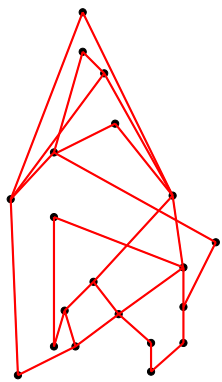}}
\caption{20 element transitive percolation universe with $p=1/4$}
\label{tranpercex}
\end{figure}
We have seen that for transitive percolation, $q_n=q^n$, where
$q=1-p$.  Using the binomial theorem, it is easy to learn from
(\ref{q_of_t}) or (\ref{TD}) that this choice of $q_n$ corresponds to
$t_n=t^n$ with $t=p/q$.  Clearly, $t$ runs from 0 to $\infty$ as $p$
runs from 0 to 1.  Figure \ref{tranpercex} shows an example.

\item{A more lifelike choice?}
\bne
        t_n = \frac{t^n}{n!}
\label{lifelike}
\ene
% We know that the transitive percolation model does not yield a
% physically realistic cosmology.  As $n \rightarrow \infty$ the
% universe tends toward a chain, it has `posts', etc.  So we would like
% to avoid this `regime'.  Plain perc demands that p=const, and p must
% be small for large n (to get a reasonable continuum limit).  Looking
% at (\ref{tp}), we see that $t \approx p$ for small values.  `To stay
% away from perc' we want $t_n$ to fall off faster than $1/t^n$.
% $1/n!$
% is a nice choice, faster than $e^{-cn}$, has potential for easier
% combinatorial interpretation...
%
% Unfortunately? simulation suggests that this also is not fast enough
% fall off to stay away from the chain regime.  Given that the number of
% causets grows like $e^{n^2}$, wouldn't we expect $t_n$ to fall off like
% this in order to stay away from the chains?  Reasoning is admittedly
% \emph{very} fuzzy here...  
%
% From studies of the scaling behavior of the transitive percolation model
% \cite{scaling}, we know that the continuum limit requires that $p$ fall
% off faster than ...
Due to its homogeneity (c.f. \S \ref{homogeneous}), we have seen that
transitive percolation with constant $p$ yields causets which could
reproduce --- at best --- only limited portions of Minkowski, de
Sitter, or anti-de Sitter\footnote
{\singlesp We have already seen that transitive percolation cannot
yield all of any homogeneous space, due to the presence of ``posts''.
It will fail to produce all of anti-de Sitter for yet another reason
--- that intervals of sufficiently large size do not have finite volume, so
that no locally finite order can embed faithfully into such a
spacetime.}
space.
It also suffers from a sort of ``scale dependent dimension'' which is
incompatible with any continuum spacetime.  This occurs because, at
finite $p$, transitive percolation generates causets with
approximately ``constant width'' ($\sim 1/p$), meaning that the Hasse
diagram looks roughly like an infinite cylinder with cross sectional
area of roughly $1/p$.
Thus, at larger and larger length scales, the transitive percolation universe looks more
and more like a one dimensional spacetime.
%, since the width is constant.  
Both of these issues suggest that 
%To do any better, 
one would have to scale $p$ so that it decreased with increasing $n$.
%\cite{contlim,scaling,AChR}.
This implies that 
$t_n$ should fall off faster than in any percolation
model, hence (by the last example) faster than exponentially in $n$.
Obviously, there are many possibilities of this sort
(e.g. $t_n\sim{}e^{-\alpha{}n^2}$), but one of the simplest is
$t_n\sim{c}/n!$.  
%Dou \cite{dou, more?} has studied the behavior of this model under the
%action of the cosmological renormalization group mentioned in \S
%\ref{cosrenormsec}.  He found that it is convenient to place a factor
%of $t^n$ in the numerator, which has no affect on the asymptotic
%behavior, whereby the
%%he found that this 
%dynamics is attracted by the renormalization flow to that of (originary)
%transitive percolation with the same value of $t$.% as in (\ref{lifelike}).
%%In some circumstances \cite{dou, more?} it is useful to
%%place an additional factor of $t^n$ in the numerator, where $t$ is a
%%positive constant, which has no affect on the asymptotic behavior.
This choice for the $t_n$ would be our candidate of the moment for a most
physically realistic choice of parameters.  
(The factor of $t^n$ in the numerator, which does not affect the
asymptotic behavior,  was added by Dou \cite{dou}, to
simplify the study of the cosmological renormalization behavior of
this dynamics.  
In fact, under the renormalization flow, this dynamics
approaches that of originary transitive percolation with a $t$
which approaches 0.
%which falls off as .)
A 20 element sample
causal set ``generated by'' this dynamics is shown in Figure \ref{20elts}.
\end{itemize}

\section{Originary dynamics}% and its generalizations}
%[its generalizations are quite silly; not physically important, so I
%don't want to give them implied significance by mentioning them in a
%section title]

It is possible that we can arrive at
%One probably does not obtain 
every possible solution of our conditions by taking limits of
the generic solution.
%and the special theories which result from
%taking certain transition probabilities to vanish must be treated
%separately.\footnote%
%%
%{\singlesp Indeed, the requirement of Bell causality itself must be given
% an unambiguous interpretation when some of the transition probabilities
% involved are zero.}
%% But originary percolation seems to work fine.  What is happening here?
One such special theory is the \emph{originary percolation} model,
which was introduced in \S \ref{originarysec}.  
It is the same as the transitive percolation model, but with the added
restriction that each element except the original one must have at
least one ancestor among the previous elements.  Algorithmically, we
generate potential elements one by one, exactly as for plain
percolation (by ``joining'' to each existing element with probability
$p$, then adding relations required by transitivity) but discard any
such element which would be unrelated to all previous elements.
Causets formed with this algorithm always have a single minimal
element, an ``origin''.
The
transition probabilities for originary percolation are just those of
ordinary transitive percolation with an added factor of $(1-q^n)$ in
the denominator at stage $n$.

This can be generalized to a non-percolation dynamics.  Here the
causal set grows as usual for a general dynamics, but with the added
restriction that it must always possess a 1-chain as a full stem.  We
call this an \emph{originary dynamics}.
The poset of
originary causets can be transformed into the poset of all causets
(exactly) by removing the origin from every originary causet.  

Further generalizations are also possible, in which a
more complex stem of the causet is enforced,
e.g. the restriction that after the first two elements form, the
causet must always have a 2-chain as a full stem (or partial
stem).
However, not every poset can be used as a stem in this manner, for
many choices are incompatible with Bell causality.

%: limits of dynamics
%In addition to this generic family, there
%are some exceptional families of solutions, but 
We conjecture that each of these exceptional families of solutions
are singular limits of the generic family. %  We have checked
For example, originary percolation is the $A\to\infty$ limit of
the dynamics given by $t_0=1$, $t_n=At^n$, $n=1,2,3,\ldots$
% [ Explain this in detail?  At the next run through...  I don't feel
% like it -- it is an exercise for the reader...]
% [[ When I get a reply from Joohan I may be able to improve this paragraph.]]

\section{The stochastic growth process as such}
\label{class_measure}

% Thus far we have spoken of the dynamics in terms of transition
% probabilities from $\C_n$ to $\C_{n+1}$.  A more physical question is
% the probability of forming a particular causet, ``starting from
% nothing''. 
We have seen that, associated with every {\it labeled} causet
$\hat{C}$ of size $N$, is a net ``probability of formation''
$\Pr(\hat{C})$ which is the product of the transition probabilities
$\alpha_i$ of the individual births described by the labeling:
\bne
   \Pr(\hat{C}) = \prod\limits_{i=0}^{N-1} \alpha_i \label{prodalpha}
\ene
where $\alpha_i=\alpha(i,\pp_i,m_i)$ is given by
(\ref{gen_trans_prob}) or (\ref{alpha_of_t}), $\pp_i$ and $m_i$ are
respectively the size and number of maximal elements in the precursor
($\equiv$ past) of the element labeled $i$.  Using (\ref{alpha_of_t}),
we can write this more explicitly as
\bne
\Pr(\hat{C}) = \frac{\prod_{i=0}^{N-1} \lambda(\pp_i,m_i)}
                {\prod_{j=0}^{N-1} \lambda(j,0)} \:.
\label{PolC}
\ene
We have also
seen that $\Pr(\hat{C})$ is in fact independent of the labeling,
i.e. $\Pr(\hat{C})=\Pr(\breve{C})$ where $\breve{C}$ is the same
causet as $\hat{C}$, but with a different labeling.
The net probability of arriving at an \emph{unlabeled} causet $C$, at
stage $N$ of the growth process is
%not $P(C)$ but
\bne
  \Pr_N(C) = W(C) \, \Pr(\hat{C}) \label{prnc0}
\ene
where $\hat{C}$ is the causet $C$ endowed with any (natural) labeling,
$N=|C|$, and $W(C)$ is the number of inequivalent\footnote%
{Two labelings of $C$ are equivalent iff they are related by an
automorphism of $C$.}
labelings of $C$,
or in other words, the total number of paths through
$\poscau$ that arrive at $C$, each link being taken with its proper
multiplicity. 
%[ Is this sentence to long, and thus too confusing? ]
Expressing (\ref{PolC}) more intrinsically, we can write (\ref{prnc0}) as
\bne
  \Pr_N(C) = W(C) \: {
                \prod\limits_{x\in C} \lambda(\pp(x),m(x)) 
                \over
                \prod\limits_{j=0}^{|C|-1} \lambda(j,0) } \,,
\label{PC0}
\ene
where $\varpi(x)=|\past{x}|$ and $m(x)=|{\rm maximal}(\past x)|$.
This expression, as far as it goes, is manifestly %``causal'' and
``covariant'' in the sense explained above.  Also causality is
manifest in the sense that it is expressed as a product of factors,
one for each element (save the $W(C)$), each of which depends only on that element's past.
% [[ Might it be hard to see ``manifest causality'' in the presence of
% W(C)? ]]
However, as explained in
\S \ref{gen_cov} and \S \ref{gencov}, it has no direct physical
meaning.  
Here we briefly discuss some probabilities which {\it do}
have a fully covariant meaning and show how, in simple cases, they are
related to $N\to\infty$ limits of probabilities like (\ref{PC0}).

As a rudimentary example of a truly covariant question, let us take 
``Does the two-chain ever occur as a partial stem of $C$?''.
The answer to this question will be a probability, $P$, which 
it is natural to identify as
\bne
         P = \lim_{N\to\infty} \Pr_N (X_N)  \,,
\label{measure}
\ene
where $X_N$ is the event that ``at stage $N$'', $C$ possesses a partial
stem which is a two-chain.
% I think the following is wrong in the sense that a element of some
% field precedes each term on the right, but I'm not sure what field
% to use.  Actually no, I think that it is o.k.  A sigma algebra seems
% to be a collection of subsets of a set.  Is it really an algebra?
To state this more precisely, define $X_N$ to be the set of $N$-orders
which satisfy some criterion, e.g. that the order possess the causet
$S$ as a partial stem ($S$ was a two-chain in the above example), and
%generalize %(\ref{PC0}) 
define %$\Pr_N (X_N)$ to mean
\bne
\Pr_N (X_N) = \sum_{C\in X_N} \Pr_N(C) \:.
\label{classical_measure}
\ene
%which is simply the statement that this is a probability measure.
% [[ Can I somehow ressurect a comment like this? ]]
%the measure defined in (\ref{PC0})
%satisfies a Kolmogorov sum rule, 
Of course it is not %obviously
guaranteed that $P$ will be well defined in the $N\to\infty$ limit.
%Does P converge in limit \sim is measure well defined, did we choose
%the questions well.
%
% Could include some portion of my discussion with Rafael about this
% -- look at my notes to see if there is anything.  (discussion about
% proving completeness, which we had in the Noble room of the chapel
% (there was a foul smell of solvent or something in the department))
% Naah.  too much trouble
In this connection, 
we conjecture that the questions ``Does $S$ occur
as a partial stem of $C$?'' furnish a physically complete set, when
$S$ ranges over all (isomorphism equivalence classes of) finite
causets.
As a simple example of an answer to a question of this form, consider
``What is the probability $P$ that a 1-chain is a full stem of the
universe $C$?''.  Clearly this is equivalent to the question ``Does
$C$ have a unique minimal element?''.
% or ``Is it originary?''.   [Slight confusion may result --
% originary dynamics arises from putting this product in the
% denominator at the outset.  But there is some subtlety here about how
% do do this, so I will not mention this here...]
In
terms of partial stems, this question is equivalent to demanding that
the 2-antichain not be a partial stem of $C$.  The answer is simple to
formulate by thinking in terms of the growth process, as follows.  At
stage 0 of the process, $C$ is a 1-chain, which occurs with
probability 1.  At stage 1, $C$ must not become a 2-antichain, which
occurs with probability $1-q_1$.  At stage 2, the new element must not
be born with no ancestors, which occurs with probability $1-q_2$.  At
stage 3, the same condition occurs with probability $1-q_3$, and so
on.  Thus in the limit $N\to\infty$, the answer becomes
\bne
P = \prod_{i=1}^\infty (1-q_i) \;. \label{originary}
\ene
Expressed in terms of the $t_n$, the $q_i$ in (\ref{originary}) are
simply replaced with $1/\sum_{k=0}^i {i \choose k} t_k$.
% This is the denominator which is added to generate originary
% dynamics.  Should I state this.  Now I'll say no, because of the slight
% uncertainty. 

\section{Two Ising-like state-models}
\label{Ising}
\def\R{\mathcal{R}}
% \L is defined in LaTeX itself.  Can redefine.

%exhibit a pair of state-models for the dynamics that illustrate how
%not only geometry, but other matter might arise implicitly from
%order.
%
%In this section, we present two Ising-like state-models from which
%$P(C)$ of equation (\ref{PC0}) can be obtained.  
The dynamics, as written in (\ref{prodalpha}) (say) can be expressed
in terms of either of a pair of Ising-like state models.  
%The models involve $\Z_2$-valued ``spin'' variables living on the
%relations of the causal set. 
% [ Skipped because mentioned again below.  O.k.?]
To derive
one such model, consider $\alpha_n$ as given in
(\ref{gen_trans_prob}).  Inserting this form into (\ref{prodalpha})
(and discarding the labeling decoration on $C$)
\bne
\Pr(C) = \prod_{j=0}^{N-1} q_j \sum_{k=0}^{m_j} (-)^k {m_j \choose k}
\frac{1}{q_{\pp_j-k}} \:,
\label{I1}
\ene
where we have placed an index $j$ on $\pp$ and $m$ to indicate that
they refer to the transition at stage $j$.  %clearly worded?
For a given $j$, the sum $\sum_{k=0}^{m_j} {m_j \choose k}$ can be
regarded as a sum over subsets of the $m_j$ maximal elements to the
past of element $j$ (where $k$ can be regarded as the cardinality of
each subset), or equivalently a sum over the $m_j$ links whose future
endpoint is %element 
$j$.  
This suggests an interpretation of this %sum 
as a sum over $\Z_2$
valued ``spin configurations'' on these links.  Before describing in
detail how this construction follows, it will be helpful % necessary 
to state a
number of definitions.
%as follows.

The following paragraph %definitions  constructions
 will %all 
refer to a given causal set $C$.
%, define the following 
First define $\R$ as the set of all relations in $C$
\be
\R = \{(x,y) \in C\times C \,|\, x\prec y \} \:.
\ee
$\R_j$ will denote the set of relations in $C$ whose future
endpoint is the element labeled $j$
\be
\R_j = \{ (x,x_j) \,|\, x\prec x_j \;\forall\, x \in C\} \:.
\ee
%It is convenient to think of 
%To each relation of the causal set $C$, assign a $\Z_2$ $(=\{0,1\}$
%valued ``spin''
Now define %a map
$\phi$ as a map which assigns a $\Z_2$ ($=\{0,1\}$) valued ``spin'' to each relation of the causal set $C$,
i.e.
\be
\phi : \R \to \Z_2 \:.
\ee
Let the set of all such maps, for the given causal set $C$, be denoted by
$\Phi$.  Furthermore, define a restriction $\phi_j$ of a map $\phi$ to an
element $j$
%, $\phi_j$, 
by restricting the domain to %be 
only those
relations which have $j$ as a future endpoint, i.e.
\be
\phi_j : \R_j \to \Z_2 \:.
\ee
We denote the set of such
restricted maps by $\Phi_j$.
We also need to define two quantities associated with a map $\phi$,
$a(\phi)$ and $r(\phi)$.  The former counts the number of ``absent
spins'' in $\phi$, i.e. it represents the number of relations which
map to 0
\be
a(\phi) = |\{ x\in\R \,|\, \phi(x)=0 \}| \:.
\ee
The latter counts the number of ``present spins'' in $\phi$, i.e.
\be
r(\phi) = |\{ x\in\R \,|\, \phi(x)=1 \}| \:.
\ee
It is to be understood that a restricted map $\phi_j$ can be used in the
place of $\phi$, with $\R$ replaced by $\R_j$ in the above two
definitions.

In order to express (\ref{I1}) using these definitions, it is
necessary to place a %further (?)
restriction on all maps $\phi$ introduced
in the previous paragraph, namely
that each relation which is not a link (i.e. pairs $(x,y)$ such that
$\interval[x,y]\neq\emptyset$) map to 1.  Such maps (and sets of such maps)
will be decorated with a ``$\,\tilde{~}\,$''. %$\tilde{\ }\,$.
%To remind the reader of this
%condition on the $\phi$, each is (and sets of each are) decorated with
%a $\tilde{\ }$.
With these definitions in place, (\ref{I1}) can be written as
%Writing this sum as $\sum\limits_{\phi_j}$,
\bne
\Pr(C) = \left(\prod_{j=0}^{N-1} q_j\right) \prod_{j=0}^{N-1}
\:\sum_{\tilde{\phi_j}\in \tilde{\Phi_j}} (-)^{a(\tilde{\phi_j})} 
\frac{1}{q_{r(\tilde{\phi_j})}} \:,
%\pp_j-a(\tilde{\phi_j})}} \:,
\label{I2}
\ene
%It is convenient to think of $\phi$ as an assignment of a $\Z_2$
%$(=\{0,1\}$ valued ``spin'' to each relation of the causal set $C$.  
where we have interpreted the index $k$ in (\ref{I1}) as counting the
number of ``zero spins'' on links ``pointing to'' $j$.  In writing
(\ref{I2}), we have noted that $\pp_j$, the number of elements to the
past of $j$, counts the number of relations in the domain of $\phi_j$.
Thus, after subtracting the ``zero spins'', the subscript in which it
appears becomes simply $r(\tilde{\phi_j})$.

% [ paragraph break here?]
Equation (\ref{I2}), save an initial coefficient $\left(\prod_{j=0}^{N-1}
q_j\right)$, is written as a product, one factor for each element $j$,
of sums of terms, one for each ``spin configuration'' at $j$.  (A
``spin configuration at $j$'' being an assignment of ones or zeros to
each past-directed link at $j$.)  Expanding, we arrive at a sum of
terms, each of which contains one factor from each element $j$, for
one choice of spin configuration at $j$.  The sum contains a term for
each possible spin configuration at each $j$.  Thus, after expanding,
(\ref{I2}) becomes
\be
\Pr(C) = \left(\prod_{j=0}^{N-1} q_j\right) 
\sum_{\tilde{\phi}\in \tilde{\Phi}} \prod_{j=0}^{N-1}
 (-)^{a(\tilde{\phi_j})} \frac{1}{q_{r(\tilde{\phi_j})}} \:,
\ee
% Do I need to explain better how \Phi_j --> \Phi ?
or, using a more covariant notation,
%Since the dynamics is independent of labeling, this can be written in
%a more covariant form as
\bne
%\fbox{$ ... squishes it! :( -- need to use /displaystyle
\Pr(C) = \left(\prod_{j=0}^{N-1} q_j\right) \,
\sum_{\tilde{\phi}\in \tilde{\Phi}} \, \prod_{x\in C}
 (-)^{a(\tilde{\phi_x})} \frac{1}{q_{r(\tilde{\phi_x})}} \:. %$ .}
\label{spinmod1}
\ene

Equation (\ref{spinmod1}) writes the probability of arriving at a
particular causal set $C$, at stage $N$ of the growth process, as a
sum over spin configurations on the causal set, where only the spins
on the links are permitted to vary.  For each such spin configuration,
each element $x$ of $C$ contributes a ``vertex factor'' of
$(-)^{a(\tilde{\phi_x})} \frac{1}{q_{r(\tilde{\phi_x})}}$.  If these
vertex factors are to be interpreted as Boltzman weights, then the
negative values for odd numbers of ``present'' past-links are
%admittedly
 a bit peculiar.

% An equivalent state model exists, in which the spins sit on the `non-links'
% (i.e. a relation that is not a link) rather than the links.  
A second model arises by inserting (\ref{aot}) into
(\ref{prodalpha}).
\bne
\Pr(C) = \prod_{j=0}^{N-1} \frac{\sum_{l=m_j}^{\pp_j} 
{\pp_j-m_j \choose \pp_j-l} t_l}{\sum_{k=0}^j {j \choose k} t_k}
\label{It0}
\ene
From (\ref{q_of_t}) the factors in the denominator are easily seen to
be simply the overall product $\left(\prod_{j=0}^{N-1} 1/q_j\right)$.
As before, the sum in the numerator, $\sum_{l=m_j}^{\pp_j} {\pp_j-m_j
\choose \pp_j-l} t_l$, can be interpreted as a sum over subsets of
relations.  This time, however, the sum is over subsets of relations
which are not links.  To express this in terms of spin configurations
$\phi$, constrain all such maps to yield 1 on links, but allow them to
vary freely on the non-link relations.  We decorate maps which respect
such a constraint with a ``$\,\hat{~}\,$''.  Then
\be
\Pr(C) = \left(\prod_{j=0}^{N-1} q_j\right) \prod_{j=0}^{N-1} 
\sum_{\widehat{\phi_j}\in\widehat{\Phi_j}} t_{r(\widehat{\phi_j})} \:,
\ee
where we have interpreted the index $j$ in (\ref{It0}) as the number
of ``present'' relations in each $\widehat{\phi_j}$.
% terminology is not quite correct here, bit I will probably get away
% with it.
Expanding this product, as before, leads to a sum of terms, one for
each ``spin configuration'' on the entire causal set $C$, of a product of factors, one for each
element.  Thus
\be
\Pr(C) = \left(\prod_{j=0}^{N-1} q_j\right) 
\sum_{\widehat{\phi}\in\widehat{\Phi}}
\prod_{j=0}^{N-1} t_{r(\widehat{\phi_j})} \:,
\ee
or, in a more covariant notation
\bne
\Pr(C) = \left(\prod_{j=0}^{N-1} q_j\right) 
\sum_{\widehat{\phi}\in\widehat{\Phi}}
\prod_{x\in C} t_{r(\widehat{\phi_x})} \:.
\label{spinmod2}
\ene

Equation (\ref{spinmod2}) represents a second way to express the
probability of arriving at a causet $C$ in terms of a model of
Ising-like spins on its relations.  Here the spins on the links are
fixed at 1, while those on the other relations are free to vary.  This
time all vertex factors are positive, in closer agreement with what
one would expect from physical Boltzmann weights.
% The last line expresses the measure/dynamics as an effective action
% for another $\Integers_2$ valued object living on the non-links of the
% causet.  i.e. we have a sum over every possible `field configuration
% of a yes/no on each non-link of the causet' of a 
%`Boltzmann factor'
% which is a product of $t$'s, one for each element, 
%  There are no longer any negative terms, however,
% as the $t_i$ are always positive, which one would expect for a
% Boltzmann factor.  The physical significance of these state model
% interpretations of the dynamics is still very much an open question.

These two models (and especially the second) show that the sequential
growth dynamics can be viewed as a form of ``induced gravity''
obtained by summing over (``integrating out'') the values of 
underlying spin variables $\sigma$.  This underlying ``matter'' theory
may or may not be physically reasonable (Does it obey its own version
of Bell causality, for example?), % Is it local in an appropriate sense?),
but at a minimum, it serves to 
illustrate how a theory of non-gravitational matter can be hidden
within a theory that one might think to be limited to gravity
alone.\footnote%
{\singlesp In this connection, it bears remembering that Ising matter can
 produce fermionic as well as bosonic fields, at least in certain
 circumstances. \cite{id89,ple97}}
\footnote%
{\singlesp References \cite{kazakov} and \cite{staud} 
%(for which we thank an anonymous referee) 
describe a similar example of ``hidden'' matter
fields in the context of 2-dimensional random surfaces (Euclidean
% Riemannian ?
signature quantum gravity) and the associated matrix models in the
continuum limit.  Unfortunately, the matter fields used (Ising spins
or ``hard dimers'') were unphysical in the sense that the partition
function was a sum of Boltzmann weights which were not in general real
and positive.  This is much like our first state model described
above.  To the extent that the analogy between these two, rather
different, situations holds good, our results here suggest that there
might be, in addition to the matter fields employed in \cite{staud},
another set of fields with physical choices of the coupling constants,
which could reproduce the same effective dynamics for the random
surface.}
It should be noted that these ``spin models'' are ``non-interacting''
in that each ``lattice site'' has its own ``reserved'' set of spins
which affect the value of only its vertex factor, with no two lattice sites
``sharing'' any spins.
%the value of vertex factor at each point in the lattice is
%completely independent of the values of 
In order for these spin models to give non-trivial results, an
effective interaction must emerge from the gravitational dynamics in
the sum over causal sets.
%(``lattice configurations'').

% Note that this model gives a recursive relation for the
% probabilities.  Given the probability of forming a causet with $l$
% links, the probability with $l+1$ links is just that with $l$ links
% plus $l$ additional terms for the $l$ new subsets of links which are
% introduced.
% [RDS: I no longer am able to understand this remark about a
% recursive relation (though I know it was in my notes!).  For now,
% I've delete it. ] 
%[ In fact the statement about one more link adds l new spin configs
% seems wrong : 1==>1 while 2==>4
% So maybe the whole statement is wrong.  Maybe ask Rafael one more time.]
%[[ I think the recursion is too complicated to be useful... ]]

% [For the forest cosmology this counts the
% number of spanning trees.  How does this work?  How is this useful?
% A forest has one spanning tree, anyway...]

\section{Further Work}

%\subsection{simulating dynamics}
The sequential growth dynamics %of chapter (\ref{dynamics}) 
can be
simulated directly on a computer, but only for very small $N$.  For
$t_n=1/n!$ it takes a minute or so to generate a 64 element causet on
a DEC Alpha 600 workstation.
%Results are inconclusive at the moment.  
%Can e.g. study dimension of
% resulting causets, to see if they resemble a manifold.  Could think
% about how to compare them with causets which arise from sprinkling
% into Minkowski space.  Should one think of this as a cosmological
% model, or as a microscopic description of local spacetime...?
Because the number of partial stems, and hence the number of possible
precursors for a new element, of a $N$ element causal set grows like
$2^N$, it is difficult to simulate the growth process directly.  A
workaround may involve using something like a metropolis algorithm at
each stage to select a precursor.  Issues such as detailed balance for
stepping through precursors would have to be sorted out.
% I could also mention my idea of pre-storing all values for lambda,
% but this won't be so useful as N gets large (?)
%
% I could explain better what I mean here, but I am too lazy right
% now.  Essentially just keep near the most probable ones, as
% discussed below, as Eric does, etc.
% [ See also Daughton thesis, for more discussion? ]
%\be
%\int e^{-S} d?
%\ee
%Move the $e^{-S}$ to domain of $\int$ --- metropolis algorithm.
%This makes the integration feasible.  Usually a random element of the
%integration set has negligible contribution to the integral since it
%has ``measure'' $e^{-S}$.  Thus just stay on peak of $e^{-S}$, these
%are the only points which contribute significantly to the integral.
%$\sim$ only configurations consistent with the dynamics.

Analytic results, so far, are available only for the special case of
transitive percolation.  
An important question, of course, is whether some choice
of the $t_n$ can reproduce general relativity, or at least reproduce a
Lorentzian manifold for some range of $t$'s and of $n=|C|$.

%\subsection{Ising matter}

% The two Ising-like state models give some indication of how matter may
% arise in a pure theory of causal sets.  This could be understood
% better.  Does this give some insight into how Rob calculates actions
% for matter fields on a causet?
Another question is whether the ``Ising matter'' introduced in \S
\ref{Ising} gives rise to an
interesting effective field theory, and what relation it has with the
local scalar matter on a background causal set studied in
\cite{daughton,salgado}.

%Also, as Rafael mentioned, I have been imagining that it may be natural to
%attach a positive integer to each element of the causet, representing a
%causal loop.  Then at each stage of the growth process there would be
%several additional transitions possible, corresponding to increasing the
%"degeneracy"  of one of the maximal elements (i.e. increasing its integer by
%one).  Thus a sort of scalar field could arise naturally from the discrete
%relation alone, without introducing some "external" mathematical structure. 
%(In this case, the poset would be replaced with a "preorder", a transitive,
%reflexive relation, to allow for closed causal loops, which is equivalent
%to a poset with a positive integer attached to each element.)
Another possibility for obtaining the behavior of a scalar field on a
causal set arises if we are willing to drop the acyclicity property of
the order (i.e. replace the transitive, irreflexive order with a
transitive, reflexive ``pseudo-order''), as mentioned in \S \ref{ctcs}.  
This relaxation allows the
possibility for causal cycles to exist, but with the property
that each element of a cycle has the same causal relations with
the remainder of the pseudo-order as every other point in the cycle.
Thus at the level of the pseudo-order, these cycles hold
no more information than that of one element in an ordinary partial
order, along with a positive integer representing the ``degeneracy''
of the cycle.  This indicates that a pseudo-order is %essentially
equivalent to an order with a positive integer attached to each
element, where each integer represents the size of a cycle which
exists at that point. % in the order.
A generalization of our dynamics to allow such a
possibility may be achieved simply by allowing, for a causal set with
$m$ maximal elements at some stage of the growth process, $m$
additional transitions,
%at each stage of the growth process, 
corresponding to incrementing the
integer attached to one of the maximal elements by 1.
%Actually it
%would be rather easy to implement, as $m$ and $\pp$ would carry over in an
%obvious way.  (I think.)

%\subsection{covariant formulations}

% Can we re-express all this in terms of `gauge invariant' properties
% of the measure?  By gauge we refer to the labeling.  This is another
% avenue for investigation.  (More could be said about how one might do this.)
% One issue, here, and also when considering the quantum theory, is how
% to reformulate Bell causality.  It would be helpful to have an
% expression of causality which is based on gauge invariant properties
% of the causet, and which would be what we wish the quantum theory to
% satisfy.
Another set of questions concerns the possibility of a more ``manifestly
covariant'' formulation of our sequential growth dynamics -- or of more
general forms of causal set dynamics.  Can Bell causality be formulated
in a gauge invariant manner, without reference to a choice of birth
sequence?  Is our conjecture correct that all meaningful assertions are
logical combinations of assertions about the occurrence of partial
stems/past sets?  

%\subsection{non-generic dynamics, indistinguishable dynamics}

Also, there are the special cases we left unstudied, for example the
originary dynamics and its generalizations.
Are there other
special, non-generic cases of interest?  In addition, it may be
interesting to see what sort of dynamics may arise from omitting the
combinatorial coefficients in the Markov sum rule, which corresponds
to imparting an indistinguishability condition on the individual
causal set elements.  Such a condition is perhaps not unreasonable
physically. 

%We might continue multiplying questions, but let's finish with the
%question of how to discover a quantum generalization of our dynamics.

\chapter{Conclusions} % and open questions}

\section{Summary}

The Causal Set approach to quantum gravity selects a very sparse
framework for the ``substance'' of the theory, and seeks to
``recover'' most of existing physical theory in some appropriate
limit.  A disadvantage of this philosophy 
%toward constructing a physical theory 
is the difficulty it raises by abandoning many of the
``typical'' structures and methods that are familiar to us.  In a way
this works itself out as an advantage, however, in that it forces us
to think critically about many issues at the foundation of physical
theory.  In doing so, the structure which arises %from this approach 
is
robust in the sense that issues such as Lorentz invariance, background
independence, general covariance, discreteness, non-perturbative
formulation, Lorentzian signature, have been addressed at the outset,
rather than being put off until the theory is more developed.
% could refer to forks paper here
Also the clarity of the fundamental structure provides a fertile
ground for addressing philosophical issues %involving 
such as the nature of
quantum causality in a closed system.  % o.k.?
%such as the nature of
%causality respected by a quantum theory, formulations

Another unique aspect of Causal Sets is the theory's departure from some
conventional intuition in a number of respects.  Most obvious is the
rejection of the continuum as fundamental.  Another is the complete
departure from determinism --- even at the fundamental scale, when all
aspects of the ``state of the system'' are ``known'', the classical
limit of the theory is postulated to be stochastic in nature.  (As
opposed to the philosophical attitude of kinetic theory, which
assumes that only the incomplete knowledge of the state leads to
indeterminism, c.f. \cite{openrose})
% describe in terms of Brownian motion
Unitarity will likely have to be abandoned to formulate the quantum
theory in a discrete setting. 
Qualitatively, the theory predicts a non-zero cosmological constant
%energy density for the vacuum 
\cite{forks,Ng_vanDam}.
%Manifest locality seems to be abandoned, in that there is no obvious, simple
%way to recover spatial distances from the discrete order.
Locality as a fundamental physical principle seems to be abandoned.
  The problem
here essentially extends from the fact that the links, which are the
analogs of nearest neighbors in a Euclidean signature lattice, extend
arbitrarily far into the past and to arbitrary spatially distant points
in some reference frame.  The work of \cite{daughton, salgado}
indicates how effective locality is preserved in the causal set, but,
when looking at the discrete order alone,
it arises in a complicated and non-intuitive manner.

% [ The following would be good, but I'll skip it for now to save time ]
%not quite as unusual, but...
%histories
%measure theoretic
%one true history philosophy
%concrete (what is the philosophical term for this?)

Before beginning this work, the Causal Set theory was at a stage where
a lot was known about kinematical issues, but there seemed to be many
obstacles to the construction of a dynamics for the theory.  Even if
some reasonable guess for the action was made, the issue of how to do
the sum over histories to compute the measure was difficult, because of
the sum over an enormous discrete sample space.
% Should I say less here?  too much copying of previous stuff?
% transfer some to that discussion (beginning of tranperc chapter)
This led to the search for an algorithm to sample the space of causal
sets.  The proposed algorithm, transitive percolation, failed to
perform the desired task, but did suggest itself as a toy dynamics.
%issues of monte carlo sum led us to toy dynamics
As a dynamical model, it has a number of promising features, and the
possibility that it reproduces a region of continuum spacetime has not
yet been excluded.
%but it fails to reproduce spacetime.  (It possesses ``posts'' (big
%bang/crunch events) but is completely homogeneous, which is
%inconsistent with any Einstein spacetime.)
%in the sense that, as $N\to\infty$, for constant $p$, 

%\footnote  %[copied from comment in dynamics' gen cov section]
%{This last condition guarantees that the ``parameter time'' of our
% stochastic process {\it is compatible with} physical temporality, as
% recorded in the order relation $\prec$ that gives the causal set its
% structure.  In a broader sense, general covariance itself is also an
% aspect of internal temporality, since it guarantees that the
% parameter time {\it adds nothing to} the relation $\prec$.}
%% Seems not so important...  Perhaps this should be in the bulk,
%% rather than the conclusion?
By thinking in terms of a stochastic growth process, and positing some
very basic principles,
%which should be appropriate for any theory of gravity, 
%These principles 
we were led almost uniquely to a family of dynamical
laws (stochastic processes) parameterized by a countable sequence of
coupling constants $q_n$ (or equivalently the $t_n$).
This result is quite encouraging in that we 
%Promising that 
now know how to speak of dynamics for a theory with discrete time.
% Are there other theories with discrete time?  e.g. Ambjorn?
%now in this previously
%ununderstood regime, for a theory with discrete time.
In addition, these results are encouraging in that there exists a
natural way to transform this theory, which is expressed in terms of a
classical probability measure, to a quantum measure.  A sketch as to
how one might proceed in doing this is provided below %\S \ref{quantum}.
%formulation of QM provides framework for expressing quantum theory
Thus there is good reason to expect that we are close to constructing
a background independent theory of quantum gravity.
% with rests on a
%firm philosophical foundation.
% very promising for hope of constructing a background independent
%theory of quantum
%gravity with rests on a firm philosophical foundation.
% [ Maybe above should be last sentence ? ]

\section{Quantum Dynamics}

%One must satisfy causality, general covariance, and unitarity
%(maybe).  Then we may have a well defined quantum measure.  Again
%there will be the convergence issues.  Since the growth process must
%be taken to infinity, Do the corresponding infinite sums converge?
%What free parameters will the theory have?

Since our theory is formulated as a type of Markov process, and since a
Markov process mathematically is a probability measure on a suitable
sample space, the natural quantum generalization would seem to be a
{\it quantum} measure \cite{qmeasure} (or equivalent ``decoherence
functional'') on the same sample space.
The question then would be whether
one could find appropriate quantum analogs of Bell causality and general
covariance formulated in terms of such a quantum measure.
If so, we could hope that, just as in the classical theory treated herein,
these two principles would lead us to a relatively unique quantum
causal set dynamics,\footnote
{See \cite{criscuolo} for a promising first step toward such a dynamics.}
or rather to a family of them among which a potential
quantum theory of gravity would be recognizable.
Let me sketch briefly how one might go about constructing this quantal
generalization.

The quantum dynamics for causal sets will be expressed in terms of a
quantum measure, which is a generalization of a classical probability
measure.  For a more detailed discussion, see
\cite{quantum_measure,qmeasure,salgado1}.
It is helpful to express the measure in terms of a
%, which can be thought of as a
decoherence functional $D(C', C'')$ %defined for 
which assigns a complex number to
pairs of histories
$C'$ and $C''$.  In the context of causal sets, a ``completed'' causal
set $C$ is regarded as a history.  (Recall that a completed causal
set is one which has infinite cardinality; it has ``run to completion''.)
% Did I state this in the introduction?  Yes.  Should I restate it?
% (I dunno, why not.)
More correctly, the decoherence functional will be defined for pairs
of \emph{sets} of histories, which has the following properties: (for
any (disjoint) sets of histories $R,S,T$)
\bnee
\hbox{positivity:} & D(S,S) \geq 0 \nonumber\\
\hbox{additivity in each argument:} & D(R\sqcup S,T) = D(R,T) + D(S,T)
\label{additivity}\\
%\hbox{non-interference of null set} & ?? \nonumber\\
\hbox{hermitian:} & D(R,S) = D(S,R)^* \label{hermitean}
\enee
($\sqcup$ indicates the union of disjoint sets.)
The quantum measure of a set $S$, $\mu(S)$ is given by the diagonal elements of
$D$, i.e. $\mu(S)=D(S,S)$.
%Note that, for sets of measure zero $\mu(S)=D(S,S)=0$,
%(\ref{additivity}) implies that $D(A,S)=0$ for any set $A$.  (To see
%this, set $T=R\sqcup S$ in (\ref{additivity}).)
Two additional properties are that for a set $S$ of measure zero
$\mu(S)=0$, $D(A,S)=0$ for any set $A$, and a quantum generalization
of (\ref{classical_measure})
\be
\mu(A\sqcup B\sqcup C) - \mu(A\sqcup B) - \mu(A\sqcup C) 
- \mu(B\sqcup C) + \mu(A) + \mu(B) + \mu(C) = 0 \:.
\ee
%it has the property $I_3=0$ (see \cite{quantum_measure}). 

In practice, since the expressions we know how to write down for
causal set ``amplitudes'' are written in terms of finite causal sets,
as mentioned earlier, we conjecture that the appropriate sets to
consider are cylinder sets, defined by specifying a labeled causet as
a partial stem, and including in that set all possible extensions of
that labeled stem into the future.  Thus with each pair of labeled
finite causal sets $C', C''$ we can associate a complex number
$D(C',C'')$, which may look like some quantum generalization of
(\ref{PolC}).  
%(Of course, to maintain general covariance, we will
%probably have to include all relabelings of the stem in the cylinder
%set as well.)
% mention that really we will deal with unions and intersections of
% cylinder sets? -- I have mentioned sigma algebra elsewhere.  Lets
% just live with that as being good enough for now.  Should I explain
% what one is somewhere?

To construct a dynamics for such an object, this language needs to be
re-expressed in terms of ``transition amplitudes'', so that the growth
process construction can be carried over to the quantum dynamics.
This may be accomplished by defining a transition amplitude
$T(C'',C\to C')$ by
\be
T(C'',C\to C') = \frac{D(C'',C')}{D(C'',C)} \:.
\ee
(Note that there is some danger here if we allow some of the $D$ to
vanish, which should not be uncommon in the quantum theory due to 
%the phenomenon of 
complete destructive interference, but we'll evade this
issue for now.  It is likely that this problem can be overcome using
limits.)% or some such thing.)

With this definition in place, to derive the quantum dynamics, we need
to generalize our physical conditions of \S \ref{requirements}.
Internal temporality should carry over directly, since it is %already
encoded into the definition of the growth process itself.  General
covariance retains its obvious meaning, that products of transition
amplitudes leading to a given finite causal set in the sequential
growth process 
be independent of labeling.  (This translates into the statement that
$D(A,B)$ is invariant under relabeling of its arguments.) % $A$ and $B$.)
The Markov sum rule
generalizes to additivity in each argument (\ref{additivity}).  For Bell
causality, a promising generalization is to demand that for one
argument $S$ fixed, $D(S,\cdot)$ obeys the classical Bell causality condition.
%explain better?
This version of quantum Bell causality seems to be obeyed by
non-relativistic quantum mechanics, assuming that there are no
correlations existing in the initial data. 
%, and perhaps for relativistic field
%theory as well.  [[I'm not too sure what I can say here...]]

Phrased in this manner, this definition for quantum causality makes 
the quantum theory look a lot like %two copies of 
``the classical measure, squared''.  However, the hermitian condition (\ref{hermitean}) provides a
constraint on the theory which may considerably limit this freedom.
%but in a manner which must satisfy (\ref{hermitean}).
The hope is that one can follow the general methodology used in
deriving $\Pr(C)$ to construct a quantum dynamical law for causal sets,
phrased in terms of the $D(C',C'')$.

\appendix
\chapter{Consistency of physical conditions}
%of the conditions used to define the sequential
%growth process}
% Consistence of Bell causality and discrete general covariance ?
\label{consistency}

Our analysis of the conditions of Bell causality et al. unfolded
in the form of several lemmas.  Here we present some similar lemmas
which strictly speaking are not needed in the present context, but which
further elucidate the relationships among our conditions.  We expect
these lemmas can be useful in any attempt to formulate generalizations
of our scheme, in particular quantal generalizations.

\begin{lemma}
The Bell causality equations are mutually consistent.
\label{bc_consis}
\end{lemma}

\begin{figure}[htbp]
\center
\scalebox{.9}{\includegraphics{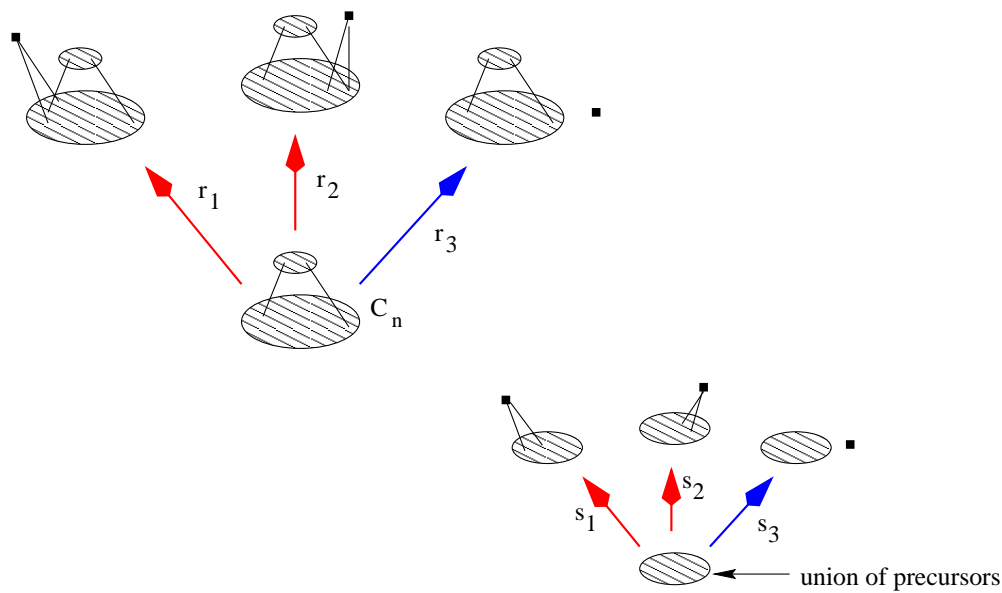}}
\caption{Two families related by Bell causality}
\label{BCconsis}
\end{figure}

\noindent \textbf{Proof:} 
The top of Figure \ref{BCconsis} shows three children of an arbitrary
causal set $C_n$.  The shaded ellipses represent  portions of
$C_n$.  The small square indicates the new element whose birth
transforms $C_n$ into a causal set $C_{n+1}$ of the next stage.  The
smaller ellipse ``stacked on top of'' the larger ellipse represents a
subcauset of $C_n$ which does not intersect the precursor set of any
of the transitions being considered (i.e. none of its elements lie to
the past of any of the new elements).  This small ellipse thus consists
entirely of ``spectators'' to the transitions under consideration.  The
bottom part of Figure \ref{BCconsis} shows the corresponding parent and
children when these spectators are removed.

Notice that one of the three children is the gregarious child.  We
will show that the Bell causality equations between
this child and each of the others imply all remaining Bell causality
equations within this family.  Since no Bell causality equation
reaches outside a single family (and since, within a family, the Bell
causality equations that involve the gregarious child obviously always
possess a solution --- in fact they determine all ratios of transition
probabilities except for that to the timid child), this will prove the
lemma.

In the figure $r_1$ and $r_2$ represent a general pair of
transitions related by a Bell causality equation, namely
\bne 
    \frac{r_1}{r_2} = \frac{s_1}{s_2} \,.
\label{BCE}
\ene 
But, as illustrated,
each of these is also related by a Bell causality equation to the
gregarious child, to wit:
\bne 
    \frac{r_1}{r_3} = \frac{s_1}{s_3} \quad \mbox{and} \quad 
    \frac{r_2}{r_3} = \frac{s_2}{s_3} 
\label{BCE2}
\ene
Since (\ref{BCE}) follows immediately from (\ref{BCE2}),
no inconsistencies can arise at stage $n$, and
the lemma follows by induction on $n$. $\Box$

\begin{lemma}
Given Bell causality and the further
consequences of general covariance that are
embodied in Lemma \ref{q's}, all the remaining general covariance
equations reduce to identities, i.e. they place no further restriction
on the parameters of the theory.
\end{lemma}

\noindent \textbf{Proof:}\\
\begin{figure}[htbp]
\center
\scalebox{.9}{\includegraphics{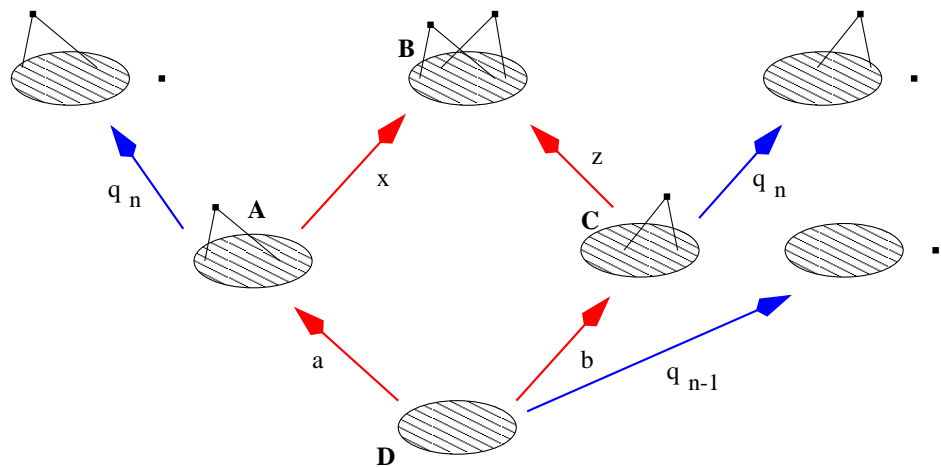}}
\caption{Consistency of remaining general covariance conditions}
\label{consis}
\end{figure}

\noindent Discrete general covariance states that the probability of forming a
causet is independent of the order in which the elements arise,
i.e. it is independent of the corresponding path through the poset of
finite causets.  
Now, general covariance relations always can be taken
to come from `diamonds' in the poset of causets, for the following
reason.  As illustrated in Figure \ref{consis}, any pair of children
of a causet (siblings) will have a common child obtained by adjoining
both new elements of the two siblings, i.e. adding to the
``grandparent'' both the new element which defines one sibling and the
new element which defines the other sibling.  (For example, consider
the case where the 2-antichain \twoach is the grandparent 
and it has the child \threeach 
(by adding a disconnected element) 
and the child \wedge
(by adding an element to the future of both elements of \twoach). 
% the 2-antichain).  
To find their common child \wedgeo add a disconnected
element to \wedge, or an element to the future of two of the elements
of \threeach.)  Now, still referring to Figure \ref{consis}, let
$|D|=n$ and suppose inductively that all the general covariance
relations are satisfied up through stage $n$.  A new condition arising
at stage $n+1$ says that some path arriving at $B$ via $x$ has the
same probability as some other path arriving via $z$.  But, by our
inductive assumption, each of these paths can be modified to go
through $D$ without affecting its probability.  Thus, the equality of
our two path probabilities reduces simply to $ax=bz$.

Now by Bell causality and lemma \ref{q's},
\be
   \frac{x}{q_n} = \frac{b}{q_{n-1}} \,,
\ee
whence
\be
   ax = ab \frac{q_n}{q_{n-1}} \,.
\ee
But by symmetry, we also have
\be
   bz = ba \frac{q_n}{q_{n-1}}  \,;
\ee
therefore $ax=bz$, as required.   $\Box$

\renewcommand{\baselinestretch}{1}
\small\normalsize
\bibliography{references}

\setlength{\parindent}{0mm}
\thispagestyle{empty}
\renewcommand{\baselinestretch}{1.5}

\large
\centerline{\textbf{CURRICULUM VIT\AE}}
\vspace{7mm}

\normalsize
NAME OF AUTHOR: David Porter Rideout\\[1mm]

PLACE OF BIRTH: Huntingdon, PA\\[1mm]

DATE OF BIRTH: April 8, 1970\\[1mm]

GRADUATE AND UNDERGRADUATE SCHOOLS ATTENDED:\\[1mm]
Syracuse University, Syracuse, NY\\
Georgia Institute of Technology, Atlanta, GA\\[1mm]

DEGREES AWARDED:\\[1mm]
Master of Science in Physics, 1995, Syracuse University\\
Bachelor of Aerospace Engineering, 1992, Georgia Institute of
Technology\\[1mm]

AWARDS AND HONORS:\\[1mm]
Sigma Xi Scientific Research Society\\
Phi Kappa Phi National Honor Society\\
Tau Beta Pi National Engineering Honor Society\\
Sigma Gamma Tau National Honor Society in Aerospace Engineering\\
Golden Key National Honor Society\\[1mm]

PROFESSIONAL EXPERIENCE:\\[1mm]
Lecturer, Department of Physics, State University of New York,
Cortland 2001, 1999\\
Graduate Research Assistant, Department of Physics, Syracuse University,
1997--2001\\
Adjunct Professor, Department of Physics, Le Moyne College, 1999, 1998\\
Teaching Assistant, Department of Physics, Syracuse University, 1992--1998\\
Engineering Intern, Robinson Industries, Zelienople, PA, 1990--1991

\end{document}